\documentclass[twocolumn,prd,floatfix,nofootinbib,longbibliography,notitlepage,superscriptaddress]{revtex4-2}
\usepackage{array}
\usepackage{autobreak}
\usepackage{epsfig}
\usepackage{euscript}
\usepackage{xcolor}
\usepackage{hyperref}
\usepackage{subfigure}
\usepackage{graphicx}
\usepackage{dcolumn}
\usepackage{bm}
\usepackage{comment}
\usepackage{float}
\usepackage[normalem]{ulem}
\usepackage{url}
\allowdisplaybreaks
\usepackage{aas_macros}
 
\usepackage{amsfonts}

\usepackage{mathrsfs}
\usepackage{orcidlink}

\allowdisplaybreaks
\allowdisplaybreaks  
\def\no{\nonumber \\}
 
\def\bes{\begin{subequations}}
\def\ens{\end{subequations}}
\newcommand{\bea}{\begin{eqnarray}}
\newcommand{\eea}{\end{eqnarray}}
\def\be{\begin{equation}}
\def\ee{\end{equation}}

\newcommand{\bs}{\begin{subequations}}
\newcommand{\es}{\end{subequations}}

\newcommand{\pd}{\partial}

\newcommand{\pp}{({\hat{\bf p}}^2)}
\newcommand{\ppn}{{\hat{\bf p}}^2}
\newcommand{\np}{({\bf{n}}\cdot\hat{\bf{p}})}

\newcommand{\cI}{\mathcal{I}}

\begin{document}

\date{\today}

\title{Generalized quasi-Keplerian solution for eccentric, non-spinning compact binaries at 4PN order and the associated IMR waveform}

\author{Gihyuk Cho\,\orcidlink{0000-0001-8813-270X}}
\email{gihyuk.cho@desy.de}
\affiliation{Deutsches Elektronen-Synchrotron DESY, Notkestr. 85, 22607 Hamburg, Germany}
\author{Sashwat Tanay\,\orcidlink{0000-0002-2964-7102}}
\email{stanay@go.olemiss.edu}
\affiliation{Department of Physics and Astronomy, The University of Mississippi, University, MS 38677, USA}
\author{Achamveedu Gopakumar}
\affiliation{Department of Astronomy and Astrophysics,
 Tata Institute of Fundamental Research, Mumbai 400005, India}
\author{Hyung Mok Lee}
\affiliation{Department of Physics and Astronomy, Seoul National University, Seoul 151-742, Korea}

\hypersetup{pdfauthor={Cho, Tanay, Gopakumar and Lee}}

\begin{abstract}
We derive fourth post-Newtonian (4PN) contributions to
the Keplerian-type parametric solution associated 
with the conservative dynamics of eccentric,
non-spinning compact binaries. The solution has been computed while ignoring
certain zero-average, oscillatory terms arising due to 4PN tail effects.
We provide explicit expressions for the parametric solution
and various orbital elements in terms of the conserved energy,
angular momentum and symmetric mass ratio.
Canonical perturbation theory
(along with the technique of Pad{\'e} approximant) is used to incorporate the 4PN
nonlocal-in-time tail effects within the action-angles framework.
We then employ the resulting solution to obtain an updated
inspiral-merger-ringdown (IMR) waveform that models the
coalescence of non-spinning, moderately eccentric black hole binaries, influenced by
Ref.~[I.~Hinder \textit{et al.}, \prd~98, 044015 (2018)].
Our updated waveform is expected to be valid over similar parameter range as the above reference.
We also present a related waveform which makes use of only the post-Newtonian equations and thus is 
valid only for the inspiral stage. This waveform is expected to work for a 
much larger range of eccentricity ($e_t \lesssim 0.85 $) than our full IMR waveform
(which assumes circularization of the binaries close to merger).
We finally pursue preliminary data analysis studies to probe the importance of including the
4PN contributions to the binary dynamics while constructing gravitational waveform templates for eccentric mergers.
\end{abstract}

\pacs{04.30.-w, 04.80.Nn, 97.60.Lf}

\maketitle

\section{Introduction}      \label{sec:Intro}

The routine detection of transient gravitational waves (GWs) from merging binary black holes (BHs) and the observations of both GWs and multi-wavelength electromagnetic radiation from
the coalescence of a neutron star binary, 
GW170817/EM170817, are establishing the era of GW astronomy
\cite{LIGOScientific:2018mvr, Abbott:2020niy, TheLIGOScientific:2017qsa,2019mbhe.confE..19P,Venu}.
More importantly, these events  allow us to do astrophysics, 
cosmology  and test general relativity with GWs \cite{Abbott:2020gyp, 2017Natur.551...85A, Abbott:2020jks}.
Further, the upcoming observational campaigns of LIGO \cite{TheLIGOScientific:2014jea}, Virgo\cite{TheVirgo:2014hva} and KAGRA \cite{Akutsu:2018axf}
 should 
 provide  astrophysical evidences for the dominant 
formation channel for the ubiquitous binary BH (BBH) events \cite{Mapelli:2018uds}.
This is because the so far observed BBH events are thought to have
originated from two distinct
formation channels \cite{2018arXiv180605820M}.
 The first scenario involves BH binaries which are formed in galactic fields via isolated binary stellar evolution 
 and therefore are expected to have 
tiny orbital eccentricities of around $10^{-4}$ when their GWs enter aLIGO frequency window \cite{2006LRR96P, 2018MNRAS.481.1908K, Kowalska:2010qg}.
The second possibility involves  
 dynamical formation of 
  BH binaries
 in dense stellar environments
and this is possible in globular clusters, young star clusters,
 and galactic nuclei \cite{2019MNRAS.488.4370F, Samsing:2017xmd, Kumamoto:2020wqr, OLeary:2008myb}. 
Interestingly, these two BBH formation scenarios lead to 
 distinct distributions for the masses and spins of binary constituents \cite{Farr:2017uvj,Sedda:2020vwo,Park_2017}.
Additionally, accurate  measurements of orbital eccentricities when BBHs enter
terrestrial GW observatory frequency windows should allow us to constrain 
their likely formation channel as dynamical formation scenarios tend to support 
non-zero orbital eccentricities \cite{Hong_2015,Samsing:2017xmd, 2018PhRvD..98l3005R}.

Ready-to-use templates that model GWs from BH binaries merging along general relativistic 
eccentric orbits are crucial to detect such events and to extract crucial astrophysical information 
they carry \cite{2019MNRAS.490.5210R, 2020arXiv201102507G}.
Indeed, there are many ongoing efforts to construct eccentric Inspiral-Merger-Ringdown (IMR) 
 waveform families, both in the frequency and time domains \cite{Hinder2017, Huerta_18, 2017PhRvD..96j4048H, 2017PhRvD..96d4028C,CN20,2019PhRvD..99l4008T, RTHH,2021arXiv210108624N,2020MNRAS.496L..64R,2020ApJ...903L...5R,2014PhRvD..90h4016H,2016PhRvD..93l4061M,2021arXiv210111033S, Huerta:2016rwp}.
It is customary to employ post-Newtonian (PN) approximation for describing the near-zone BBH inspiral
 dynamics and their far-zone inspiral GWs.
This approximation provides  general relativistic description for the 
orbital dynamics of 
 comparable mass 
compact binaries in terms of $(v/c)^2$ 
corrections to their Newtonian dynamics with $v$ and $c$ being the 
orbital speed and the speed of light in vacuum, respectively.
In this terminology, the 4PN order investigation presented in this paper deals with  
the derivation of general relativistic corrections up to the  order $(v/c)^{8}$ to the Newtonian Keplerian parametric solution. 
Further, it is customary to employ the GW phasing
approach of Refs.~\cite{Damour2004,KG06}
to describe temporally evolving PN-accurate  GW polarization states $h_{\times,+}(t)$ associated 
with compact binaries inspiralling along general relativistic eccentric orbits.
This approach is required as it provides an accurate and efficient way to incorporate
the orbital, periapsis advance and gravitational radiation reaction time scale variations, inherent in the 
dynamics of such binaries, on to $h_{\times,+}(t)$ in a PN-accurate manner.

It turns out that PN-accurate Keplerian type parametric solution is a key ingredient to implement GW phasing for eccentric inspirals.
Such solutions were presented in Refs. \cite{SW93} and \cite{MGS} at 2PN and 3PN orders, respectively.
The solution of Ref.~\cite{MGS} that solves 3PN-accurate conservative orbital dynamics, detailed in Refs.~\cite{JS_LR, LB_LR},
is crucial to model the inspiral part of two IMR families, available in Refs.~\cite{Hinder2017, Huerta_18}.
Further, the Keplerian type parametric solution is also important to compute GW emission induced secular evolution of the
orbital elements \cite{1989MNRAS.239..845B, JS92, 1997PhRvD..56.7708G, Arun2009}.
This paper computes 4PN order corrections to the 3PN-accurate 
generalized quasi-Keplerian parametric solution of Ref.~\cite{MGS}
in the Arnowitt, Deser,
and Misner (ADM) coordinates.
We have dropped certain zero-average 
oscillatory terms from our solution arising due to 4PN tail effects
due to them being relatively unimportant in GW data analysis.
It was demonstrated that the total 4PN-accurate conservative
Hamiltonian is the sum of instantaneous (local-in-time)
near-zone Hamiltonian \cite{JS_15} and time-symmetric but
nonlocal-in-time tail Hamiltonian \cite{DJS_14}.
These instantaneous contributions in the center-of-mass frame
are similar in structure to 3PN-accurate Hamiltonian,
as is evident from Eqs.~(8.40) and (8.41) of Ref.~\cite{JS_15}.
Therefore, it is reasonable to expect that this part should admit
Keplerian type  parametric solution.
Unfortunately, 4PN contributions to the time-symmetric but 
nonlocal-in-time tail Hamiltonian, given by Eq.~(8.32) of Ref.~\cite{JS_15},
do not support any closed-form exact-in-$e_t$ ($e_t$ stands for the eccentricity) expression.
Instead, we compute the 
approximate-in-$e_t$ (valid up to $e_t \lesssim 0.9$), secular dynamics 
(ignoring the oscillatory ones) of the tail effect 
by employing canonical perturbation theory.
This is reflected in corrections to the mean motion
and the periastron advance, using Pad{\'e}-like approximants.
We perform detailed consistency checks to ensure the correctness of our lengthy solution.
Further, we have verified that our local and non-local
4PN order expressions for the rate of periastron advance
is consistent with similar expressions that are present
in Refs.~\cite{Bini2020,BMMS}.

We employ our 4PN order results to obtain an improved
version of a restricted class of time-domain eccentric 
 IMR family, detailed in Ref.~\cite{Hinder2017}.
This IMR family invokes PN approximation and employs temporally evolving quadrupolar order $h_{\times,+}$   associated 
with compact binaries inspiralling along 3PN-accurate eccentric orbits while being under
the effects  of 4.5PN-accurate (or relative 2PN order)  GW damping.
As for the merger-ringdown part, we implement the approach of  Ref.~\cite{Hinder2017}
exactly (as a blackbox), explained in some detail in Sec.~\ref{subsection: stitch}.
This approach rests on identifying certain epochs during the inspiral-merger of the binary
such as $ t_\text{blend}  < t_\text{circ} < t_\text{peak}$. For $t< t_\text{blend} $,
PN equations of motion are used and for $t > t_{\text{circ}}$
we use the circular merger model (CMM) which is based on the assumption
that the binary is essentially circular for $t > t_{\text{circ}}$. The CMM model is got 
via interpolation between circular numerical relativity (NR) waveforms for different $q$ (mass ratio) values.
To build the waveform between
$t_{\text{blend}}$ and $t_{\text{circ}}$, a certain 
`blending' procedure is employed which just like the CMM model, rests on
the ideas of interpolation although of a different nature than the one employed to construct the CMM model.
Ref.~\cite{Hinder2017} used 23 NR simulations (with various mass ratios and
initial eccentricities) as a basis for these interpolations which
finally give the CMM and the blended waveforms.

We adapt the publicly available \textsc{Mathematica} package
 \cite{MMA1} of Ref.~\cite{Hinder2017} to obtain
an updated eccentric IMR family using our 4PN-order parametric solution which can be found at
Ref.~\cite{MMA2, MMA3}. Moreover, we improve their inspiral
description in a number of ways which results in a
computationally more efficient (and higher PN order accurate)
implementation of the eccentric inspiral dynamics of the binary.
This includes the use of closed-form expressions
to model orbital time scale variations and
an improved way to tackle the PN-accurate Kepler equation
(Mikkola method in conjuction with an `auxiliary eccentric anomaly').
We  also pursue preliminary data analysis implications of our approximant
to assess the importance of the 4PN order contributions to the inspiral part.

We expect our IMR waveform to be 
valid across a slightly larger parameter range than that for the waveform of
Ref.~\cite{Hinder2017},
although we have not checked it; see the
end of Sec.~\ref{4PNimr} for more details.
Apart from our main IMR  \textsc{Mathematica} package,
we also present a derived package which makes use of
only the PN equations of motion to produce the waveforms and 
hence can be trusted only for the 
inspiral part. Because this PN package has nothing to do with the CMM, it should be
valid for a much higher range of $e_t$ ($e_t \lesssim 0.85$) than our main IMR package.

The plan of the paper is as follows. In Sec.~\ref{Kep_review}, we detail the 
derivation of 4PN order generalized quasi-Keplerian parametric solution after
 giving a brief introduction to the approach. We then discuss how to incorporate the tail effect into it.
In Sec.~\ref{4PNimr}, we describe 
how we employed our parametric solution to develop an
accurate and efficient eccentric inspiral waveform and obtain its 
IMR version, influenced by Ref.~\cite{Hinder2017}.
Some preliminary data analysis explorations are also discussed.
Some of the computational details and lengthy expressions are provided in the appendices.

\textbf{Convention:} From Sec.~\ref{4PNimr} onwards, $t$ denotes the ADM 
coordinate time, whereas in Sec.~\ref{Kep_review}, it denotes the 
ADM coordinate time scaled down by a factor of $G M$,
with $G$ and $M$ standing 
for the gravitational constant and the total mass of the binary; 
more details below.

\section{Keplerian type solution at 4PN order}
\label{Kep_review}

 We begin by summarizing Keplerian type parametric solution that describes efficiently the Newtonian
 dynamics of point mass binaries in eccentric orbits and its PN-accurate extensions.
 These extensions, detailed in Refs.~\cite{DD85,DS88,SW93,MGS},
may be referred to as the 
`generalized quasi-Keplerian' parametric solution for describing 
PN-accurate orbital dynamics of compact binaries in eccentric orbits.
How we derive Keplerian type parametric solution associated 
with the 4PN-accurate  near-zone local-in-time Hamiltonian 
${\cal H}^{\rm local}_{\rm 4PN}$, given by Eq.~(8.41) of Ref.~\cite{JS_15}, is detailed in Sec.~\ref{4PN_ADM}.

\subsection{Keplerian type solution and its 3PN extensions}
\label{Sec2A}

The classical Keplerian parametric solution provides a semi-analytic 
description for the temporal evolution of a  point mass binary 
in non-circular orbits under the influence of Newtonian dynamics \cite{CM_SK}.
It other words, it provides a parametric description for the relative separation vector 
${\bf R} \equiv  R ( \cos \phi, \sin \phi, 0)$ in the usual 
center-of-mass reference frame of the binary. These angular and radial variables 
 describe the position of the reduced mass 
$\mu= m_1\,m_2/M$ around  the total mass $M= m_1 +m_2$ 
($m_1$ and $m_2$ being the individual masses). We  parametrize $R$ and $\phi$ by  
\bs
\bea                  \label{Newtonian_KE_r_phi}
r&=& a(1-e \cos u )\,,
\\
\phi -\phi_0 &=&v\equiv 2 \arctan \biggl [ \biggl ( \frac{ 1 + e}{ 1 - e}
\biggr )^{1/2} \, \tan \frac{u}{2} \biggr ]\,,
\eea
\es
where $r=R/(GM)$ and the auxiliary angles $u$ and $v$ are 
called  eccentric and true anomalies, respectively.
Further,  $a$ and $e$ denote
the semi-major axis and orbital eccentricity of the Newtonian closed orbit of 
$\mu$ around $M$.
The explicit temporal evolution for ${\bf R}$ is specified by the classical 
Kepler equation, namely 
\be                         \label{Newtonian_KE_l}
l \equiv n (t - t_0)  = u - e\,\sin u\,,
\ee
where $l$ and $n$ are usually referred to as 
the mean anomaly and 
the mean motion, respectively. Further, 
$ n = {2\,\pi}/{P}$ with $P$ being the orbital period and 
$t_0$ and $\phi_0$ stand for some initial coordinate time and initial
orbital phase, respectively. 
We reserve the symbol $t$ for the re-scaled coordinate time
$t = t'/(GM)$, where $t'$ represents the ADM time coordinate 
whose unit is seconds. The explicit expressions for these 
orbital elements $a, e$ and 
$n$ are given by
\bs
\bea
a &=& \frac{1}{{(-2\,E)}}\,,\\
e^2 &=& 1 + 2\,E\,h^2\,, \\
n&=& {{(-2\,E)}}^{3/2}\,, 
\eea
\es
where $E$ is the orbital energy 
per unit reduced mass
while the reduced angular momentum $h$ is given by $h = {J}/(G\,M)$ with 
$J$ being the orbital angular momentum per unit reduced mass. 
It is customary to employ $J_{s}(s e)$,
the Bessel functions of the first kind,
to express $u$ in terms $l$ as \cite{Colwell92}
\begin{align}
\label{eq: N_KE_Solution}
	u = l + \sum_{s =1}^{\infty} \frac{2}{s} J_{s }(s e) \sin(s\, l)\,.
\end{align}

Remarkably, it is also possible to find Keplerian type parametric solution to the
conservative orbital dynamics of compact binaries moving in relativistic orbits in the PN approximation.
It was Damour and Deruelle  who first proposed certain quasi-Keplerian 
parametrization to tackle 1PN accurate orbital dynamics of 
non-spinning compact binaries \cite{DD85}.
Thereafter, Sch\"afer and his collaborators developed the generalized 
quasi-Keplerian parametric solution to tackle both the 2PN and 3PN-accurate
orbital dynamics of compact binaries \cite{DS88,SW93,MGS}.
The fully 3PN-accurate 
generalized quasi-Keplerian parametrization for a compact binary
 in an eccentric orbit may be written as 
\begin{subequations}      \label{pn_KE}
\begin{align}                 
\label{Eq_gqkp3}
&~~~~~~~~~~~~~~~~ ~~ r = a_r \left ( 1 -e_r\,\cos u \right )\,,\\
\label{Eq_3PN_KE}
&   l   = n \left ( t' - t'_0 \right ) =
u -e_t\,\sin u +
\left ( \frac{g_{4t}}{c^4} + \frac{g_{6t}}{c^6} \right )\,(v -u)      \nonumber  \\
 & +
  \left ( \frac{f_{4t}}{c^4} + \frac{f_{6t}}{c^6} \right )\,\sin v
  + \frac{i_{6t}}{c^6} \, \sin 2\, v
  + \frac{h_{6t}}{c^6} \, \sin 3\, v \,,\\
 &   \frac{ 2\,\pi}{ \Phi}   \left (\phi - \phi_{0} \right )=   v +
\left ( \frac{f_{4\phi}}{c^4} + \frac{f_{6\phi}}{c^6} \right )\,\sin 2v              \nonumber  \\
&
+
\left ( \frac{g_{4\phi}}{c^4} + \frac{g_{6\phi}}{c^6} \right )\, \sin 3v       + \frac{i_{6\phi}}{c^6}\, \sin 4v
+ \frac{h_{6\phi}}{c^6}\, \sin 5v \,,
\end{align}
\end{subequations}
where $ v = 2\arctan   [  \sqrt{(1+e_{\phi})/(1-e_{\phi})}~ \tan (u/2)  ]$(see Ref.~\cite{MGS}).
A casual comparison of Eqs.~(\ref{pn_KE}) with Eqs.~(\ref{Newtonian_KE_r_phi}) and (\ref{Newtonian_KE_l})
reveals that at  PN orders there are three different
eccentricities which are denoted by $e_r, e_t$ and $e_{\phi}$
These radial, time and angular eccentricity parameters are introduced 
to ensure that the resulting PN-accurate parametric solution looks Keplerian
at the first post-Newtonian order
(at higher PN orders it gains additional terms as can be seen in Eqs.~(\ref{pn_KE})).
The orbital elements $a_r$ and $n$ provide certain PN-accurate semi-major axis and mean motion 
while the factor ${2\,\pi}/{\Phi}$ 
gives the angle of advance of the pericenter 
per orbital revolution. Further, we have many orbital functions 
that appear at 2PN and 3PN orders and these are denoted by 
$  g_{4t},  g_{6t}, f_{4t}, f_{6t}, i_{6t}, h_{6t},
 f_{4\phi}, f_{6\phi}, g_{4\phi}, g_{6\phi},   i_{6\phi}, $
 and $h_{6\phi}$.
 The explicit 3PN-accurate expressions for these orbital elements and functions 
 in terms of 3PN accurate orbital energy
$E$, angular  momentum $h$ and 
 the symmetric mass ratio $\eta = {\mu}/{M}$
 are provided in Ref.~\cite{MGS}.
 Additionally, Ref.~\cite{Boetzel_17} derived 3PN-accurate extension of 
Eq.~(\ref{eq: N_KE_Solution}) thereby providing a closed-form solution of the 3PN Kepler equation, namely Eq.~(\ref{Eq_3PN_KE}).
Further, Ref.~\cite{GG_18} provided hyperbolic extension of 
Eq.~(\ref{Eq_gqkp3}) to model PN-accurate GWs from compact binaries in hyperbolic passages.
We note in passing that the 1PN-accurate parametric solution is usually referred to as the 
quasi-Keplerian parameterization as it looks functionally similar to its Newtonian counterpart.
However, the parametric solutions at higher PN orders are termed the 
generalized quasi-Keplerian parameterizations and this is mainly due to the appearances of PN-accurate orbital functions 
in the expressions for $(\phi- \phi_0)$ and the Kepler equation.

These PN-accurate parametric solutions are of definite interest from observational points of view.
For example, the 1PN-accurate Keplerian type parametric solution is crucial to
operationalize the widely employed Damour-Deruelle timing formula to time relativistic binary pulsars \cite{DD_86,DT92}.
Moreover, the above  parametrization is also employed to construct  accurate and efficient GW  templates for compact binaries inspiralling along 
  PN-accurate eccentric orbits \cite{Damour2004,KG06}.
Very recently, the above solution was also invoked to model pulsar timing array residuals
induced by nano-Hz GWs from massive BH binaries in relativistic eccentric orbits \cite{SGHT_20}.
In what follows, we extend the computations of Ref.~\cite{MGS} to obtain 4PN order 
Keplerian type parametric solution for eccentric compact binaries.

\subsection{Incorporating local-in-time 4PN Hamiltonian: generalized quasi-Keplerian solution}
\label{4PN_ADM}


We employ the local-in-time part of the 4PN accurate ADM Hamiltonian for non-spinning compact binaries,
given by  Eq.~(8.41) of Ref.~\cite{JS_15}, for computing 4PN order Keplerian parametric
solution\footnote{A masterly treatise on the 
PN computations for  compact binary dynamics in the Hamiltonian approach to general relativity is
available in Ref.~\cite{JS_LR}}.
This  Hamiltonian was crucial to complete the program to derive the 4PN accurate 
compact binary dynamics 
that incorporates  all the general relativity based  $(v/c)^8$ corrections as detailed in Ref.~\cite{DJS_14,DJS_15}.
The above Hamiltonian included certain time-symmetric 
 nonlocal-in-time interactions, which
are connected to the dominant order tail effects in the 
gravitational radiation reaction \cite{Galley_15}.
The resulting compact binary dynamics extends the 3PN-accurate conservative orbital
dynamics, presented in Ref.~\cite{DJS_01},
with the help of  lengthy  4PN order computations that employ 
dimensional regularization and detailed  far-zone matching \cite{JS_12, JS_13, JS_15, DJS_14}.
Very recently, Ref.~\cite{BMMS} 
provided an independent check for the 4PN-accurate Hamiltonian of  
Sch\"afer and his collaborators by recomputing it 
within an effective field theory (EFT) approach in harmonic coordinates.
We note that there are independent efforts to obtain 4PN accurate orbital dynamics using the 
EFT approach \cite{Foffa2019, Foffa2019a} and Fokker action  computations \cite{Marchand18}.
We give below 
the local-in-time near-zone 
4PN-accurate reduced Hamiltonian in ADM-type coordinates 
and in the center-of-mass frame, given by Eq.~(8.41) of Ref.~\cite{JS_15}, as (see footnote~\ref{footnote_ord-red})
\begin{widetext}
\bea
 \label{H_3}
{\cal{H}}^{\rm local}_{\rm 4PN}({\bf r},{\hat {\bf p}})
= 
\frac{{\hat{\bf p}}^2}{2} - \frac{1}{r}
+ \frac{1}{c^2}{\cal{H}}_{1}({\bf r},{\hat{\bf p}})
+\frac{1}{c^4} {\cal{H}}_{2}({\bf r},{\hat{\bf p}})
+\frac{1}{c^6} {\cal{H}}_{3}({\bf r},{\hat{\bf p}})
+\frac{1}{c^8} {\cal{H}}_{4}({\bf r},{\hat{\bf p}})
\,,
\eea                        
where the explicit expressions for the 1PN, 2PN and 3PN 
contributions are given in Eq.~(7) of Ref.~\cite{MGS}.
This reduced Hamiltonian is connected to the 
Hamiltonian by $ {\cal{H}}^{\rm local}_{\rm 4PN}({\bf r},{\hat {\bf p}}) = {H}^{\rm local}_{\rm 4PN}({\bf r},{\hat {\bf p}}) / \mu$.
The 4PN order contributions, extracted from 
Eq.~(8.41) of Ref.~\cite{JS_15}, read
\newline 
\begin{align}
{\cal{H}}_{4}({\bf r},{\hat{\bf p}})   &= 
\bigg( \frac{7}{256} - \frac{63}{256}\eta +\frac{189}{256}\eta^2 - \frac{105}{128}\eta^3 + \frac{63}{256}\eta^4 \bigg) \pp^5
\nonumber\\[1ex]&\kern-5ex
+ \Bigg\{
\frac{45}{128}\pp^4 - \frac{45}{16}\pp^4\,\eta
+\left( \frac{423}{64}\pp^4 -\frac{3}{32}\np^2\pp^3 - \frac{9}{64}\np^4\pp^2 \right)\,\eta^2
\nonumber\\[1ex]&\kern-5ex
+ \left( -\frac{1013}{256}\pp^4 + \frac{23}{64}\np^2\pp^3 + \frac{69}{128}\np^4\pp^2
- \frac{5}{64}\np^6\pp + \frac{35}{256}\np^8 \right)\,\eta^3
\nonumber\\[1ex]&\kern-5ex
+ \left( -\frac{35}{128}\pp^4 - \frac{5}{32}\np^2\pp^3 - \frac{9}{64}\np^4\pp^2 -\frac{5}{32}\np^6\ppn
- \frac{35}{128}\np^8 \right)\,\eta^4
\Bigg\}\frac{1}{r}
\nonumber\\[1ex]&\kern-5ex
+ \Bigg\{ \frac{13}{8}\pp^3
+ \left( -\frac{791}{64}\pp^3 + \frac{49}{16}\np^2\pp^2 -\frac{889}{192}\np^4\ppn + \frac{369}{160}\np^6 \right)\,\eta
\nonumber\\[1ex]&\kern-5ex
+\left( \frac{4857}{256}\pp^3 -\frac{545}{64}\np^2\pp^2 +\frac{9475}{768}\np^4\ppn - \frac{1151}{128}\np^6 \right)\,\eta^2
\nonumber\\[1ex]&\kern-5ex
+ \left( \frac{2335}{256}\pp^3 + \frac{1135}{256}\np^2\pp^2 - \frac{1649}{768}\np^4\ppn + \frac{10353}{1280}\np^6 \right)\,\eta^3
\Bigg\}\frac{1}{r^2}
\nonumber\\[1ex]&\kern-5ex
+ \Bigg\{ \frac{105}{32}\pp^2
+ \bigglb( \left(\frac{2749\pi^2}{8192}-\frac{589189}{19200}\right)\pp^2
+ \left(\frac{63347}{1600}-\frac{1059\pi^2}{1024}\right)\np^2\ppn
+ \left(\frac{375\pi^2}{8192}-\frac{23533}{1280}\right)\np^4 \biggrb)\,\eta
\nonumber\\[1ex]&\kern-5ex
+ \bigglb( \left(\frac{18491\pi^2}{16384}-\frac{1189789}{28800}\right)\pp^2
+ \left(-\frac{127}{3}-\frac{4035\pi^2}{2048}\right)\np^2\ppn
+ \left(\frac{57563}{1920}-\frac{38655\pi^2}{16384}\right)\np^4 \biggrb)\,\eta^2
\nonumber\\[1ex]&\kern-5ex
+\left( -\frac{553}{128}\pp^2 -\frac{225}{64}\np^2\ppn -\frac{381}{128}\np^4 \right)\,\eta^3
\Bigg\}\frac{1}{r^3}
\nonumber\\[1ex]&\kern-5ex
+ \Bigg\{
\frac{105}{32}\ppn
+ \bigglb( \left(\frac{185761}{19200}-\frac{21837\pi^2}{8192}\right) \ppn
+ \left(\frac{3401779}{57600}-\frac{28691\pi^2}{24576}\right) \,\np^2 \biggrb)\,\eta
\nonumber\\[1ex]&\kern-5ex
+ \bigglb( \left(\frac{672811}{19200}-\frac{158177\pi^2}{49152}\right) \ppn
+ \left(\frac{110099\pi^2}{49152}-\frac{21827}{3840}\right) \np^2 \biggrb)\,\eta^2 \Bigg\}\frac{1}{r^4}
\nonumber\\[1ex]&\kern-5ex
+ \Bigg\{ -\frac{1}{16} + \left(\frac{6237\pi^2}{1024}-\frac{169199}{2400}\right)\,\eta 
+ \left(\frac{7403\pi^2}{3072}-\frac{1256}{45} \right)\,\eta^2 \Bigg\}\frac{1}{r^5}.
\end{align}
\end{widetext}
where $ {\bf r} = {\bf R}/(GM)$, $r= |{\bf r}|$, ${\bf n}={\bf r}/r$ and $ \hat {\bf p} = {\bf P}/\mu$;
${\bf R}$ and ${\bf P}$ are the relative separation vector and
its conjugate momentum vector respectively.
It is easy to show that the above ${\cal H}({\bf r},{\hat{\bf p}})$ admits two conserved quantities,
namely the 4PN order reduced 
energy $E={\cal H}^{\it local}_{\rm 4PN}$ and the reduced
angular momentum $\hat {\bf J} = {\bf r} \times \hat {\bf p}$
of the binary in the center-of-mass frame due to its invariance 
 under time translations and spatial rotations.
These considerations allow us to restrict the motion of our non-spinning
compact binary to a plane and employ 
polar coordinates 
such that ${\bf r} = r ( \cos \phi, \sin \phi)$.
Naturally, the relative motion follows the following differential equations 
arising from the Hamiltonian equations
\begin{subequations}       \label{H_EOM} 
\begin{align}
\dot r &= \left.{\bf n}\cdot \frac{\partial {\cal H}}{\partial\hat {\bf p}}\right.\,,\\
r^2\, \dot \phi &= \left|{\bf r} \times \frac{\partial {\cal H}}{\partial \hat {\bf p}}\right|  ,
\end{align}    
\end{subequations}
where $\dot r = dr/dt, \dot \phi = d \phi/dt $ . 
It is convenient  to introduce a variable  $s=1/r$ 
such that the $ \dot r^2$  expression at the Newtonian
order becomes a quadratic polynomial in $s$.
 In terms of $s$, we have 
$ \dot r^2 = ( d s/dt)^2/s^4 = \dot s^2/s^4$
which leads to a 4PN order expression for 
$\dot s^2$ in terms of $(-2\,E),~ h=|\hat {\bf J}|,~ \eta$
and $s$. 
This allows us to obtain PN-accurate expressions for the two  
turning points of an eccentric orbit, defined by Eqs.\ \eqref{H_EOM}.
Further, we also compute 4PN order differential equation for  
$ d \phi/ds = \dot \phi/\dot s$ using Eqs.\ \eqref{H_EOM} to tackle the 
angular part of our 4PN order Keplerian type parametric solution.

 We first focus on the 4PN order expression for $\dot r^2$
 and it turns out to be a $9$th degree polynomial in $s$.
This expression may be written  symbolically  as
\begin{align}
\dot r ^2   =   \frac{1}{s^4}
{\left({\frac{ds}{ dt}}\right)}^{2}   &= a_{{0}}+a_{{1}}\,s+
a_{{2}}\,{s}^{2}+a_{{3}}\,{s}^{3}+a_{{4}}\,{s}^{4}   +   a_{{5}}\,{s}^{5}   \nonumber     \\
    &  +a_{{6}}\,{s}
^{6}+a_{{7}}\,{s}^{7} +a_{{8}}\,{s}^{8}+a_{{9}}\,{s}^{9}  \,.
\label{Eq_4PN_rdotS}
\end{align}
The explicit  4PN order contributions to these coefficients are provided
in the accompanying \textsc{Mathematica} file \texttt{Lengthy\_Expressions.nb}~\cite{MMA2, MMA3},
while their 3PN-accurate contributions are available as Eqs.~(A1) in Ref.~\cite{MGS}.
Further, the coefficients $a_{{8}}$ and $a_{{9}}$  contain only 4PN order contributions.
To obtain parametric solution to Eq.~(\ref{Eq_4PN_rdotS}), we need to follow a couple of steps.
First, we  compute the two positive roots of the RHS of Eq.~\eqref{Eq_4PN_rdotS}
having finite limits as
$1/c \rightarrow 0$ by demanding
$\dot r^2=0$ (other roots are pushed to $\pm \infty$ in this limit). We label these 4PN order
roots as $s_-$ (pericenter) and $s_+$ (apocenter). They 
correspond to the turning points of our PN-accurate eccentric orbits
and are functions of $E$, $h$ and $\eta$.
For illustration, 
we display below their 1PN accurate 
expressions
\begin{widetext}
\begin{align}
s_{\pm}  =  \frac{1  \pm \sqrt{1+2\,h^2\,E}}{h^2}  \mp \frac{1}{c^2}\frac{\left(1 \pm \sqrt{1+2 \,h^2\,E}\right)^2
 \left[-\eta-9 \pm 2 \,(\eta -7)\,\sqrt{1+2 \,h^2\, E}+(3 \eta -1)\,\left(1+2 \,h^2\, E\right)\right]}{8\, h^4 \,\sqrt{1+2\, h^2 \,E}}      
\,.                      \label{two_roots}
\end{align}
\end{widetext} 
The explicit 4PN order expressions for these two roots are available
in the accompanying \textsc{Mathematica} notebook \texttt{Lengthy\_Expressions.nb}~\cite{MMA2, MMA3}.
We now parametrize the 4PN order 
 radial motion with the help of the following  ansatz:
\be
r = a_r \,( 1 -e_r\, \cos u),
\ee
where $a_r$ and $e_r$ are some 4PN order semi-major axis and radial eccentricity, respectively.
This ansatz allows us to express both $a_r$ and $e_r$ 
in terms of $s_-$ and $s_+$ as
\be
a_r = \frac{1}{2} \frac{s_- + s_+}{ s_-\,s_+}\,,
\hspace*{1cm}\,\, e_r = \frac{s_- -s_+}{s_-+\,s_+}\,. 
\ee
This leads in a straightforward manner to the 4PN order expressions  
for $a_r$ and $e_r^2$ in terms of $E, h$ and $\eta$.

We now move on to obtain an integral connecting $t$ and 
$s$ after factorizing the above $( ds/dt)^2/s^4$ expression using
4PN order $s_-$ and $s_+$ expressions.
The resulting 4PN order integral may be written as 
\begin{widetext}
\be
t - t_0 = \int_s^{s_-}\frac{A_0+A_1\bar s+A_2\bar s^2
+A_3\bar s^3+A_4\bar s^4+A_5\bar s^5+A_6\bar s^6+A_7\bar s^7}{\sqrt{(s_--\bar s)(\bar s-s_+)}\,\bar s^2}d \bar s \,,
\label{Eq_4PN_t_t0_N}
\ee
\end{widetext}
and how we obtain the above integral 
from our PN-accuate expression for $ds/dt$ is explained in Appendix~\ref{AppA}.
Additionally, we gather from 
the structure of the expressions of $(ds/dt)^2,~s_-$ and $s_+$ that 
the coefficients $A_i$ (with $i=1,..,7$) should be some PN-accurate 
functions of $E, h$ and $\eta$.
We now compute the radial orbital period as the 
value of the above integral between $s_-$ and $s_+$, multiplied by two.
In other words, 4PN order expression for the radial period reads
\begin{widetext}
\be
P = 2\, \int^{s_-}_{s_+}\frac{A_0+A_1\bar s+A_2\bar s^2
+A_3\bar s^3+A_4\bar s^4+A_5\bar s^5+A_6\bar s^6+A_7\bar s^7}{\sqrt{(s_--\bar s)(\bar s-s_+)}\,\bar s^2}d \bar s \,.
\label{Eq_4PN_T}
\ee
\end{widetext}
The explicit expression for $P$ will be displayed when we present 4PN order
Keplerian type solution. Note that PN-accurate mean motion $n= 2\, \pi/P$.


We now have all the necessary ingredients to obtain the 4PN order Kepler equation. This requires us to express  the mean anomaly $ l \equiv n(t -t_0) $ as a function of eccentric anomaly $u$ 
with the help of our Eqs.~(\ref{Eq_4PN_t_t0_N}) and (\ref{Eq_4PN_T})
while employing our parametric equation for $r= a_r ( 1 -e_r\, \cos u )$.
It is convenient to 
introduce an auxiliary variable 
$\tilde  v = 2 \arctan 
[    \sqrt{(1+e_r)/(1-e_r)} 
\, \tan (u/2) ]$
and with the help of a few 
 trigonometric relations
involving $\tilde v$, we obtain 
the following provisional  parametrization for $l$ as
\begin{align}
l    &  \equiv n\,(t - t_0)  =
u+ \kappa_0 \sin u+
 \frac{\kappa_1}{c^4} (\Tilde v-u)
 +\frac{\kappa_2}{c^4} \sin \Tilde v     \nonumber \\ 
 &  +\frac{\kappa_3}{c^6} \sin 2 \Tilde v
+\frac{\kappa_4}{c^6} \sin 3 \Tilde v  +\frac{\kappa_5}{c^8} \sin 4 \Tilde v+\frac{\kappa_6}{c^8} \sin 5 \Tilde v \,.
\label{Eq_t_t0_temp}
\end{align}
The steps required to obtain the above expression  from Eq.~(\ref{Eq_4PN_t_t0_N}) are 
sketched in the Appendix~B of Ref.~\cite{MGS}.
Note that these $\kappa_i$ coefficients are some PN accurate 
functions of $E,h$ and $\eta$ and they can be had from
the accompanying \textsc{Mathematica} notebook \texttt{Lengthy\_Expressions.nb}~\cite{MMA2, MMA3} and some details of the
underlying computations are 
provided in Appendices \ref{AppA} and \ref{AppB}.
We treat the above expression as a provisional one as it contains
1PN order corrections to the classical Kepler equation due to the presence 
of PN accurate $\kappa_0$ expression that multiplies $\sin u$. 
Recall that there exists 1PN-accurate Kepler equation that is structurally similar
(by ``structurally similar'' we mean having no explicit 1PN additive correction terms)
to the classical Kepler equation, obtained by invoking certain 
conchoidal transformation \cite{DD85}.
It will be desirable to keep such a structure while computing 4PN order 
Kepler equation and this will be taken up later.

We move on to tackle the angular part by first computing 
our 4PN order expression for $ d \phi/ds $ with the help of
$ d \phi/ds  = \dot \phi/ \dot s$, where 
$\dot \phi $  expression arises from the usual Hamiltonian equations of motion.
Influenced by our approach to tackle the radial motion, we 
obtain an expression for $d \phi/ds $ which involves a similar factorization
based on $s_+$ and $s_-$ as in Eqs.~\eqref{Eq_4PN_t_t0_N} and \eqref{Eq_4PN_T}.
The resulting expression may be written as
\begin{widetext}
\be
\frac{d \phi}{ds}=
\frac{B_0+B_1\, s+B_2\, s^2+B_3\,s^3+B_4\, s^4+B_5\, s^5 +B_6\, s^6+B_7\, s^7}{\sqrt{(s_-- s)( s-s_+)}}
\label{dphids_4PN}
\ee                
\end{widetext} 
where the coefficients $B_i~(i=1,...,7)$, as expected, are some
4PN order functions of $E, h $ and $\eta $ (the explicit expressions for these 
coefficients are listed in the accompanying \textsc{Mathematica} notebook).
Additionally, we sketch how to obtain 1PN-accurate $d \phi/ds$ expression 
from $ \dot \phi$ and $dt/ds$ in Appendix~\ref{AppA}.
The above equation also allows us to 
compute the amount by which periastron (or pericenter)
advances during the above computed 4PN order radial period $P$. 
This is obtained by integrating the above equation 
between our 4PN order roots
$s_+$ and $s_-$ and multiplying the result by two.
In other words, the amount of periastron advance during one radial period is
\begin{widetext}
\be
\Phi = 2\, \int^{s_-}_{s_+}
\frac{B_0+B_1\bar s+B_2\bar s^2+B_3\bar s^3+B_4\bar s^4+B_5\,\bar s^5 +B_6\, \bar s^6+B_7\, \bar s^7}{\sqrt{(s_--\bar s)(\bar s-s_+)}}d \bar s\,.
\label{Phi_3PN}
\ee
It should be obvious that the resulting 4PN order $\Phi$ expression depends on $E, h $ and $\eta $ and 
we have verified that our expression is consistent with Eq.~(20-k) in Ref.~\cite{MGS}.
We now invoke 4PN order expressions for 
$\Phi$ (got by evaluating the integral in Eq.~\eqref{Phi_3PN}) and  $d \phi/ds$ to obtain $ ( \phi - \phi_0) \times ( 2 \, \pi/ \Phi) $ which 
we symbolically write as  
\be
\frac{2\,\pi}{\Phi} \times (\phi -\phi_0 ) =\int_s^{s_-}
\frac{B'_0+B_1\bar s+B'_2\bar s^2+B'_3\bar s^3+B'_4\bar s^4+B'_5\bar s^5 +B'_6\bar s^6 +B'_7\, \bar s^7}{\sqrt{(s_--\bar s)(\bar s-s_+)}}d \bar s\,   ,
\label{phi_3PN}
\ee     
\end{widetext}   
where the primed $B'_i$ coefficients are got from the unprimed 
$B_i$'s. It is possible to evaluate the above integral with the help of certain 
trigonometric relations and  steps as detailed in Appendix~\ref{AppB}.
This results in the following provisional parametric expression 
for the 4PN order angular motion
\begin{align}
&  \frac{2\,\pi}{\Phi}\,( \phi -   \phi_0) =  \Tilde v
+\frac{\lambda_1}{c^2}\sin \Tilde v
+
\frac{\lambda_2}{c^4}\sin 2 \Tilde v
+
\frac{\lambda_3}{c^4} \sin 3\Tilde v    \nonumber \\
 &  +
\frac{\lambda_4}{c^6} \sin 4\Tilde v
+
\frac{\lambda_5}{c^6} \sin 5\Tilde v
+ 
\frac{\lambda_6}{c^8} \sin 6\Tilde v
+\frac{\lambda_7}{c^8} \sin 7\Tilde v\,   ,    
\label{temp_phi}
\end{align}
where $\lambda_i$ are some PN accurate functions, expressible
in terms of $E, h$ and $\eta$.

  Following Ref.~\cite{MGS},
   we obtain our  final parametrization for $l$ and $\phi$ equations with the help of some true anomaly variable
$ v = 2 \arctan [    \sqrt{(1+e_\phi)/(1-e_\phi)}  \tan (u/2) ]$
that involves a new angular eccentricity parameter $e_{\phi}$.
The plan is to write Eq.~(\ref{temp_phi}) in terms of $v$ rather than $\tilde{v}$
so that there are no explicit, additive $1/c^2$ corrections, while allowing 
$e_{\phi}$ to differ from $e_r$ by some yet to be determined PN corrections.
It is possible to write our $\tilde v$ in terms of $v$
 \begin{align}            
\tilde v  & =  v   +\frac{y}{c^2} \sin v
 + \frac{{y}^{2}}{4c^4}\left(-2\sin v+\sin 2v   \right)     \nonumber  \\
   &  +\frac{{y}^{3}}{12c^6}\left( 3\sin v  
-3\sin 2v +\sin 3v \right)      \nonumber  \\
   &  +  \frac{{y}^{4}}{32c^8}\left( -4\sin v+6\sin 2v-4\sin 3v+\sin 4v \right)                   \label{tildevofv}      
 \end{align}                  
where $y$ connects $e_{\phi}$ and $e_r$ to 4PN order and 
it is natural to introduce $y$ such that 
\begin{align}
y= \frac{      \sqrt{(1+e_r)/(1-e_r)}   }{  \sqrt{(1+e_\phi)/(1-e_\phi)} }-1    .
\end{align}
We now  express ${2\,\pi}
\, \left ( \phi - \phi_0 \right )/\Phi $, given by Eq.~(\ref{temp_phi}),
in terms of $v$ and demand that there are no {\it $\sin v$} terms up to 4PN 
order.
This requirement uniquely determines $y$ as a PN series which 
connects $e_{\phi}$ to $e_r$. For example, the dominant 1PN contribution of $y$ may be written as 
\be
y=-\frac{\eta  \sqrt{1+2\,E\,h^2}}{2\,c^2\, h^2}+{\cal O}\left(\frac{1}{c^4}\right)\,.
\ee
It should be noted that we imposed such a restriction because 1PN-accurate parametric 
solution, derived in Ref.~\cite{DD85}, supported a {\it Keplerian} like 
parametrization for the angular part with the help of $v$.
This leads to the following parametric solution 
for the angular motion while incorporating 4PN order contributions:
\begin{widetext}
\begin{align}
\frac{2\,\pi}{\Phi}\,( \phi -\phi_0 )&= v +
\left ( \frac{f_{4\phi}}{c^4} + \frac{f_{6\phi}}{c^6} + \frac{f_{8\phi}}{c^8} \right )\,\sin 2v
+
\left ( \frac{g_{4\phi}}{c^4} + \frac{g_{6\phi}}{c^6} + \frac{g_{8\phi}}{c^8}\right )\, \sin 3v     \nonumber  \\
& + \left(\frac{i_{6\phi}}{c^6}+\frac{i_{8\phi}}{c^8}\right)\, \sin 4v
+  \left(\frac{h_{6\phi}}{c^6}+\frac{h_{8\phi}}{c^8}\right)\, \sin 5v   +\frac{k_{8\phi}}{c^8}\,\sin 6\,v+\frac{j_{8\phi}}{c^8}\,\sin7\,v\,,
\label{fin_phi}
\end{align}   
\end{widetext}
where $ v = 2 \arctan [    \sqrt{(1+e_\phi)/(1-e_\phi)}  \tan (u/2) ]$.
Interestingly, the contributions  
at 2PN, 3PN and 4PN orders are 
supplemented by other trigonometric functions of $v$ and
this is why we term the resulting solution 
as the generalized quasi-Keplerian parametric solution.
We will display shortly 
the explicit 4PN order expressions for these orbital elements and functions.

 We now move to finalize the provisional expression for our 4PN order Kepler equation,
 given by Eq.~(\ref{Eq_t_t0_temp}). The idea is to express $\tilde v$ in terms of $v$ 
with the help of the above listed PN-accurate relation of Eq.~(\ref{tildevofv}).
This leads to the following Kepler equation that includes 4PN order contributions
in terms of $u, e_t$, $v(u)$ and its trigonometric functions as 
\begin{widetext}
\begin{align}
l  =  n\,(t -t_0) &= u -e_t\,\sin u +
\left ( \frac{g_{4t}}{c^4} + \frac{g_{6t}}{c^6} + \frac{g_{8t}}{c^8}  \right )\,(v -u)
  + \left ( \frac{f_{4t}}{c^4} + \frac{f_{6t}}{c^6}+ \frac{f_{8t}}{c^8} \right )\,\sin v        \nonumber  \\
&  + \left(\frac{i_{6t}}{c^6}+\frac{i_{8t}}{c^8}\right) \, \sin 2\, v
  + \left(\frac{h_{6t}}{c^6}+\frac{h_{8t}}{c^8}\right)\, \sin 3\, v    +\frac{k_{8t}}{c^8}\,\sin 4\,v+\frac{j_{8t}}{c^8}\,\sin 5\,v\,.
\label{K_Eqn_4PN}
\end{align} 
The PN accurate expressions for $n, e_t$ and the orbital functions 
appearing in the above PN-accurate Kepler equation will be listed below.

We now have all the parts to display, in its entirety, the fourth post-Newtonian order
generalized quasi-Keplerian parametrization for an eccentric compact binary
 in ADM-type coordinates as
\begin{align}
r &=    a_r \left ( 1 -e_r\,\cos u \right )\,,    \label{e:FinalParamADM_1}   \\
l   =   n \left ( t - t_0 \right )   &  =
u -e_t\,\sin u +
\left ( \frac{g_{4t}}{c^4} + \frac{g_{6t}}{c^6} + \frac{g_{8t}}{c^8}  \right )\,(v -u)
  +  \left ( \frac{f_{4t}}{c^4} + \frac{f_{6t}}{c^6}+ \frac{f_{8t}}{c^8} \right )\,\sin v           \nonumber  \\
   &    +   \left(\frac{i_{6t}}{c^6}+\frac{i_{8t}}{c^8}\right) \, \sin 2\, v  + \left(\frac{h_{6t}}{c^6}+\frac{h_{8t}}{c^8}\right)\, \sin 3\, v 
  +\frac{k_{8t}}{c^8}\,\sin 4\,v+\frac{j_{8t}}{c^8}\,\sin 5\,v\,,    \label{e:FinalParamADM_2}  \\
\frac{ 2\,\pi}{ \Phi} \left (\phi - \phi_{0} \right )
&= v +
\left ( \frac{f_{4\phi}}{c^4} + \frac{f_{6\phi}}{c^6} + \frac{f_{8\phi}}{c^8} \right )\,\sin 2v
+
\left ( \frac{g_{4\phi}}{c^4} + \frac{g_{6\phi}}{c^6} + \frac{g_{8\phi}}{c^8}\right )\, \sin 3v               \nonumber  \\
   &  + \left(\frac{i_{6\phi}}{c^6}+\frac{i_{8\phi}}{c^8}\right)\, \sin 4v
+  \left(\frac{h_{6\phi}}{c^6}+\frac{h_{8\phi}}{c^8}\right)\, \sin 5v \,    +\frac{k_{8\phi}}{c^8}\,\sin 6\,v+\frac{j_{8\phi}}{c^8}\,\sin7\,v\,,         \label{e:FinalParamADM_3}
\end{align} 
where  $ v = 2 \arctan [    \sqrt{(1+e_\phi)/(1-e_\phi)}  \tan (u/2) ]$.
In what follows, we display 
the 4PN order expressions for the orbital elements 
$a_r,n, \Phi,$ 
and the post-Newtonian orbital
functions that appear at 2PN, 3PN and 4PN orders 
in terms of the conserved quantities: 
\bes
\label{e:CoeffKP}
\bea
a_r &=&  \frac{1}{{(-2\,E)}}\bigg\{ 1+\frac{ (-2\, E )}{4\,c^2} \left( -7+\eta \right) +
\frac{{{ (-2\, E) }}^{2}}{16 c^4}\,\bigg[ 
(1+10\,\eta+{\eta}^{2})
 \no&&
 +\frac {1}{(-2\,E\,h^2)}
(  -68+44\,\eta)
\bigg] 
+{\frac {{{ (-2\,E) }}^{3}}{192\,c^6}}\, 
\biggl [ 
3-9\,\eta-6\,{\eta}^{2}
\no&&
+3\,{\eta}^{3}+\frac{1}{(-2\,E\,h^2)}
\biggl (
864+ \left( -3\,{\pi}^{2}-2212 \right) \eta+432\,{\eta}^
{2}\biggr)
\no 
&&
+
\frac{1}
{ (-2\,E\, h^2)^2 }
\biggl (
-6432+ \left( 13488-240\,{\pi}^{2} \right) \eta
-768\,{\eta}^{2}\biggr )
\biggr ]
\no&& 
+\frac{(-2\,E)^4}{3686400\, c^8}\,\biggr[14400 - 57600 \,\eta + 28800 \,\eta^2 - 158400 \,\eta^3 + 14400 \,\eta^4\no&&
+\frac{1}{(-2\,E\,h^2)}\biggl(-4147200 + (-38071488 + 1280250 \pi^2) \,\eta 
\no&&
+ \left(19038208 +  4030875\, \pi^2\right) \,\eta^2 + 4262400  \,\eta^3\biggr)
\no&&
+\frac{1}{(-2\,E\,h^2)^2}\,\biggl(316800000 + \left(-661398528 + 21132000 \pi^2\right)\,\eta 
\no&&
+ \left(363371776 - 
    26908200 \pi^2\right)\,\eta^2 - 20160000 \,\eta^3\biggr)
\no&&
+\frac{1}{(-2\,E\,h^2)^3}\,\biggl(-1228492800 + \left(2644664832 - 59785200\, \pi^2\right) \,\eta
\no&&
+ \left(-826707456 + 
    34613400 \,\pi^2\right)\,\eta^2 + 13824000 \,\eta^3\biggr)   \biggr]
\bigg\}\,,
\\
n&=&{{(-2\,E)}}^{3/2} \bigg\{ 1+{\frac {{(-2\,E)}} {8\,{c}^{2}}}\, 
\left( -15+\eta \right)
+{\frac {{{(-2\,E)}}^{2} }{128{c}^{4}}} 
\biggl [ 555 
+30\,\eta
\no&&
+11\,{ \eta}^{2}
+ \frac{192}{ \sqrt{(-2\,E\,h^2)}}
\left( -5+2\,\eta \right )
\biggr ]
+
{\frac {{{(-2\,E)}}^{3}}{3072\,{c}^{6}}}
\biggl [  -29385 
\no&&
-4995\,\eta-315\,{\eta}^{2}+135
\,{\eta}^{3}
-
\frac {16}{( -2\,E\,h^2)^{3/2}}
\bigg(
10080+123\,\eta\,{\pi}^{2}
\no&&
-13952\,\eta+1440\,{\eta}^{2}\bigg)
+ \frac{5760}{ \sqrt{ (-2\,E\,h^2)}}
\left(17 -9\,\eta+2\,{\eta}^{2} \right )
\biggr ] 
\no&&
+\frac{(-2\,E)^4}{1474560 \,c^8}\,\biggl[\,\frac{3317760\, (-5+2 \,\eta )^2}{-2\,E\,h^2}+\frac{138240}{\sqrt{(-2\,E\,h^2)}}\,\no&&\times \left(-1125+550\, \eta -175\, \eta ^2+38\, \eta ^3\right)+135\, \biggl(232881\no&&
+65300 \,\eta +4070 \,\eta ^2-460 \,\eta ^3+241\, \eta ^4\biggr)\no&&
-\frac{80}{(-2\,E\,h^2)^{3/2}}\,\biggl(-5443200+\left(10467328-150987 \pi ^2\right) \,\eta\no&&
+\left(-3959808+32472 \pi ^2\right) \,\eta ^2+311040\, \eta ^3\biggr)\no&&
+\frac{48}{(-2\,E\,h^2)^{5/2}}\biggl(-17297280+\left(37556864-771585 \,\pi ^2\right) \,\eta\no&&
+\left(-13464960+236160\, \pi ^2\right) \,\eta ^2+403200\, \eta ^3\biggr)\biggr]
\biggr\} \,,              \label{eqn_n} 
\\
g_{{4\,t}} &=&\frac{3\,(-2\,E)^{2}}{2}\,\biggl \{ \frac{5 -2\,\eta }{ \sqrt{ (-2\,E\,h^2)}}
 \biggr \}\,, 
\\
g_{{6\,t}} &=&{\frac {{{(-2\,E)}}^{3}}{192}}
\biggl \{
\frac{1}{( -2\,E\,h^2)^{3/2} }
\bigg(
10080+123\,\eta\,{\pi }^{2}-13952\,\eta
\no&&
+1440\,{\eta}^{2}
\bigg)
+ \frac{1}{\sqrt{(-2\,E\,h^2)}}
\left ( -3420
+1980\,\eta-648\,{\eta}^{2}
\right )
\biggr \}\,,
\\ 
g_{{8\,t}} &=&-{\frac {{{(-2\,E)}}^{4}}{92160}}
\biggl \{
\frac{3}{( -2\,E\,h^2)^{5/2} }
\bigg(
-17297280+\big(37556864\no&&-771585 \pi ^2\big) \eta
+1920 \left(-7013+123\,\pi ^2\right) \eta ^2+403200 \eta ^3
\bigg)\no&&
-\frac{5}{( -2\,E\,h^2)^{3/2} }\,
\bigg(
-3628800+\left(7835008-128847 \pi ^2\right) \eta \no&&
+36 \left(-98144+861 \pi ^2\right) \eta ^2+293760 \eta ^3
\bigg)
+\frac{207360}{( -2\,E\,h^2)}\,(5-2\,\eta)^2\,
\no&&
+\frac{1080}{\sqrt{( -2\,E\,h^2)}}\,(-3375+1600 \eta -755 \eta ^2+246 \eta ^3)\,
\biggr \}\,,
\\ 
f_{{4\,t}} &=& -\frac{1}{8}\,
\frac{ (-2\,E)^2}{ \sqrt{ (-2\,E\,h^2)}}
\biggl \{ (4 + \eta)\,\eta \, \sqrt{(1 +2\,E\,h^2)}
\biggr \}\,,
\\
f_{{6\,t}}&=&{\frac {{{(-2\,E)}}^{3}}{192}}
\bigg\{
\frac{1}{(-2\,E\,h^2)^{3/2} } \,
\frac{1}{ \sqrt{1 +2\,E\,h^2 }}
\bigg(
1728-4148\, \eta +3\,\eta\,{\pi }^{2}
\no&&
+600\,{\eta}^{2}+33\,{\eta}^{3}\bigg)
+3\,
\frac{\sqrt{(-2\,E\,h^2)}}{ \sqrt{ ( 1+2\,E\,h^2)}}
\eta\, \left(-64-4\,\eta
+ 23\,{\eta}^{2} \right)
\no&& 
+ \frac{1}{ \sqrt{ ( -2\,E\,h^2)\,(1 +2\,E\,h^2 )}}
\biggl ( 
-1728
+
4232\,\eta-3\,\eta\,{\pi }^{2}
\no&&
-627\,{\eta}^{2}-105\,{\eta}^{3}
\biggr )
\bigg\}\,,
\\
f_{{8\,t}}&=&-\frac {(-2\,E)^4}{14745600}\,\frac{(-2\,E\,h^2)^{3/2}}{(1+2\,E\,h^2)^{3/2}}\,
\bigg\{7200\, \eta\,  \big(4672+912\, \eta \no&&
-303\, \eta ^2+902 \,\eta ^3\big)+\frac{2764800}{\sqrt{(-2\,E\,h^2)}}\,\eta \, (4+\eta )\, (-5+2 \eta )\no&&
+\frac{1}{(-2\,E\,h^2)}\biggl(331776000+1350 \left(-919776+2377 \pi ^2\right) \eta\no&&
+\left(568404992+2468925 \pi ^2\right) \eta ^2-94248000 \eta ^3-16128000 \eta ^4\biggr)\no&&
-\frac{5529600}{(-2\,E\,h^2)^{3/2}}\,\eta \, (4+\eta )\, (-5+2 \eta )\no&&+\frac{1}{(-2\,E\,h^2)^{2}}\,
\biggl(-2226585600+\left(10348301504-252478050 \pi ^2\right) \eta \no&&
+9 \left(-614377024+9064225 \pi ^2\right) \eta ^2+383328000 \eta ^3+10411200 \eta ^4\biggr)\no&&
+\frac{2764800}{(-2\,E\,h^2)^{5/2}}\,\eta  \,\left(-20+3 \eta +2 \eta ^2\right)\no&&
+\frac{1}{(-2\,E\,h^2)^{3}}\,  \biggl(3607142400+2 \left(-8729633504+247794225 \,\pi ^2\right) \eta\no&&
+\left(9340505856-170534025 \,\pi ^2\right) \,\eta ^2-471441600 \,\eta ^3+1152000\, \eta ^4\biggr)\no&&
+\frac{1}{(-2\,E\,h^2)^{4}} \,\biggl(-1712332800+\left(8314359104-246319350 \pi ^2\right) \eta\no&& +\left(-4388287232+86487075 \pi ^2\right) \eta ^2+184226400 \eta ^3-1944000 \eta ^4\biggr)
\bigg\}\,,
\\
h_{{6\,t}} &=&\frac{{{(-2\,E)}}^{3}}{32}\,\eta
\biggl \{ 
\frac{(1 +2\,E\,h^2)}{ (-2\,E\,h^2)^{3/2} }
\left(23+12\,\eta+ 6\,{\eta}^{2} \right) 
\biggr \}\,,
\\
h_{{8\,t}} &=&\frac{{{(-2\,E)}}^{4}}{921600}
\biggl \{ 
-\frac{300}{\sqrt{-2\,E\,h^2}}\,\eta \, \biggl(-8904+12207 \eta +2356 \eta ^2+864 \eta ^3\biggr)\no&&+\frac{1}{(-2\,E\,h^2)^{3/2}}\,\biggl(-1857600+\left(10986256-1072425 \pi ^2\right) \eta \no&&
+\left(-38708632+3152775 \pi ^2\right) \eta ^2+4891200 \eta ^3+176400 \eta ^4\biggr)\no&&+\frac{1}{(-2\,E\,h^2)^{5/2}}\,\biggl(1857600+\left(-12167056+1072425 \pi ^2\right) \eta\no&&
+\left(43313932-3152775 \pi ^2\right) \eta ^2-3709200 \eta ^3+126000 \eta ^4\biggr)
\biggr \}\,,
\\
i_{{6\,t}} &=&
{\frac {13\,{{(-2\,E)}}^{3}}{192}}
\eta^3
\biggl ( 
\frac{  1 + 2\,E\,h^2 }{ -2\,E\,h^2}  
\biggr )^{3/2}\,,
\\
i_{{8\,t}} &=&
{\frac {{{(-2\,E)}}^{4}}{14745600}}\,\biggl ( 
\frac{  1 + 2\,E\,h^2 }{ -2\,E\,h^2}  
\biggr )^{1/2}\,
\eta\,
\,\biggl \{
-3600\, \eta ^2 \,(-839+526\, \eta )\no&&
+\frac{1}{(-2\,E\,h^2)}\biggl(6 \,\left(-9586592+405075\, \pi ^2\right)\no&&
+5 \,\left(-21746432+2673315\, \pi ^2\right) \,\eta +23416800\, \eta ^2+1368000 \,\eta ^3\biggr)\no&&
+\frac{1}{(-2\,E\,h^2)^2}\biggl(57519552-2430450\, \pi ^2\no&&+\left(108732160-13366575 \,\pi ^2\right)\, \eta -23067600\, \eta ^2
+900000\, \eta ^3\biggr)
\biggr\}\,,
\\
k_{{8\,t}} &=&
-\frac{(-2\,E)^{3/2} \,\eta\, (150 \eta^3+2444 \eta^2-3303 \eta+516) (1+2\,E\,h^2)^2}{6144 h^5}\,,
\\
j_{{8\,t}} &=&
-\frac{(-2\,E)^{3/2} \eta^3 (66 \eta-25) \left(1+2\,E\,h^2\right)^{5/2}}{4096 h^5}\,,
\\
\Phi&=&2\,\pi \, \bigg\{ 1+{\frac {3}{{c}^{2}{h}^{2}}}+
\frac{{{(-2\,E)}}^{2}}{4\,{c}^{4}}
\biggl [  \frac{3}{(-2\,E\,h^2)} \left ( -5+2\,\eta \right )
\no&&
+ \frac{15}{(-2\,E\,h^2)^2} \left ( 7 -2\,\eta \right )
\biggr ] 
+{\frac {\,{{(-2\,E)}}^{3}}{128\,{c}^{6}}}
\biggl [  
 \frac{24}{ (-2\,E\,h^2)}
( 5 -5\eta 
\no&&
+ 4\eta^2)
- \frac{1}{ (-2\,E\,h^2)^2}
\biggl ( 10080
 -13952\,\eta+123 \,\eta\,{\pi }^{2}+1440\,{\eta}^{2}
\biggr )
\no&&
+ \frac{5}{(-2\,E\,h^2)^3}
\biggl (7392-8000\,\eta+  123\,\eta\,{\pi }^{2}
+ 336\,{\eta}^{2} \biggr  )
\biggr ]\no&&
-\frac{(-2\,E)^4}{73728\,c^8}\,\frac{1}{(-2\,E\,h^2)}\biggl[-6912 \,\eta ^2\, (-5+4 \eta )\no&&
+\frac{3}{(-2\,E\,h^2)}\biggl(-1814400+\left(5202688-106707 \pi ^2\right) \eta\no&&
+240 \left(-12944+123 \pi ^2\right) \eta ^2+276480 \eta ^3\biggr)\no&&
-\frac{6}{(-2\,E\,h^2)^2}\biggl(-17297280+\left(37556864-771585 \pi ^2\right) \eta\no&&
+1920 \left(-7013+123 \pi ^2\right) \eta ^2+403200 \eta ^3\biggr)\biggr]\no&&
+\frac{7}{(-2\,E\,h^2)^3}\biggl(-37065600+\left(63502592-1275315 \pi ^2\right) \eta\no&& +2400 \left(-6056+123 \pi ^2\right) \eta ^2+207360 \eta ^3\biggr)\biggr]
 \bigg\}\,,     \label{QKP-for-Phi}
\\
f_{{4\,\phi}} &=&
\frac{{{(-2\,E)}}^{2}}{8}
\,\frac{( 1+2\,E\,h^2)}{(-2\,E\,h^2)^2}\,
\eta \, (1 -3\,\eta)\,,
\\
f_{{6\,\phi}}&=&{\frac {{{(-2\,E)}}^{3}}{256}}
\bigg\{
\frac{4\,\eta}{(-2\,E\,h^2)}
\left( -11-40\,\eta+24\, {\eta}^{2} \right)
\no&& 
+ \frac{1}{{(-2\,E\,h^2)}^2} \biggl (
-256
+1192\,\eta-49\,\eta\,{\pi }^{2}
+336\,{\eta}^ {2}
-80\,{\eta}^{3}
\biggr )
\no&&
+ \frac{1}{(-2\,E\,h^2)^3}
\biggl ( 
256+49\,\eta\,{\pi }^{2}-1076\,\eta-384\,{\eta}^{2}-40\,{\eta}^{3}
\biggr )
\bigg\}\,,
\\
f_{{8\,\phi}}&=&{\frac {{{(-2\,E)}}^{4}}{7372800}}\,
\bigg\{
\frac{900\,\eta}{(-2\,E\,h^2)^{}}\,\biggl(6844-13989 \eta -1530 \eta ^2+1888 \eta ^3\biggr)\no&&
+\frac{1}{(-2\,E\,h^2)^{2}}\,\biggl(9273600+2 \left(-303923464+7907025 \pi ^2\right) \eta \no&&
+\left(567130588+8219475 \pi ^2\right) \eta ^2-26411400 \eta ^3+1180800 \eta ^4\biggr)\no&&
-\frac{2}{(-2\,E\,h^2)^{3}}\,\biggl(84844800+10 \left(-149381636+4263405 \pi ^2\right) \eta \no&&
-19 \left(-67975466+173325 \pi ^2\right) \eta ^2-54814500 \eta ^3+2980800 \eta ^4\biggr)\no&&
\frac{1}{(-2\,E\,h^2)^{4}}\,\biggl(177004800+\left(-2446310192+72629250 \pi ^2\right) \eta \no&&
-15 \left(-132716108+963535 \pi ^2\right) \eta ^2-86679000 \eta ^3+1929600 \eta ^4\biggr)
\bigg\}\,,
\\
g_{{4\,\phi}} &=&
-{\frac {3{{(-2\,E)}}^{2}}{32}}
\frac{\,\eta^2\,}{(-2\,E\,h^2)^2}
 ( 1 +2\,E\,h^2)^{3/2}\,,
\\
g_{{6\,\phi}}&=&
\frac{ (-2\,E)^3}{768}\, \sqrt{(1 +2\,E\,h^2)}\,
\bigg\{
-\frac{3}{ (-2\,E\,h^2)}\,\eta^2\, \left( 9-26 \,\eta \right )
\no   &&
- \frac{1}{(-2\,E\,h^2)^2} \,\eta
\biggl ( 220+3\,{\pi }^
{2}
+312\,\eta+150\,{\eta}^{2} \biggr )
\no&&
+ \frac{1}{(-2\,E\,h^2)^3}\,\eta
 \left( 220+3\,{\pi }^{2}+96\,\eta+45\,{\eta}^{2}
\right )
\bigg\}\,,
\\
g_{{8\,\phi}}&=&
\frac{ (-2\,E)^4}{176947200 }\, \frac{1}{\sqrt{(1 +2\,E\,h^2)}}\,
\bigg\{-10800 \eta ^2 \left(36-95 \eta +1226 \eta ^2\right)\no&&
+\frac{3\,\eta}{(-2\,E\,h^2)}\,\biggl(-404533824+5453550 \pi ^2\no&&+\left(731023360+381825 \,\pi ^2\right) \eta 
-115070400 \,\eta ^2-9129600\, \eta ^3\biggr)\no&&
+\frac{1}{(-2\,E\,h^2)^2}\,\biggl(44236800+2 \left(-6842155424+127907475 \pi ^2\right) \eta \no&&
-87 \left(-253902848+3210525 \,\pi ^2\right) \,\eta ^2\no&&
-2262477600\, \eta ^3+39096000 \,\eta ^4\biggr)\no&&
+\frac{1}{(-2\,E\,h^2)^3}\,\biggl(-88473600+\left(23556745280-464880750 \,\pi ^2\right) \eta\no&&
+\left(-37649997312+561808575 \,\pi ^2\right) \eta ^2\no&&+3436488000 \eta ^3-103766400 \,\eta ^4\biggr)\no&&
+\frac{1}{(-2\,E\,h^2)^4}\,\biggl(44236800+\left(-11086035904+225426450 \pi ^2\right) \eta 
\no&&+\left(17722415616-281347425 \pi ^2\right) \eta ^2\no&&-1527246000 \eta ^3+50133600 \eta ^4\biggr)
\bigg\}\,,
\\
i_{{6\,\phi}} &=&{\frac {{{(-2\,E)}}^{3}}{128}}
\,\frac{{(1 +2\,E\,h^2)}^2}{(-2\,E\,h^2)^3}\,\eta 
\left( 5+28\,\eta+10\,{\eta}^{2} \right)\,,
\\
i_{{8\,\phi}} &=&{\frac {{{(-2\,E)}}^{4}}{14745600}}
\,\frac{\sqrt{(1 +2\,E\,h^2)}}{(-2\,E\,h^2)^2}\,\biggl\{
-7200 \left(440-1330 \eta +700 \eta ^2+173 \eta ^3\right)\no&&
+\frac{1}{(-2\,E\,h^2)}\biggl(175308224+1767300 \pi ^2\no&&
+\left(-407514720-9062175 \pi ^2\right) \eta+70257600 \eta ^2-1713600 \eta ^3\biggr)\no&&
+\frac{1}{(-2\,E\,h^2)^2}\biggl(-169548224+1767300 \pi ^2\no&&
+\left(412741920-9062175 \pi ^2\right) \eta 
-58420800 \eta ^2+3535200 \eta ^3\biggr)
\biggr\}\,,
\\
h_{{6\,\phi}} &=&
\frac{5\, (-2\,E)^3}{256}\, \frac{\eta^3}{ (-2\,E\,h^2)^3}
\, (1 +2\,E\,h^2)^{5/2} \,,
\\
h_{{8\,\phi}} &=&
\frac{ (-2\,E)^4}{6553600}\, \frac{\eta}{ (-2\,E\,h^2)^2}
\, (1 +2\,E\,h^2)^{3/2} \,\biggl\{78000 \eta ^2\no&&
-172000 \eta ^3+\frac{1}{(-2\,E\,h^2)}\,\biggl(8273856+11250 \pi ^2\no&&
+\left(-24254464-579825 \pi ^2\right) \eta +7604000 \eta ^2-238400 \eta ^3\biggr)\no&&
+\frac{1}{(-2\,E\,h^2)^2}\,\biggl(-8273856+11250 \pi ^2\no&&
+\left(24254464-579825 \pi ^2\right)\, \eta -6962000 \,\eta ^2+490400 \,\eta ^3\biggr)
\biggr\}\,,
\\
k_{{8\,\phi}} &=&
-\frac{\eta\left(150 \eta^3+4154 \eta^2-5755 \eta+1476\right) \left(1+2\,E\, h^2 \right)^3}{24576\, h^8}\,,
\\
j_{{8\,\phi}} &=&
-\frac{35 (2 \eta-1)\,\eta^3 \left(1+2\,E\, h^2\right)^{7/2}}{16384 \,h^8}\,.
\eea
\ens 
Appendix~\ref{AppC} provides 
the explicit expressions for $e_t$
and 4PN-order relations that connect 
$e_r$ and $e_\phi$ to $e_t$. This is influenced by the GW phasing approach
that usually employs the time eccentricity to characterize PN-accurate 
eccentric orbits. Borrowing from Ref.~\cite{Damour2004}, we refer to as `phasing',
the task of specifying the time dependencies $r(t), \dot{r}(t), \phi(t)$ and $\dot{\phi}(t)$.
In the limit of circular motion (\textit{i.e.} $e_t=e_r=e_\phi=0$), we display relations between some gauge invariant quantities
\begin{subequations}
\begin{align}
\begin{autobreak}
\MoveEqLeft
(-2\,E)\,|_\text{circ}
   =  \frac{1}{h^2}
+\frac{\eta+9}{4\,c^2\, h^4}
+\frac{\eta^2-7 \,\eta+81 }{8\,c^4\, h^6}
+\frac{15 \,\eta^3-30 \,\eta^2+246 \pi ^2 \,\eta-8833 \,\eta+11583}{192\,c^6\, h^8}
+\frac{420 \,\eta^4-360 \,\eta^3-20 \left(246 \pi ^2-8875\right) \,\eta^2+\left(98715 \pi ^2-3959644\right) \,\eta+3222180}{7680 \,c^8\,h^{10}}\,,
\end{autobreak}
\end{align}
\begin{align}
\begin{autobreak}
\MoveEqLeft
h\,|_\text{circ}
=\frac{1}{c \sqrt{x}}\,\Big(1+\frac{1}{6} (\,\eta+9) x
+\frac{1}{24} \left(\,\eta^2-57 \,\eta+81\right) x^2
+\frac{\left(7 \,\eta^3+1674 \,\eta^2+2214 \pi ^2 \,\eta-62001 \,\eta+10935\right) x^3}{1296}
+\frac{\left(-220 \,\eta^4-15480 \,\eta^3-487080 \pi ^2 \,\eta^2+12817260 \,\eta^2-522855 \pi ^2 \,\eta+2666412 \,\eta+2755620\right) x^4}{124416}\Big)\,,
\end{autobreak}
\end{align}
\begin{align}
\begin{autobreak}
\MoveEqLeft
n|_\text{circ}
=\frac{1}{h^3}
+\frac{(\,\eta+3) }{2 \,c^2\,h^5}
+\frac{3 \left(\,\eta^2+5 \,\eta+9\right)}{8 \,c^4\,h^7}
+\frac{\left(30 \,\eta^3-168 \,\eta^2+\left(970+123 \pi ^2\right) \,\eta-486\right) }{96\,c^6\, h^9}
+\frac{\left(2520 \,\eta^4-3888 \,\eta^3+24 \left(1968 \pi ^2-74129\right) \,\eta^2+\left(4753624-53571 \pi ^2\right) \,\eta-2221992\right) }{9216\,c^8\, h^{11}}\,,
\end{autobreak}
\end{align}
\begin{align}
\begin{autobreak}
\MoveEqLeft
\frac{\Phi}{2\,\pi}|_\text{circ}
=1+\frac{3}{c^2\,h^2}+\frac{(45-12 \,\eta)}{2\,c^4\, h^4}
+\frac{96 \,\eta^2+123 \pi ^2 \,\eta-6464 \,\eta+6480}{32 \,c^6\,h^6}
+\frac{-768 \left(123 \pi ^2-4958\right) \,\eta^2+\left(557301 \pi ^2-27133696\right) \,\eta+18195840}{9216\,c^8\,h^8}\,,
\end{autobreak}
\end{align}
\end{subequations}
\end{widetext}
where $x= ( \Phi\, n/(2 \pi c^3) )^{2/3}$.
As mentioned ealier, we have verified that our
$(-2\,E)\,|_\text{circ}$ and $ (\Phi/(2 \pi))|_\text{circ}$ are matched with the expression $E^\text{loc,f,circ}(j)$ from Eq.~(8.26) and $K^\text{loc,f,circ}(j)$ from Eq.~(8.29) of Ref.~\cite{Bini2020}, respectively. 
We also tried to verify the relation between $x$ and $h$ in the circular limit as given in Ref.~\cite{Bini2020}.
Unfortunately, it turned out to be rather difficult to do so as 
 the local and nonlocal parts are not separated in Ref.~\cite{Bini2020}.

Clearly, it is important to go through a consistency check in order to
 ensure the correctness of these lengthy expressions for the 
4PN order orbital elements and functions.
The plan is to adapt two consistency checks, detailed in Ref.~\cite{MGS}.
This requires  us to express PN-accurate expressions for 
$\dot r^2 $ and
$\dot \phi^2$, derived using the Hamiltonian equations of motion and given
by Eqs.\ (\ref{H_EOM}), in terms of $E, h, \eta$ and $(1-e_r\,\cos u)$ 
while using the fact that  $ r=a_r\, ( 1-e_r\, \cos u)  $.
Note that the expressions for $a_r$ and $e_r^2$ were obtained 
from the PN accurate roots $s_-$ and $s_+$, and therefore, do  
not involve any of our complicated integrals.
In the first part of our check, we compare such an expression 
with the one that explicitly employed our parametric solution, namely
$\dot r^2 = \left ( dr/du\, \times du/dt \right )^2$.
This expression for $\dot r^2$ is
found to be in full agreement 
with our earlier $\dot r^2$ expression
up to 4PN order after some elaborate simplifications.
Thereafter, we performed a similar check on the angular part by computing 
${\dot \phi}^2 = {\left ( {d \phi}/{dv}\, \times {d v}/{du}\, \times {du}/{dt}
\right )}^2 $ in terms of $E, h, \eta$ and $(1-e_r\,\cos u)$.
We have verified  that such an expression is identical to 
our Hamiltonian equations of motion based $\dot \phi^2$ expression to 4PN order.
Additionally, we have also performed the above two checks using $\tilde v$ variable 
based parametric solution.
These computations provided us with two powerful checks on our 
4PN order generalized quasi-Keplerian parametrization.

\subsection{Incorporating non-local in time 4PN Hamiltonian}

Having dealt with the local-in-time component of the 4PN Hamiltonian
in the previous subsection, we turn our attention to the non-local one.
Our strategy will be to treat the 4PN tail effect as a
perturbation to the basic Newtonian Kepler problem
within the framework of the action-angle formalism.
We will employ canonical (or classical) perturbation theory, a perturbation
technique that is tailor-made for the action-angle framework \cite{jose1998classical},
following the application this technique as was done in Ref.~\cite{DJS_15}.

\subsubsection{Action-angles of the Newtonian system}

We start with a small review of the action-angle (Delaunay) picture of the Newtonian system
Ref.~\cite{DJS_15, goldstein2002classical, jose1998classical, ferraz2007canonical}.
We will consider the planar (two actions, two angles) version of the picture
because the system is confined to a plane.
The action-angle pairs $ \left\lbrace  (\mathcal{L} , l),( \mathcal{G}, g)    \right\rbrace $ written in 
terms of more familiar quantities become \cite{DJS_15}
\begin{align}
  \mathcal{L}   & =  a^{1/2}      ,~~~~~~~~~~~~ \mathcal{G}    =   (a(1-e^2))^{1/2}   ,   \label{Newt-actions}   \\
    l   & =  u - e \sin u                   ,   ~~~~~            g = \text{arugment~of~periastron} ,                                         
\end{align}
where $a$ and $e$ are Newtonian versions of $a_r$ and $e_r$;
$u$ is as usual the eccentric anomaly.
The Hamiltonian is a function of only one of the actions
\begin{align}
\mathcal{H}_N   =  - \frac{1}{2   \mathcal{L}^2}
\end{align}
The rate of change of the first angle variable is
\begin{align}
\frac{dl}{d t}  =  \frac{\partial   \mathcal{H}_N}{ \partial \mathcal{L}}  =   \frac{1}{ \mathcal{L}^3}  \equiv   \Omega(\mathcal{L}),
\end{align}
whereas $g$ does not change with time since $\partial \mathcal{H}_N / \partial \mathcal{G} = 0$.

\subsubsection{The nonlocal-in-time 4PN tail Hamiltonian}

To incorporate the effect of the nonlocal-in-time Hamiltonian $\mathcal{H}^{\text{nonlocal}}_{\text{4PN}} $ on the BBH dynamics, it suffices to
consider it as a perturbation on the Newtonian dynamics. Then the reduced Hamiltonian of interest is \cite{DJS_14}
\begin{align}
\mathcal{H} &  =  \frac{\hat{\mathbf{p}}^{2}}{2}-\frac{1}{r}   +  \mathcal{H}^{\text{nonlocal}}_{\text{4PN}} ,
\end{align}
where
\begin{align}        
\mathcal{H}^{\text{nonlocal}}_{\text{4PN}}   =     -  \frac{G^{2} }{5 \eta c^{8}} I_{i j}^{(3)}(t') 
  \operatorname{Pf}_{2 R / c} \int_{-\infty}^{+\infty} \frac{\mathrm{d} w}{|w|} I_{i j}^{(3)}(t'+w).                \label{pert-H-2}
\end{align}
In the above expression, $I^{(3)}_{ij}$ denotes the third derivative
with respect to the coordinate time $t'$ of the
center-of-mass Newtonian quadrupole moment $I_{ij}$ of the system (with $r^i$ denoting the components of $\mathbf{r}$)
\begin{align}
I_{i j}    \equiv  (G M)^{2} \mu\left(r^{i} r^{j}-\frac{1}{3}  {r}^{2} \delta^{i j}\right)    ,
\end{align}
and $R \equiv | \mathbf{R}|$.
$\text{Pf}_{T}$ stands for the Hadamard partie finie which is defined as \cite{DJS_14}
\begin{align}
\operatorname{Pf}_{T} \int_{0}^{+\infty} \frac{\mathrm{d} w}{w} g(w)  \equiv & \int_{0}^{T} \frac{\mathrm{d} w}{w}(g(w)-g(0))  +\int_{T}^{+\infty} \frac{\mathrm{d} w}{w} g(w)  ,
\end{align}
a quick application of which gives
\begin{align}
\operatorname{Pf}_{T} \int_{0}^{+\infty} \frac{\mathrm{d} w}{w} \cos (\omega w)=-\left(\gamma_{\mathrm{E}}+\ln (\omega T)\right),    \label{Pf-applied}
\end{align}
with $\gamma_E = 0.577...$ being the Euler-Mascheroni constant.
The two-sided integral in Eq.~(\ref{pert-H-2}) needs to be converted into one-sided ones so as to make it amenable to
the above definition of $\text{Pf}_T$.

Now focusing our attention on $\mathcal{H}^{\text{nonlocal}}_{\text{4PN}} $, to manipulate the
integral in Eq.~(\ref{pert-H-2}) we change
the variable of differentiation from $t'$ to $l$ and decompose $I_{ij}$ into Fourier components. We get
(with $w' \equiv w/(GM)$)
\begin{widetext}
\begin{align}
\mathcal{H}^{\text{nonlocal}}_{\text{4PN}}  &  =     -  \frac{G^{2}  }{5 \eta c^{8}}    \left( \frac{\Omega}{G M} \right)^6     \sum_{p = - \infty}^{\infty}      \frac{ d^3 ( \cI_{ij}(p) e^{i p l } )}{d l^3} 
  \operatorname{Pf}_{2 r / c} \int_{-\infty}^{+\infty} \frac{\mathrm{d} w'}{|w'|}     \sum_{q = - \infty}^{\infty}    \frac{ d^3  (  \cI_{ij}(q) e^{i q  (l+ \Omega w') } )}{d l^3}  ,   \\    
                  &  =      \frac{G^{2}  }{5 \eta c^{8}}    \left( \frac{\Omega}{G M} \right)^6  \sum_{p = - \infty}^{\infty}     p^3      \cI_{ij}(p) e^{i p l }  
  \operatorname{Pf}_{2 r / c} \int_{-\infty}^{+\infty} \frac{\mathrm{d} w'}{|w'|}     \sum_{q = - \infty , q \neq 0}^{\infty}   q^3     \cI_{ij}(q) e^{i q  (l+ \Omega w') }    ,    \\
                   &  =      \frac{G^{2} }{5 \eta c^{8}}    \left( \frac{\Omega}{G M} \right)^6  \sum_{p,q = - \infty, q \neq 0}^{\infty}     p^3  q^3     \cI_{ij}(p)   \cI_{ij}(q)   e^{i (p+q) l }  
  \operatorname{Pf}_{2 r / c} \int_{-\infty}^{+\infty} \frac{\mathrm{d} w'}{|w'|}        e^{i q   \Omega w'  }     ,    \\
    &  =   -    \frac{ 2  G^{2} }{5 \eta c^{8}}    \left( \frac{\Omega}{G M} \right)^6    \sum_{p,q = - \infty, q \neq 0}^{\infty}    p^3  q^3     \cI_{ij}(p)   \cI_{ij}(q)   e^{i (p+q) l }   (\gamma_E + \ln (|q| \Omega \times  2 r/c ))   \label{H_nl_decomp-m1} ,  \\
      &  =   -    \frac{ 2 G^{2} }{5 \eta c^{8}}    \left( \frac{\Omega}{G M} \right)^6    \bigg[  \sum_{p,q = - \infty, q \neq 0}^{\infty}    p^3  q^3     \cI_{ij}(p)   \cI_{ij}(q)   e^{i (p+q) l }   (\gamma_E + \ln (|q| \Omega  ))  -    \left( \frac{ d^3  I_{ij}}{d l^3}    \right)^2   \ln (  2 r/c ) \bigg] .                      \label{H_nl_decomp}
\end{align}
\end{widetext}

A few comments are needed to clarify the manipulations in the above lines.
$\mathcal{I}_{ij}$'s are defined as per the following Fourier decomposition
\begin{align}
I_{ij}(t')    & =   \sum_{p = - \infty}^{\infty}  \mathcal{I}_{ij}(p) e^{ i p l}    ,    \\
 \mathcal{I}_{ij}(p)    & =  \frac{1}{2 \pi} \int_{0}^{2 \pi}  I_{ij}(t') e^{-\mathrm{i} p l }    dl .
\end{align}
These Fourier coefficients of the Newtonian multipole moments
are available in terms of Bessel functions in the 
appendix of Ref.~\cite{2008PhRvD..77f4034A}
 in dimensionless form.
In the above lines and the material below, 
the summation is over all integers from $- \infty$ to $+\infty$
except where it is explicitly mentioned that $q=0$ be omitted.
We have omitted $q = 0$ from the above summations because $\ln(0)$ is undefined.
Also, with the variable of integration having changed from $w \rightarrow w'$, we also had to
change the time-scale of the $\operatorname{Pf}$ operation from $2 R/c \rightarrow 2 r/c$.
One may also notice another issue of the arguments inside the two $\ln$'s in Eq.~\eqref{H_nl_decomp}
not being dimensionless, due to breaking the $\ln$ in Eq.~\eqref{H_nl_decomp-m1} into two $\ln$'s in the following line.
This is no reason to worry since the final result in Eq.~\eqref{avgd-H-pert} combines back
the two $\ln$'s into one with an argument which is indeed dimensionless.
And finally, in Eq.~\eqref{H_nl_decomp-m1}, use has been made of 
the result in Eq.~\eqref{Pf-applied},
and a similar result for sine being inside the integrand instead of cosine
to evaluate the integral in terms of the logarithm.

Due to the definite integral having been performed above
and the application of order-reduction (eliminating the derivatives of variables in the Hamiltonian
with the equations of motion of a lower order \cite{DJS_15}),
we now have to deal only with the local 
dynamics\footnote{See Refs.~\cite{Damour:1990jh,
Damour:2016abl, Damour:1999cr} for a background on order-reduction.
The procedure of order-reducing
the Hamiltonian leads to shifts in canonical coordinates. Apart from the tail part~\cite{DJS_15}, the instantaneous
Hamiltonian of Eq.~\eqref{H_3} is also order-reduced \cite{Damour:1999cr}. 
Since these shifts lead to   
amplitude and zero-average oscillatory phase corrections
 (rather than secular phase corrections), we ignore such shifts, as our IMR waveform is meant to
be only leading order accurate in amplitude. \label{footnote_ord-red}}.
The 4PN tail Hamiltonian
has been decomposed in this particular way of Eq.~\eqref{H_nl_decomp}
to render it amenable to the averaging procedure, to be carried out below.

\subsubsection{4PN tail effect as a perturbation to the Newtonian system}      \label{4PN-tail-pert-Newt}

To deal with the 4PN tail Hamiltonian of Eq.~\eqref{H_nl_decomp}, we employ the so called
``Delaunay technique'' of averaging the perturbation Hamiltonian which is discussed in Appendix \ref{Delaunay technique}.
The method dictates the following: take the perturbation part $\epsilon H_1(\phi_0, J_0)$ of the full Hamiltonian
$H = H_0(J_0) +  \epsilon H_1(\phi_0, J_0)$ and average it over $\phi_0$, thus yielding $\overline{H}_1 (J_0)$.
Then there exists a generating function $S (\phi_0, J)$ such that it gives new action-angles $(J, \phi)$ which are connected to
the old ones via Eqs.~\eqref{new_AAVs_1} and \eqref{new_AAVs_2},
and the total Hamiltonian depends only on the new action $J$
as $E(J) = E_0 (J) + \epsilon E_1(J)$, where the functional dependence of $E_0$ and $E_1$ on $J$ is the same as
that of $H_0$ and $\overline{H}_1$ on $J_0$. 
For our system in consideration, $H_1 = \mathcal{H}^{\text{nonlocal}}_{\text{4PN}}$ and $\phi_0 = l$.
We now proceed to average $H_1$ over $l$.

The two additive terms in Eq.~(\ref{H_nl_decomp}) will be averaged using different methods.
This is so because they possess different structures and hence their averaging is tractable via different means.
To average the first term of Eq.~(\ref{H_nl_decomp}) (the one involving the summation), we note that
the averaged value will be equal to the sum of all the terms in this double summation which
correspond to $p+q=0$. Hence, we substitute $q = -p$, thereby turning the double summation
into a single summation over $p$ and we choose to retain all the term with $|p| \lesssim 500$ terms in this summation.
This is so because we found by inspection that these many 
terms are enough for $e \lesssim 0.9 $.
We then try to come up with a certain
Pad\'e-like approximant for this series using a method that we
now briefly sketch.

  Using the Fourier coefficients of the Newtonian multipole moments
as given in the appendix of Ref.~\cite{2008PhRvD..77f4034A},
it can be shown that the summation in the first term of Eq.~\eqref{H_nl_decomp} (without the prefactor
and $(\gamma_E + \ln (|q| \Omega  )$) with $q=-p$ can be written as
\begin{align}    \label{PM63}
&  \sum_{ p,q  = -\infty , q \neq 0}^{\infty} p^{3} q^{3} \mathcal{I}_{i j}(p) \mathcal{I}_{i j}(q) e^{i(p+q) l}   \bigg\rvert_{q=-p}    \\ 
  =  &  - \sum_{ p = -\infty}^{\infty} p^{6}  |\mathcal{I}_{i j}(p)|^2     =    - 32   \mu^2  (a G M)^4   \frac{1 + \frac{73 {e}^{2}}{24}+\frac{37 {e}^{4}}{96}}{(1- e^2)^{7/2} } .   \nonumber   
\end{align}
We actually factored out $(1-e^2)^{-7/2}$ from the above sum before trying to evaluate
it to preserve the formal structure in the limit of large angular momentum
\cite{2021arXiv210608276D}\footnote{It can be checked that the average (over $l$) of the first term of
the nonlocal Hamiltonian of Eq.~\eqref{H_nl_decomp}
has the following structure in the $h~\rightarrow~\infty$ limit
$\overline{H}_{1(\text{A})}     \sim   { (E h^{2})^2 h^{-7}}    \sum_{n=0} a_n \alpha^n $,
with $\alpha = (E h^2)^{-1} $. This, along with $e^2 = 1 + 2 E h^2$, lets us see that
$\overline{H}_{1(\text{A})}$ has the form $\overline{H}_{1(\text{A})}    \sim  {F(e)}/{h^7}
   =   {F(e)}     \left(  {2 E}/{(e^2-1)} \right)^{7/2}$.
This motivated us to first factor out
$(1-e^2)^{-7/2}$ from the Bessel series whose Pad{\'e}-like approximant is sought.}.
Note that the above result appears to be exact in $e$ and coincides with the ``$f(e)$'' of Peters and Mathews \cite{PM63}.
Obtaining this nice exact in $e$ expression became possible due to 
factoring out $(1-e)^{-7/2}$ in the beginning and then series expanding the remnant.

Focusing our attention back to the summation in Eq.~\eqref{H_nl_decomp},
we see that it can further be decomposed as (with $q = -p$)
\begin{align}
-\sum_{p = -\infty}^{\infty} p^{6}  | \mathcal{I}_{i j}(p)|^{2} \left[  ( \gamma_E +  \ln \Omega ) + \ln |p|   \right] .
\end{align}
\\Since the first additive term in the above expression (in the parenthesis) has already been taken care of in Eq.~\eqref{PM63},
we now show how to deal with the second one (involving $\ln |p|$).
Again, after factoring out $(1-e^2)^{-7/2}$,
we first find the Pad\'e approximation 
of this second term such that both the numerator and denominator are expanded up to $\mathcal{O}(e^{10})$. 
Then by hand, we add $p_{12} e^{12} + p_{14} e^{14}$ in the numerator, where $p_{12}$ and $p_{14}$
are to be determined by numerically matching (at high eccentricities like $e = 0.85, 0.90$)
this new Pad\'e-like ansatz with the evaluated value of the series
where terms up to $|p|  \lesssim 500$ have been retained. Tentatively, we have
\begin{widetext}
\begin{align}                \label{tentative-Pade}
& -\sum_{p=-\infty}^{\infty} p^{6}\left|\mathcal{I}_{i j}(p)\right|^{2}   \ln |p|  \nonumber \\ 
&  = -  G^4 \mathcal{L}^8 M^6 \eta^2   \left[ \frac{{p_{14} e^{14}} + {p_{12}e^{12}}-\frac{382996272 e^{10}}{13601521}+\frac{579332351 e^{8}}{4983158}-\frac{324710645 e^{6}}{8433524}-\frac{758231515 e^{4}}{3359177}+\frac{263415291 e^{2}}{1639996}+\frac{286746937}{12927762}}{\left(1-e^{2}\right)^{7 / 2}\left(\frac{4447985 e^{10}}{4076572203}+\frac{49804512 e^{8}}{1158420851}-\frac{105413189 e^{6}}{194334558}+\frac{103729937 e^{4}}{57112735}-\frac{56374811 e^{2}}{24380301}+1\right)}   \right] .
\end{align}
\end{widetext}
We call the resulting approximant ``Pad\'e-like'', since it is 
a result of combining the methods of Pad\'e approximation and numerical fitting. Such a procedure, combined with
the idea of factoring out $(1-e^2)^{-7/2}$ gives us
approximants which are valid up to higher $e$'s than
would have been possible with the standard Pad\'e approximants.
Interestingly, these approximants capture the eccentricity effects up to $e \lesssim 0.9$, despite the 
existence of Laplace limit of $e<0.66...$ beyond which the series solution
of Kepler equation in $e$ diverges \cite{Colwell92,
moulton1970introduction, finch2003mathematical}.

Finally, averaging the second term term 
of Eq.~\eqref{H_nl_decomp} (involving $\ln 2\,r/c$) over $l$ in closed-form is possible,
although lengthy. Using the order-reduced Newtonian equations
\begin{subequations}
\begin{align}
x     &=   a(\cos u-e)    ,\\
y    &=  a \sqrt{1-e^{2}} \sin u   , \\
z  & = 0   \\
l &= u - e \sin u      ,  \\   
r   & = a (1- e \cos u)      ,    \\
I_{ij} &=(G M)^{2} \mu\left(r^{i} r^{j}-\frac{1}{3} {r}^{2} \delta^{i j}\right)  ,
\end{align}
\end{subequations}
(with $x,y,z$ being the components of $\mathbf{r}$) in the second term of Eq.~\eqref{H_nl_decomp}, 
we arrive at \\ 
\begin{widetext}
\begin{align}                  \label{last-piece-H-tail}
 &  \frac{ 1}{2\pi\,a^4\,G^4\,M^4\,\mu^2} \int^{2\pi}_0  \left( \frac{ d^3  I_{ij}}{d l^3}    \right)^2   \ln (  2 r/c ) \,dl    =   \frac{ 2 }{3 \pi} \int_0^{2 \pi} \frac{( 24 - 23 e^2 - e^2 \cos 2 u)  \left( \ln \left[ \frac{2 a(1 - e \cos u) }{c}  \right]  \right)}{(1- e \cos u)^5} du    \\
 =  & \frac{ 12 \left(37 e^4+292 e^2+96\right) \ln \left[ \frac{4 a \left(1-e^2\right) \left(1-\sqrt{1-e^2}\right)}{c\,e^2}\right]  - \left(255 e^4+3792 e^2+2408\right) }{36 \left(1-e^2\right)^{7/2}}+\frac{\left(673 e^2+602\right)}{9 \left(1-e^2\right)^3}  .
\end{align}
Finally, Eqs.~\eqref{PM63}, \eqref{tentative-Pade} and \eqref{last-piece-H-tail} 
culminate in the expression of order-reduced averaged 4PN tail Hamiltonian
\begin{align}
\overline{\mathcal{H}}^{\text{nonlocal}}_{\text{4PN}} & = \frac{2\, \eta }{5\, c^8 \mathcal{L}^{10}}\, \Bigg[ \frac{ 12 \left(37 e^4+292 e^2+96\right) \ln \left[ \frac{4  \left(1-e^2\right) \left(1-\sqrt{1-e^2} \right) \exp({\gamma_E} )  }{\mathcal{L} c\,e^2}\right]  - \left(255 e^4+3792 e^2+2408\right) }{36 \left(1-e^2\right)^{7/2}}+\frac{\left(673 e^2+602\right)}{9 \left(1-e^2\right)^3}   \nonumber  \\
  &  +   \mathcal{H}_\text{Pad\'e} \Bigg]  \,,        \label{avgd-H-pert}
\end{align}
where 
\begin{align}
\mathcal{H}_\text{Pad\'e}  =  \frac{\frac{14262437 e^{14}}{328008227}-\frac{6775509 e^{12}}{248174614}-\frac{382996272 e^{10}}{13601521}+\frac{579332351 e^8}{4983158}-\frac{324710645 e^6}{8433524}-\frac{758231515 e^4}{3359177}+\frac{263415291 e^2}{1639996}+\frac{286746937}{12927762}}{\left(1-e^2\right)^{7/2} \left(\frac{4447985 e^{10}}{4076572203}+\frac{49804512 e^8}{1158420851}-\frac{105413189 e^6}{194334558}+\frac{103729937 e^4}{57112735}-\frac{56374811 e^2}{24380301}+1\right)}\, ,
\end{align}
is the Pad\'e-like approximant that we constructed above.
Eliminating $e$ using $e=\sqrt{1- \mathcal{G}^2/\mathcal{L}^2}$ expresses
this averaged Hamiltonian in terms of only the actions.

Now, in the action-angles formalism, one can compute the ``frequencies'' corresponding to all
the angle variables by partially differentiating the Hamiltonian with respect to the corresponding action
variables. Application of this to our system in consideration yields the 4PN tail correction to the 
frequencies $n$ and $k n$ (where $n$ denotes the mean motion and $k$ is the periastron advance parameter)~\cite{DS88}.
After partially differentiating with respect to $\mathcal{L}$ and $\mathcal{G}$, we write the 
results in terms of $\mathcal{L}$ and $e$. For illustration, we give their small $e$ expanded expressions 
\begin{align} 
n_\text{tail}& \equiv \frac{\partial \overline{\mathcal{H}}^{\text{nonlocal}}_{\text{4PN}} ( \mathcal{G},  \mathcal{L} )}{\partial \mathcal{L}}                  \label{new-freq-1}               \\[0.5ex]               
&=\frac{19289894530752349825264 \,\eta }{323061509079949009095 \,\mathcal{L}^{11}\,c^8}+\frac{592 \eta  \left(\ln \left(\frac{1}{\mathcal{L}c}\right)+\gamma_E +\ln (2)\right)}{15\, \mathcal{L}^{11}\,c^8}\notag\\[0.5ex]
&+\frac{e^2 \,\eta\,  \left(-184 \ln \left(\frac{1}{\mathcal{L}c}\right)-184 \gamma_E +\frac{176764157117697232864351125327585955400510029}{604435714840557267408843920105441163950610}-184 \ln (2)\right)}{5 \mathcal{L}^{11}\,c^8}+ \mathcal{O}\left(e^4\right)\,,
\\[1ex]
(k\, n)_\text{tail}&  \equiv \frac{\partial \overline{\mathcal{H}}^{\text{nonlocal}}_{\text{4PN}} ( \mathcal{G},  \mathcal{L} )}{\partial \mathcal{G}}\,  ,    \label{new-freq-2}\\[1ex]
k_\text{tail}&=-\frac{74831546379478710201598\,\eta }{323061509079949009095 \mathcal{L}^8\,c^8}+\frac{352 \,\eta }{5 \mathcal{L}^8\,c^8}-\frac{2512 \eta  \left(\ln \left(\frac{1}{\mathcal{L}c}\right)+\gamma_E +\ln (2)\right)}{15 \mathcal{L}^8\,c^8}\\[0.5ex]
&+\frac{e^2 \eta  \left(-\frac{5164100568419311748345089878114388585281349219}{201478571613519089136281306701813721316870}-13264 \ln \left(\frac{1}{\mathcal{L}c}\right)-13264 \gamma_E +10196-13264 \ln (2)\right)}{15 \mathcal{L}^8\,c^8}+ \mathcal{O}\left(e^4\right)\, \notag  .
\end{align}
\end{widetext}
Note that $k = \Delta \Phi/2 \pi$, where $\Delta \Phi$ is the angle of periastron advance 
in the radial period (periastron to periastron) $P$, which means that $k = \Phi/(2 \pi) -1$.
This means that among the earlier presented set of equations for the 4PN local-in-time 
quasi-Keplerian parameterization (QKP), 
it is Eqs. \eqref{eqn_n} and \eqref{QKP-for-Phi} which should include the effects encoded in
Eqs. \eqref{new-freq-1} and \eqref{new-freq-2}, but they do not since
 Eqs. \eqref{eqn_n} and \eqref{QKP-for-Phi} are meant to include only the local-in-time
 contributions only. We have checked that Eqs. \eqref{new-freq-1} and \eqref{new-freq-2}
 agree with the results of Ref.~\cite{Bini2020} in the circular limit.

Also, note that the argument of $\ln$ in Eqs.~\eqref{last-piece-H-tail} and \eqref{avgd-H-pert} 
becomes undefined at $e=0$ although its $e \rightarrow 0$ limit is well defined. Therefore,
in our \textsc{Mathematica} package, we chose to replace this argument of $\ln$ by its
Taylor expansion so that it is valid in the range $0 \leq e \lesssim 0.85$.

To bring all the pieces together, let us have a bird's eye view of the situation.
We are trying to include the effect of 4PN tail Hamiltonian as a perturbation to the 
Newtonian one in the action-angles framework. The effect entails 
\begin{enumerate}
\item the action-angles being perturbed as per 
Eqs.~\eqref{gen-func}, \eqref{new_AAVs_1} and \eqref{new_AAVs_2}     
\item the functional dependence of the Hamiltonian on the actions being perturbed as per
Eqs.~\eqref{pert-0}, \eqref{pert-1b} and Eq.~\eqref{avgd-H-pert}
\end{enumerate}
At this point, we invoke the ``semi-perturbation'' scheme detailed in Appendix~\ref{neglect}, whereby
we don't use the information contained in 
Eqs.~\eqref{gen-func}, \eqref{new_AAVs_1} and \eqref{new_AAVs_2},
but rather only make use of the information in Eqs.~\eqref{pert-0}, \eqref{pert-1b}.
This way we don't need the oscillatory corrections to the action-angles
and the generating function.
All one needs to do is to average the perturbation
Hamiltonian and find the perturbed frequencies, as has already been done in Eqs.~(\ref{new-freq-1}) and
 (\ref{new-freq-2}) above.

Now we discuss how to merge the results of Eqs.~\eqref{new-freq-1} and \eqref{new-freq-2}
into the QKP Eqs.~\eqref{e:CoeffKP} if one wants to.
Since we have chosen not to perturb the actions, there won't be any
corrections to the eccentricity and semi-major axis due to the tail effects
(see Eqs.~\eqref{Newt-actions}).
This means that in the RHSs of Eqs.~\eqref{new-freq-1} and \eqref{new-freq-2},
after evaluating the partial derivatives,
we can replace $\mathcal{L}$ and $\mathcal{G}$ with $a$ and $e$ using Eqs.~\eqref{Newt-actions}
and then make the substitutions
$ ( e, a ) \rightarrow ( (1+2 E h^2)^{1/2},  -1/(2 E))$ to write $n_{\text{tail}}$ and
$k_{\text{tail}}$ in terms of $E$ and $h$. Then with the earlier derived equation
$k = \Phi/(2 \pi) -1$, $n_{\text{tail}}$ and $k_{\text{tail}}$ can both be incorporated
as the 4PN tail effects into the
quasi-Keplerian solution (specifically Eqs.~\eqref{eqn_n} and \eqref{QKP-for-Phi})
for the 4PN local Hamiltonian presented in the previous subsection. 
We do not take it because we have reserved Eqs.~\eqref{e:CoeffKP} for local-in-time effects 
only. Note that in accordance with our decision to adopt the semi-perturbation scheme,
 the 4PN tail corrections 
 do not enter in the expressions for any other quantity except $n$ and $\Phi$ (or $k$),
 as far as the 4PN conservative quasi-Keplerian solution is concerned.

Strictly speaking, we have not worked out the full 4PN dynamics since we have
ignored the corrections to the action-angles (linear in $\epsilon$ terms on the RHSs of 
Eqs.~\eqref{new_AAVs_1} and \eqref{new_AAVs_2}) due to the 
tail effects. Adopting the semi-perturbation scheme rather than full perturbation scheme
simplifies the calculations and
 the ignored 4PN zero-average, oscillatory corrections 
 to action-angles (due to tail effects)
are less important than the 4PN secular tail effects
for GW data analysis purposes.
An alert reader may be able to point out that we have
included the oscillatory terms in the Keplerian type solution
for the 4PN local Hamiltonian 
(the ones with sinusoidal functions of integer multiples of $v$
in Eq.~\eqref{e:FinalParamADM_2}),
despite dropping the oscillatory correction terms 
to action-angles (as per the semi-perturbation scheme)
resulting from the 4PN tail effect. 
One may insist that all the oscillatory terms should be discarded.
One could do so.
We decided to include these oscillatory terms for the local Hamiltonian
in Eq.~\eqref{e:FinalParamADM_2}
just to be in line with the tradition of including all the terms,
be it oscillatory or secular, while giving the QKP solution 
of a BBH for various PN accurate local conservative Hamiltonians as has been done in Refs.~\cite{DD85, SW93, MGS}.

We finally mention beforehand that both Eqs.~\eqref{new-freq-1}
and \eqref{new-freq-2} have a bearing on
Eq.~\eqref{l-dot-equation} which gives the mean motion $n$, 
so that when it comes to the IMR waveform construction,
of all the equations in Sec. \ref{4PNimr},
only Eq.~\eqref{l-dot-equation} incorporates the 4PN tail effect
 as per our choice of adopting the semi-perturbation strategy to include the tail effects.
The accompanying \textsc{Mathematica}
notebook contains the relevant expressions for $n$ (both local and nonlocal).
We now move on to apply the above parametric solution to construct 
a time-domain eccentric IMR waveform, influenced by Ref.~\cite{Hinder2017}.

\section{An improved time-domain eccentric IMR waveform }
\label{4PNimr}

The present section details our effort to incorporate 4PN order Keplerian type parametric solution
into a \textsc{Mathematica} 
package, namely \texttt{EccentricIMR}, available as an open source software \cite{MMA1},
in an accurate and efficient manner.
This package implements an eccentric IMR model, detailed in Ref.~\cite{Hinder2017},
where an eccentric PN-accurate inspiral model was combined
with a quasi-circular merger waveform.
Detailed interpolation using several non-eccentric and non-spinning 
NR waveforms in the neighborhood of their 
merger phase ensured that the analytic 
quasi-circular merger waveform of Ref.~\cite{Hinder2017} is accurate for non-spinning 
BH binaries with mass ratio $q \equiv  m_1/m_2$  between $1$ and $4$ and for 
 arbitrary $\phi_0$ (the initial
phase of the waveform).
The IMR model of Ref.~\cite{Hinder2017} combines
the post-Newtonian inspiral waveform, adapted from Ref.~\cite{Hinder_10},
and their above described quasi-circular merger model (CMM) to obtain a time-domain eccentric IMR waveform.
This involves `blending' of the above two models in a transition region where
neither PN approximation is valid nor the binary can be assumed to have circularized.

It was noted in Ref.~\cite{Hinder2017} that incorporating higher order PN corrections should 
improve the performance of the early inspiral phase.
In fact, Ref.~\cite{Hinder2017} invoked 3PN order conservative and 2PN order reactive contributions 
to the BBH dynamics for describing the inspiral part of their IMR $h(t)$.
The present effort improves the treatment of the conservative part in the \texttt{EccentricIMR} package 
with the help of our 4PN order Keplerian type parametric solution 
while adapting PN-accurate results from Refs.~\cite{Arun2009, KBG18}
to incorporate effects of 3PN-accurate GW emission.
Moreover, we provide computationally efficient version of the $x$-model, employed in \texttt{EccentricIMR},
by adapting the GW phasing approach of Refs.~\cite{Damour2004,KG06}
during the modeling of the eccentric inspiral phase.
These changes are incorporated into the \textsc{Mathematica} package accompanying Ref.~\cite{Hinder2017}
and we treat their circular merger model as a black box \cite{MMA1}.
The resulting package is available at Ref.~\cite{MMA2, MMA3}.
In what follows, we present our approach to improve various aspects of the $x$-model.                             
In this section we will employ $t$ to denote the ADM coordinate time for the sake of convenience (unlike the previous section). 
A derivative with respect to the ADM time will be represented by an overdot.

\subsection{4PN conservative and 3PN radiative phasing}

We begin by listing the usual expression for the quadrupolar order 
(or restricted) complex gravitational waveform for compact binaries in non-circular orbits \cite{JS92, Hinder_10,Huerta:2016rwp, Damour2004}
\begin{subequations}
\begin{align}
h_{\text{sig}}  & =h_{+}-i h_{ \times}           \label{ST1}         \\  
h_{+}  &    =-\frac{G M \eta}{c^4 D}\left\{\left(\cos ^{2} \theta+1\right)\left[ \left(\frac{G M}{R} - \dot{R}^{2}+R^{2} \dot{\phi}^{2}\right)  \cos 2 \phi^{\prime}   \right.\right.  \nonumber  \\  
&   \left.\left. +2  R \dot{\phi}  \dot{R} \sin 2 \phi^{\prime} \right]  +  \left(\frac{G M}{R}-\dot{R}^{2}-R^{2} \dot{\phi}^{2}\right) \sin ^{2}{\theta}    \right\}   ,     \label{ST2}    \\      
h_{ \times} & =-\frac{2 G M \eta}{c^4  D} \cos \theta\left\{\left(\frac{G M}{R} -\dot{R}^{2}+R^{2} \dot{\phi}^{2}\right) \sin 2 \phi^{\prime}\right. \nonumber \\
 &     \left.    -2 R \dot{\phi}  \dot{R}   \cos 2 \phi^{\prime}  \right\}  ,      \label{ST3}
\end{align}
\end{subequations}
where $\phi'  \equiv  \phi - \varphi  $ and $\theta$ and $\varphi$
are the spherical polar angles 
that specify the observer in a frame centered around
the source which is a distance $D$ away from the detector.
Recall that $R$ and $\phi$ serve to specify the relative separation vector $\textbf{R}$
between the two BHs whereas $r \equiv R/(GM)$.  
Also, we have $\dot r \equiv dr/dt $ and $\dot \phi  \equiv d \phi/dt$.
The NR relevant spin-weight $-2$, $ \ell=2, m=2$ spherical harmonic 
mode of $h(t)$ reads
\begin{align}
\label{h22_N}
    \begin{aligned} h^{22} &=\int {_{-2} Y_{2}^{2^{*}}(\theta, \varphi)} h_{\text{sig}}(\theta, \varphi) d \Omega \\ &=-\frac{4 G M \eta  e^{-2 i \phi}}{ c^4 D } \sqrt{\frac{\pi}{5}}\left(\frac{ G M}{R}+(\dot{\phi} R + i \dot{R})^{2}\right) ,\end{aligned}
\end{align}
where the spherical harmonic $_{-2} Y_{2}^{2}(\theta, \varphi) = \frac{1}{2}  e^{2 i \varphi} \sqrt{5 / \pi} \cos ^{4}(\theta / 2)$.
In what follows, we describe our improved $x$-model to obtain  $h^{22}(t)$ for 
non-spinning BH binaries inspiralling along relativistic
eccentric orbits in a computationally accurate and efficient manner.

We adapt, as mentioned earlier, the GW phasing approach of Refs.~\cite{Damour2004,KG06} to model 
eccentric inspiral of BH binaries to describe 
the temporal evolution of dynamical variables that appear in the above
$h^{22}(R(t), \dot R(t), \phi(t), \dot\phi(t))$ expression.
This approach imposes numerically the effects of GW emission on the conservative  dynamics of eccentric binaries by incorporating changes in orbital configurations that occur at the orbital, periapsis advance and gravitational radiation reaction time scales.
In the GW phasing approach, the conservative BBH  dynamics is described using 
our Keplerian type parametric solution and we first focus on describing parametrically the temporal evolution of 
 $r, \dot r$ and $\dot\phi$, whereas $\phi$ will be dealt with later.
We begin by expressing the 4PN-accurate expressions for these variables as 
\begin{align}
r (-2\,E, h,u)  &= a_r(-2\,E, h) \times \Big( 1 - e_r (-2\,E, h) \times \cos u    \Big)  \,,           \label{parametric_sol_1}
\\
\dot r (-2\,E, h,u)  &= \frac{dr }{du}\times \frac{du}{dt}\,,
\\
\dot \phi (-2\,E, h,u)  &= \frac{d \phi}{d v}\times \frac{dv}{du} \times  \frac{du}{dt}\,.                 \label{parametric_sol_3}
\end{align}
where
\begin{align}
\frac{d u }{dt}   = \frac{du}{dl}    \frac{dl }{dt}    =     \frac{du}{dl} \frac{n}{G M}   .
\end{align}
Recall that $t$ in this section stands for the ADM time, whereas it stood for the 
scaled ADM time in Sec.~\ref{Kep_review}.
Employing the 4PN solution given in Eqs.~(\ref{e:FinalParamADM_1})-(\ref{e:FinalParamADM_3}), we can see that
the resulting 4PN order expressions of $r,~\dot{r}$ and $\dot{\phi}$ will be functions of
$(-2\,E), h, e_r, e_t, e_{\phi}, u$ and $\eta$.
Conventionally in the $x$-model, we rather write these expressions in terms of
$\omega, e_t$ and $u$. This is achieved with the help of the following steps.

In the first step, we define a 4PN order $\omega$ from our 4PN order expressions for $n$ and $k=\Delta \Phi/2\,\pi$ 
as $\omega   \equiv  n \times ( 1 + k)$.
It is now possible to invert 4PN order $\omega$ expression 
in a PN-accurate manner to 
express $(-2\,E)$ in terms of $\omega$ and $h$.
In the next step, we invert 4PN order $e_t$ expression (given by the first of Eqs.~(\ref{Eq_e4PN_AppC})) to get
an intermediate 4PN order expression for $h$ in terms of $e_t$ and $(-2\,E)$.
Then use the above obtained expression of $(-2E)$ (in terms of $\omega$ and $h$)
to express $h$ in terms of $e_t$ and $\omega$ up to 4PN order in a post-Newtonian perturbative way.
Using this expression of $h(e_t, \omega)$ in the above
expression of $(-2 E)$ finally gives us $(-2E)$ in terms of $e_t$ and
$\omega$. The resulting relations for $(-2\,E)$ and $h$
(in terms of $e_t$ and $\omega$) allow us to express 
$e_r$ and $e_{\phi}$ in terms of $\omega, e_t$ to 4PN order 
with the help of Eqs.~(\ref{Eq_e4PN_AppC}).
All these relations in conjunction with Eqs.\,(\ref{parametric_sol_1})-(\ref{parametric_sol_3})
lead to 4PN order parametric expressions for $r, \dot r $ and 
$\dot \phi$ in terms of $\omega, e_t,u$. We provide the expressions of $E$ and $h$ in terms of
$x \equiv  (G M \omega/c^3)^{2/3}   $ and $e_t$ in Appendix~\ref{E-h}.

An additional requirement of the GW phasing approach is to split 
the orbital phase into two parts such that (with $\lambda(t_0) = 0$)
\begin{subequations}    \label{ST4:full}
\begin{align}
    \phi(t)  & =  \lambda(t) +  W (u(l), (-2\,E), h )        ,    \\
    \lambda(t)  & =  (1+k) n (t-t_0)  \equiv \omega ( t- t_0)\,  ,      \label{ST4} 
\end{align}
\end{subequations}
as done in Ref.~\cite{Damour2004}. 
Note that Eqs.~(\ref{ST4:full}) are for conservative dynamics only and need to be 
modified by the application of the method of variation of arbitrary 
constants (as done in Ref.~\cite{Damour2004}) to include the radiation reaction
effects\footnote{See Box.~3.3 of 
Ref.~\cite{poisson2014gravity} for a pedagogical treatment of the method
of variation of arbitrary constants.}.
This will be done later in Eqs.~(\ref{lambda-dot-l-dot}).
The above equations ensure that the secular evolution
of the orbital phase depends linearly on time 
which makes it easier to track the evolution of the  argument of periapsis while 
the $W$ part provides the periodic variations, present in the $\phi$ evolution due to its  dependence on $l$.  
Additionally, we need to express the above 4PN order expression for $W$ in terms of
$\omega, e_t$ and $u$  which we write symbolically as 
\begin{align}
 W   & = (v-u) + e_t \sin u + W^{\mathrm{1PN}} x + W^{\mathrm{2PN}} x^2     \nonumber  \\   
    &  + W^{\mathrm{3PN}} x^3 + W^{\mathrm{4PN}} x^4  .
\end{align}
For the ease of implementation, it is helpful to use the 
 following 4PN order expression of Ref.~\cite{KG06} 
\begin{align} 
    v-u & =  2 \arctan   \left( \frac{\beta_{\phi} \sin u}{1 -  \beta_{\phi} \cos u }  \right) .  \label{ST10}
\end{align}
\\Our parametric solution allows us to obtain the following 4PN order expression for 
$\beta_{\phi}$ in terms of $e_t, x, u$ as
\begin{align}
\beta_{\phi} = \frac{1-\sqrt{1-e_t^2}}{e_t}  +  \beta_{\phi}^{\mathrm{1PN}} x  +  \beta_{\phi}^{\mathrm{2PN}} x^2   +  \beta_{\phi}^{\mathrm{3PN}} x^3 +  \beta_{\phi}^{\mathrm{4PN}} x^4.    
\end{align}
The explicit and lengthy expressions for these 4PN order quantities are listed in the accompanying 
\textsc{Mathematica} notebook~\cite{MMA2, MMA3}.

We now collect relevant expressions that  describe the temporal evolution of
a precessing eccentric orbit whose conservative (without radiation reaction) orbital
dynamics is specified by the 4PN order
Hamiltonian, given by Eq.~(8.41) of Ref.~\cite{JS_15}.
These equations may be symbolically written as\\
\begin{widetext}
\bes
\label{4PN_dy_var}
\begin{align} 
\frac{\dot{R}}{c} &= \frac{\sqrt{x} e_{t} \sin u}{1-e_{t} \cos u}\left\{1+\dot{R}^{1 \mathrm{PN}}\left(\eta, e_{t}\right) x  +\dot{R}^{2 \mathrm{PN}}\left(\eta, e_{t}, u\right) x^{2} +       +\dot{R}^{3 \mathrm{PN}}\left(\eta, e_{t}, u\right) x^{3}             +\dot{R}^{4 \mathrm{PN}}\left(\eta, e_{t}, u\right) x^{4}\right\},  \label{ST5} \\
\dot{\phi} &=    \frac{ x^{3/2}  c^3 \sqrt{1-e_t^2}  }  { G M (1- e_t \cos u)^2}       \left\{  1+\dot{\phi }^{1 \mathrm{PN}}\left(\eta, e_{t} , u\right) x  +\dot{\phi }^{2 \mathrm{PN}}\left(\eta, e_{t}, u\right) x^{2}     +\dot{\phi }^{3 \mathrm{PN}}\left(\eta, e_{t}, u\right) x^{3}             +\dot{\phi }^{4 \mathrm{PN}}\left(\eta, e_{t}, u\right) x^{4}\right\} ,     \label{ST6}   \\
R  &=     \frac{G M (1-e_t \cos u)}{c^2 x}        \left\{1+ R^{1 \mathrm{PN}}\left(\eta, e_{t}, u \right) x  + R^{2 \mathrm{PN}}\left(\eta, e_{t}, u\right) x^{2} +   R^{3 \mathrm{PN}}\left(\eta, e_{t}, u\right) x^{3}     +  R^{4 \mathrm{PN}}\left(\eta, e_{t}, u\right) x^{4}\right\} ,     \label{ST7}  \\
 \phi & =  \lambda  +  W\,,    \label{ST8} \\
\lambda(t)  & = \omega ( t- t_0)\,,      \label{ST9}     \\
W &= (v-u) + e_t \sin u + W^{\mathrm{1PN}}\left(\eta, e_{t}, u\right)  x + W^{\mathrm{2PN}} \left(\eta, e_{t}, u\right)  x^2+ W^{\mathrm{3PN}}\left(\eta, e_{t}, u\right)  x^3 + W^{\mathrm{4PN}} \left(\eta, e_{t}, u\right)  x^4 \,    ,
\end{align}  
\ens
\end{widetext}
where we employ the 4PN-accurate expression for $(v-u)$ given symbolically by Eq.~(\ref{ST10}).
It should be noted that the above equations provide analytically 4PN order conservative dynamics of BH binaries
in terms of $u$ and we require to numerically solve our 4PN order Kepler equation after re-writing it
in terms of $x$ and $e_t$ to obtain $u(l(t))$. Our full system however, is not conservative and as we will soon see
that inclusion of the radiation reaction effects will modify some of the above equations.
For the purpose of illustration, we list the 1PN contributions that appear in  Eqs.~(\ref{4PN_dy_var}) as
\begin{subequations}
\begin{align}
\dot{R}^{1 \mathrm{PN}}\left(\eta, e_{t}\right)  &=\frac{-7 \eta+e_{t}^{2}(-6+7 \eta)}{6\left(1-e_{t}^{2}\right)}  , \\
\dot{\phi}^{1 \mathrm{ P N}}\left(\eta, e_{t}, u\right)  &=\frac{\left(-1+\chi+e_{t}^{2}\right)(-4+\eta)}{\chi\left(1-e_{t}^{2}\right)} ,  \\
R^{1 \mathrm{PN}}\left(\eta, e_{t}, u\right)  & =\frac{1}{6 \chi\left(1-e_{t}^{2}\right)} \left( -24+9 \eta+\chi(18-7 \eta)  \right. \nonumber \\ 
                                    &  \left. +e_{t}^{2}[24-9 \eta+\chi(-6+7 \eta)] \right)    , \\
W^{1 \mathrm{PN}}\left(\eta, e_{t}, u\right)& =3 \frac{e_{t} \sin u+(v-u)_{1 \mathrm{PN}}}{1-e_{t}^{2}} ,   \nonumber \\
 \beta_{\phi}^{1 \mathrm{PN}}\left(\eta, e_{t}\right)  & =  \frac{-4+\eta+e_{t}^{2}(8-2 \eta)+(4-\eta) \sqrt{1-e_{t}^{2}}}{e_{t} \sqrt{1-e_{t}^{2}}} \,, 
\end{align} 
\end{subequations}  
where $\chi$ stands for $1-e_t \cos u$.

We now impose the effects of GW emission and explain how we provide the full temporal
evolution for a BBH, characterized by  $m_1$ and $m_2$ and specified by 
initial values of $(x, e_t, \lambda, l )$. Clearly, we require to specify how 
these variables vary in time and Ref.~\cite{Damour2004} demonstrated that 
$\omega$ (or $x$) and $e_t$ evolve due to gravitational radiation reaction effects.
It turns out that the secular variations of $\omega$ and $e_t$ arise by employing the orbital (binding)
energy and angular momentum balance arguments \cite{BS89}.
This requires PN accurate expressions 
of $\omega$ and $e_t$ in terms of $(-2E)$ and $h$ and PN-accurate expressions for the orbit-averaged far-zone 
energy and angular momentum fluxes associated with non-spinning compact binaries in 
PN-accurate eccentric orbits \cite{JS92,gopu1997}.
We employ the following 3PN-accurate expressions for $\dot x$ and $\dot e_t$, extractable from 
Ref.~\cite{Arun2009}, and displayed symbolically as
\begin{widetext}
\bes      \label{x-dot-e-dot}
\begin{align} 
 \dot{x} & =  \frac{c^{3} x^{5} \eta}{G M} \left\{\frac{192+584 e_t^{2}+74 e_t^{4}}{15\left(1-e_t^{2}\right)^{7 / 2}}+\dot{x}^{1 \mathrm{PN}}\left(\eta, e_t\right) x+\dot{x}^{1.5 \mathrm{PN}}\left(\eta, e_t\right) x^{3 / 2} +\dot{x}^{2 \mathrm{PN}}\left(\eta, e_t\right) x^{2}       +\dot{x}^{2.5 \mathrm{PN}}\left(\eta, e_t\right) x^{2.5}   +\dot{x}^{3 \mathrm{PN}}\left(\eta, e_t\right) x^{3} \right\}\,,\\
 \dot{e_t} &=- \frac{c^{3} x^{4} \eta ~ e_t}{G M} \left\{\frac{304+121 e_t^{2}}{15\left(1-e_t^{2}\right)^{5 / 2}}+\dot{e}_{t}^{1 \mathrm{PN}}\left(\eta, e_t\right) x+\dot{e}_{t}^{1.5 \mathrm{PN}}\left(\eta, e_t \right) x^{3 / 2} +\dot{e}_{t}^{2 \mathrm{PN}}\left(\eta, e_t  \right) x^{2}+\dot{e}_{t}^{2.5 \mathrm{PN}}\left(\eta, e_t  \right) x^{2.5} +\dot{e}_{t}^{3 \mathrm{PN}}\left(\eta, e_t  \right) x^{3} \right\}\,.
\end{align}
\ens
\end{widetext}
The explicit expressions for various PN contributions are available in Refs. \cite{Arun2009, KBG18}.
We would like to point out that the contributions appearing at the 1.5PN, 2.5PN and 3PN orders 
contain certain hereditary contributions while 
the Newtonian, 1PN and 2PN terms are purely instantaneous.
Additionally, the 1.5PN and 2.5PN contributions in the above equations are purely hereditary and the relative 3PN terms 
contain both instantaneous and hereditary parts.
 The hereditary parts are expressed in terms of certain `eccentricity enhancement functions' (such as the ones given in Eqs. 6.22 of \cite{Arun2009}). We employed accurate  Pad\'e approximants for these enhancement functions and ensured that they are consistent 
 with analytic fits to these functions up to at least $e_t = 0.85$, as detailed in Ref.~\cite{KBG18}.

We now provide a prescription for the secular evolution $\lambda$ and $l$ as 
we specify the orbital configuration of our eccentric BH binary by specifying the initial values
of $(x, e_t, \lambda, l )$.
Eqs.~(\ref{e:FinalParamADM_2}) and (\ref{ST9}) which imply constant time derivatives of $l$ and $\lambda$,
are valid only when the radiation reaction is ignored. Under the radiation reaction,
the differential equations that specify the secular evolution of $\lambda$ and $l$ 
can be considered to be extensions of our 4PN order Keplerian type parametric solution and are given by
\bes          \label{lambda-dot-l-dot}
\begin{align}
 \frac{d \lambda}{d t}  & = \omega  \equiv  \frac{x^{3 / 2} c^{3}}{G M}  , \\
\frac{d l}{d t}  & = n  = \frac{ x^{3 / 2}  c^{3}}{G M }  \left\{   1+\dot{l}^{1 \mathrm{PN}}\left(e_{t}\right) x    +\dot{l}^{2 \mathrm{PN}}\left(\eta, e_{t}\right) x^{2}   \right.    \nonumber \\
 &  \left.   +\dot{l}^{3 \mathrm{PN}}\left(\eta, e_{t}\right) x^{3} +\dot{l}^{4 \mathrm{PN}}\left(\eta, e_{t},\ln x\right) x^{4}     \right\}     \label{l-dot-equation}
\end{align}
\ens
Promoting Eqs.~(\ref{ST4:full}) to Eqs.~(\ref{lambda-dot-l-dot}) to incorporate
radiation-reaction effects is basically an application of the method of variation of
arbitrary constants~\cite{Damour2004, poisson2014gravity}.
Both the 4PN local and nonlocal-in-time contributions
can be found in the accompanying \textsc{Mathematica} notebook~\cite{MMA2, MMA3}.
Note that Eq.~\eqref{l-dot-equation} arises from our
4PN order expression for $n$, given by Eq.~(\ref{eqn_n}),
and requires 4PN order expressions for $(-2\,E)$ and 3PN
order expression for $h$ in terms of $\omega$ and $e_t$.
Plus, as per our choice of ignoring the oscillatory 4PN
tail corrections to action-angles
(as per the semi-perturbation scheme) as detailed in \ref{4PN-tail-pert-Newt},
Eq.~\eqref{l-dot-equation} is the only equation in this
Sec.~\ref{4PNimr} which incorporates the 4PN tail effects
via Eqs.~\eqref{new-freq-1} and \eqref{new-freq-2},
as far as the IMR waveform construction is concerned.
Also, we are only imposing secular variations to the four variables whose initial values 
we employ to specify the BBH configurations. It is fairly straightforward to include quasi-periodic variations to these variables that occur at 2.5PN and 3.5PN orders, detailed in Ref.~\cite{KG06}.
We now give the 1PN contributions to Eqs.~\eqref{x-dot-e-dot} and \eqref{lambda-dot-l-dot} for illustration
\begin{widetext}
\begin{align}
\dot{x}^{1 \mathrm{PN}}\left(\eta, e_{t}\right)  & = \frac{-11888-14784 \eta+e_{t}^{2}(87720-159600 \eta)+e_{t}^{4}(171038-141708 \eta)+e_{t}^{6}(11717-8288 \eta)}{420\left(1-e_{t}^{2}\right)^{9 / 2}}      ,          \\
\dot{e}_{t}^{1 \mathrm{PN}}\left(\eta, e_{t}\right)  &= -\frac{67608+228704 \eta+e_{t}^{2}(-718008+651252 \eta)+e_{t}^{4}(-125361+93184 \eta)}{2520\left(1-e_{t}^{2}\right)^{7 / 2}} ,  \\
\dot{l}^{1 \mathrm{PN}}\left(e_{t}\right)  & = -\frac{3}{1-e_{t}^{2}}.
\end{align}
\end{widetext}

\begin{figure*}
  \includegraphics[width=\textwidth,height=7cm]{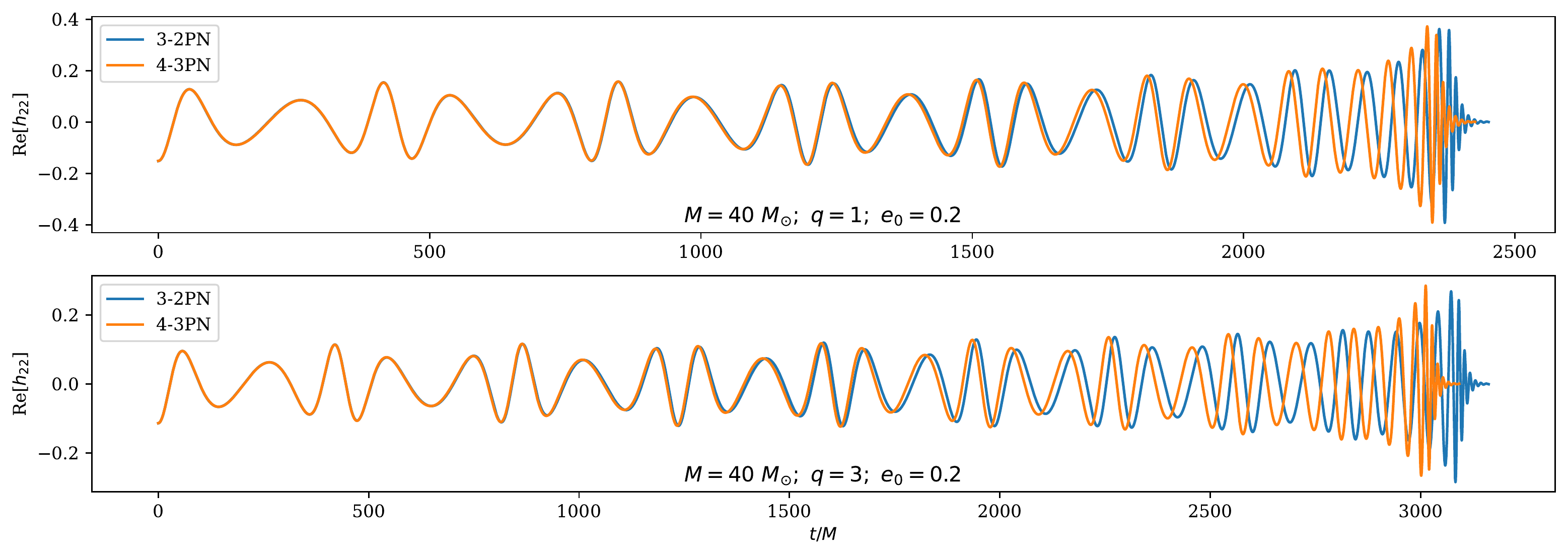}
  \caption{
Plots of GW strains that originate from  our approach (4PN-3PN) against the 
  3PN-2PN IMR family of Ref.~\cite{Hinder2017} for BH binaries with $q=1$ (top) and $q=3$ (bottom).
  Both plots begin at $F_{\text{orb}} = 15 $Hz  and we observe significant 
  de-phasing towards the later part of the inspiral. This may be attributed to the use of 
2PN order equations to describe the inspiral dynamics in  Ref.~\cite{Hinder2017}.
 ``$n$PN-$m$PN'' label stands for a waveform which includes $n(m)$PN conservative (reactive) dynamics. 
$e_0 \equiv e_t (F_{\text{orb}} = 15 $Hz) and  
 $G=c=1$ has been assumed for these plots.
}
  \label{fig:signal_7}
\end{figure*} 

\begin{figure*} 
  \includegraphics[width=\textwidth,height=17cm]{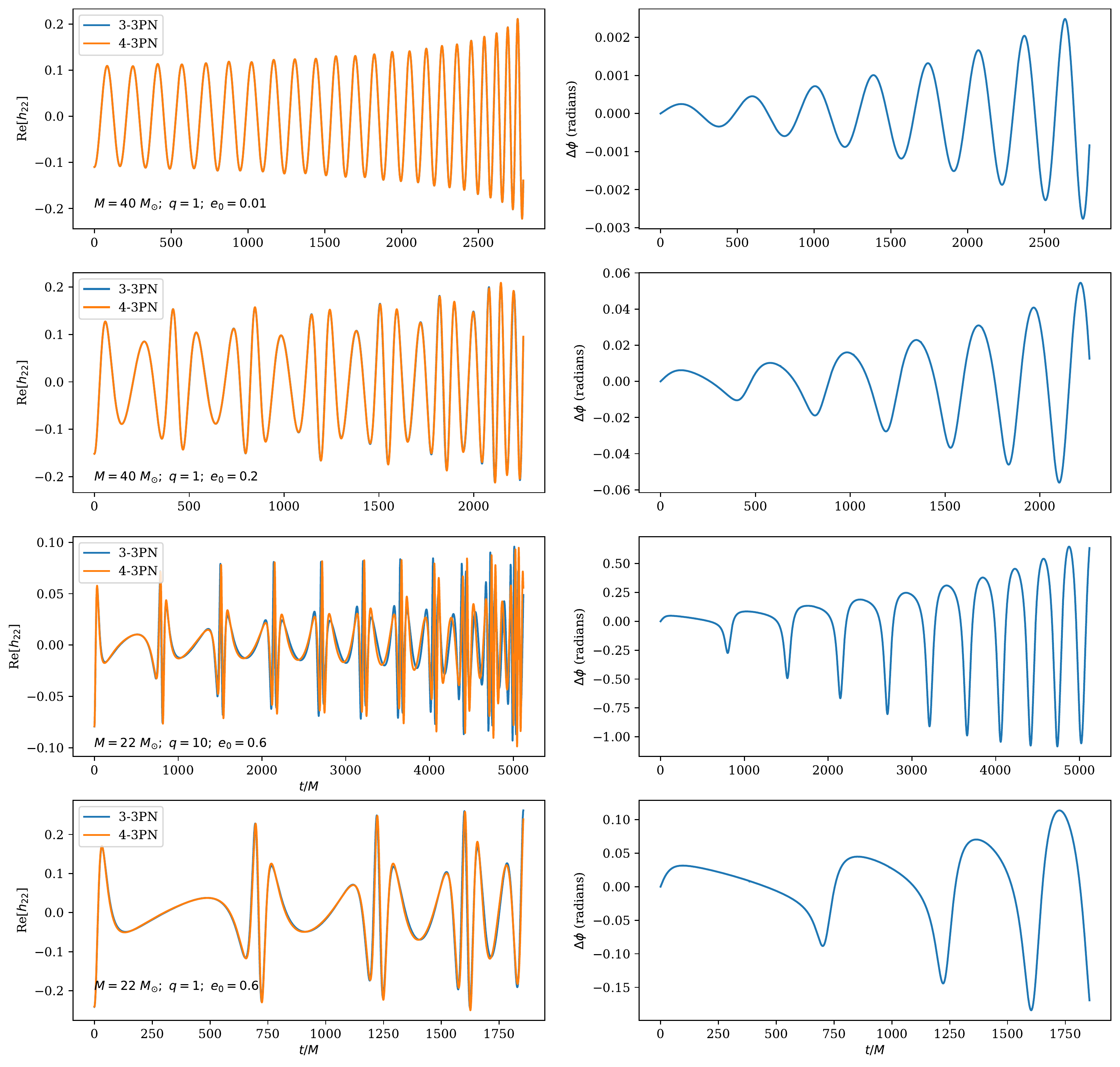}
  \caption{
We overlay plots of inspiral GW strains associated with the (4PN-3PN) and (3PN-3PN) inspiral families 
in the left panel. These plots are from $F_{\text{orb}} = 15 $Hz to $x = 1/6$ while varying 
values of the initial eccentricity and the mass ratio.
The associated differences in the accumulated orbital phases between these two inspiral families are plotted in the right panel.
The 4PN conservative contributions make noticeable differences for the moderately high and large mass ratio BH binaries.
Similar to Fig.~\ref{fig:signal_7}, ``$n$PN-$m$PN'' label stands for a waveform which includes $n(m)$PN conservative (reactive) dynamics.
$e_0 \equiv e_t (F_{\text{orb}} = 15 $Hz) and 
 $G=c=1$ has been assumed for these plots.
}
  \label{fig:signal_6}
\end{figure*}
 
\begin{figure*} 
  \includegraphics[width=\textwidth,height=6.5cm]{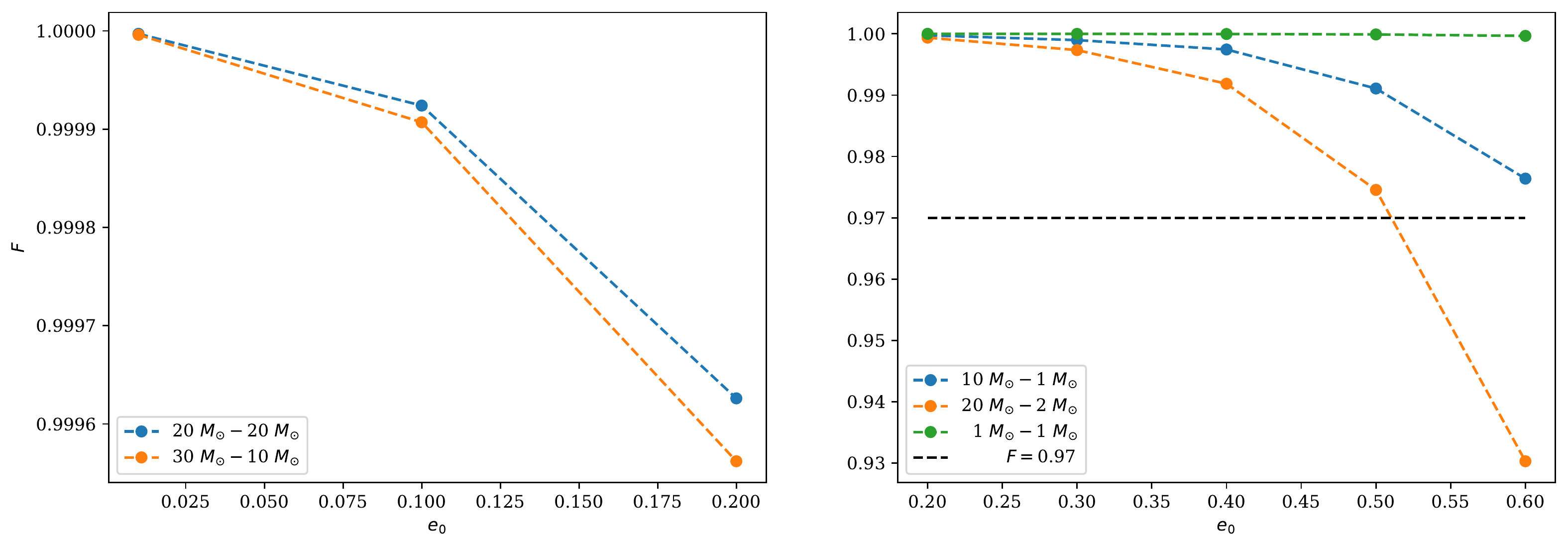}
  \caption{
Faithfulness plots between 4PN-3PN and 3PN-3PN eccentric IMR waveforms (left panel)
and the associated inspiral only waveforms (right panel) for various 
initial eccentricities at 
$F_{\text{orb}} = 15 $Hz and mass ratios with
 $e_0 \equiv e_t (F_{\text{orb}} = 15 $Hz ).
The inspiral waveforms of the right panel end at $x = 1/6$.
Small values of  $e_0$ 
and $q$  ($q \equiv m_1/m_2$) are employed in the left panel to ensure the
validity of our IMR waveforms. The 
inclusion of 4PN contributions do not affect the faithfulness estimates 
significantly. The right panel plots show that 
4PN order effects are relevant for BBHs with
higher $e_0$ and $q$.
This is consistent with 
the de-phasing plots of Fig. \ref{fig:signal_6}.
As noted earlier, the ``$n$PN-$m$PN''
 label stands for a waveform which includes $n(m)$PN conservative (reactive) dynamics.  
 }
  \label{fig:match_7}
\end{figure*}

The last ingredient, required to obtain temporal evolution for the dynamical variables 
$(r, \dot r, \dot \phi, \phi = \lambda+ W) $, is an accurate and efficient way of solving 
our 4PN order Kepler equation, given by Eq.~(\ref{K_Eqn_4PN}), which we symbolically write as
$l=\mathscr{L}(u, x,e_t)$. We employ Mikkola's method \cite{Mikkola} to numerically solve the 4PN order Kepler equation.
Since Mikkola's method was originally built and optimized to handle the classical Kepler equation (with no PN corrections),
the process of solving the 4PN order Kepler equation using this method requires the numerical inversion to be
done in a PN iterative manner as detailed in Ref.~\cite{Tanay2016}.
The need to bypass this numerical iteration procedure led us to
introduce a new `auxiliary eccentric anomaly' $\hat{u}$ such that our 4PN order Kepler equation 
takes the same form as the Newtonian one when written in terms of $\hat{u}$
\begin{align}
\label{K_eq_uhat} 
\mathscr{L}(u, x, e_t)  =  \hat{u}-e_t\,\sin \hat{u} ,
\end{align}
and therefore we can obtain $\hat{u}$ values essentially by employing Mikkola's method just once on Eq.~(\ref{K_eq_uhat}).
Thereafter, we evaluate $u$ from the PN accurate relation $u=u(\hat{u}, x,e_t)$ by 
demanding that the following equality holds
\begin{align}
l  =  \hat{u} - e_t \cos \hat{u} = u - e_t \cos u +\mathcal{O}(x^2)\,,                   \label{u-uhat}
\end{align}
where the RHS of the last equality includes corrections all the way up to 4PN
as in Eq.~(\ref{K_Eqn_4PN}).
The straightforward way to do so would be to start with an ansatz $u=  \sum  \mathcal{C}_i(\hat{u}) x^i$ 
and plug it into the RHS of the second equality of Eq.~\eqref{u-uhat} and evaluate the 
undetermined $\mathcal{C}_i(\hat{u})$ as functions of $\hat{u}$.
For illustration, we present $u$ in terms of $\hat{u}$ up to 2PN order as
\begin{widetext}
\begin{align}
u = \hat{u}  +  x^2  \left\lbrace   \frac{e_t \,\eta\,  (4 + \eta ) \, \sin \hat{u}}{8\, (1- e_t\, \cos \hat{u})^2}+\frac{3\, (5-2\, \eta) \,  }{\sqrt{1-e_t^2} \,(1- e_t \cos \hat{u})}    \arctan \left(\frac{\left(\sqrt{1-e_t^2}-1\right) \sin \hat{u}}{\left(\sqrt{1-e_t^2}-1\right) \cos  \hat{u} +e_t}\right)  \right\rbrace  +  {\cal{O}}(x^3)  , \label{u-ucap}
\end{align}
\end{widetext}
whereas the 4PN-accurate version of the above equation is provided in the accompanying \textsc{Mathematica}
notebook \texttt{Lengthy\_Expressions.nb}~\cite{MMA2, MMA3}.
We have verified that the relative difference between numerical values of $u$ and $\hat{u}$ around late inspiral
for a handful of cases is $\sim 0.0001 \%- 0.01 \%$ with the lower and upper bounds corresponding to cases with
$q \sim 1$ and $3$ respectively.

We now sketch  our procedure to generate temporally evolving eccentric inspiral $h^{22}(t)$ in   
 a computationally efficient way. The following steps are required
  \begin{enumerate}
\item  Specify an eccentric BH binary configuration by providing values 
for  $\{x,e_t, \lambda, l \}$ at an initial epoch $t_0$ along with values of fixed parameters
like $m_1, m_2$ and $D$.
\item
With the help of our numerical solution to 4PN order Kepler equation via Eq.~(\ref{K_eq_uhat})
and Eqs.~(\ref{4PN_dy_var})
for $R, \dot R, \phi $ and $ \dot{\phi} $,
we evaluate Eq.~(\ref{h22_N}) and obtain the value of $h^{22}$ at $t_0$.
\item
Thereafter, solve the coupled differential equations for $x,e_t,\lambda$ and $l$ to obtain their values at $t_0+ \Delta t$.
Find $R, \dot R, \phi, \dot{\phi}$ and $h^{22}$ at $t_0 + \Delta t$ using the same steps as before and so on.
\end{enumerate}
This gives the quadrupolar order waveform $h^{22}$ associated with  our  eccentric and non-spinning BH binary 
that inspirals due to the effect of 3PN-accurate GW emission along 4PN order orbits.

A few comments are required to contrast our approach with the $x$-model of Ref.~\cite{Hinder2017}.
A close inspection reveals that our differential equations for $x,e_t,\lambda$ and $l$ 
do not contain any orbital time scale variations.  However, the numerical 
treatment of  $\dot \phi$ equation in the $x$-model of Ref.~\cite{Hinder2017} ensures that the secular and periodic 
variations are intertwined in their approach, especially while dealing with the orbital phase evolution.
The reason behind this contrast is that we have closed-form solutions for the orbital time-scale dynamics
(in the form of 4PN generalized quasi-Keplerian parameterization) and hence numerical integration can happen
on radiation-reaction time scale.
Additionally, we use Mikkola method
to solve our PN Kepler equation without needing to iterate as in Ref.~\cite{Tanay2016}
because of writing the PN Kepler equation in terms of a new variable $\hat{u}(u)$. 
These features make our computation routine efficient,
apart from being accurate (3PN and 4PN accurate in
reactive and conservative dynamics, respectively).
It should be noted that we have not incorporated the GW emission induced orbital time scale 
variations to  $x,e_t,\lambda$ and $l$, as detailed in Refs.~\cite{Damour2004,KG06},
although this is fairly straightforward 
to do so using our prescription. However, it will be rather difficult to include such periodic
contributions in the $x$-model of Ref.~\cite{Hinder2017} due to their  use of PN-accurate $\dot \phi( x,e_t,u)$ expression.
In what follows, we explain briefly how we attach our eccentric inspiral $h^{22}$ to the circular 
merger $h^{22}$ model of Ref.~\cite{Hinder2017} while treating their approach as a black box.

\subsection{Stitching the circular merger-ringdown waveform to the eccentric inspiral waveform}     \label{subsection: stitch}

As mentioned earlier, we exclusively follow Ref.~\cite{Hinder2017} as a black box
when it comes to stitching the merger-ringdown waveform to our inspiral waveform and it involves the following steps. 
First, we mark four special time 
instants: $t_\text{ref}$, $t_\text{blend}$, $t_\text{circ}$ and $t_\text{peak}$ such that 
$t_\text{ref} < t_\text{blend}  < t_\text{circ} < t_\text{peak}$. The first two instants correspond to 
epochs when 
$x=0.11$ and $0.12$, respectively. Further, the instant $t_\text{circ}$ is the time after which the binary can be 
treated as circular and $t_\text{peak}$ is the time at which the dominant $  \ell =2, m=2$ 
mode of the GW strain $h_{22}$ reaches the maximum value in magnitude.

Importantly, Ref.~\cite{Hinder2017} demonstrated the circularization of  BBHs with 
$q<3$ and initial eccentricity $e_t \lesssim 0.2$ (measured when  $x \sim 0.07 $)
for $t> t_{\text{peak}} - 30 M $.
This implies  that $t_\text{circ} = t_\text{peak} - 30 M$.
It allowed Ref.~\cite{Hinder2017} to propose the use of their  circular merger model 
(CMM)  
for $t > t_\text{circ}$. For its construction, three circular, non-spinning BBH
waveforms from the Simulating eXtreme Spacetimes catalog \cite{Mroue:2013xna}
with $q=1,2$ and $4$ were used and an interpolating function for the amplitude and frequency
was constructed for all $q$ values in the range $1<q<4$.


 We employ our earlier described PN-accurate eccentric waveform 
 in the interval $  t < t _{\text{blend}}$ by specifying the eccentric BBH with its   
PN parameters $(x, e_t , \phi , l)$\footnote{Switching from $(x,e_t,\lambda,l)$
to $(x,e_t,\phi,l)$ is easy since $\phi = \lambda + W(u(l))$} at a certain initial epoch.
We employ the CMM for $t>t _{\text{circ}}$.
This leaves only the `blending region' to deal with, which is the interval $t_{\text{blend}} < t <t_{\text{circ}}$. 
Using 23 NR simulations, a fitting function for $\Delta t \equiv t_{\text{peak}} - t_{\text{ref}}$
was arrived at (see Eq.~(9) of Ref.~\cite{Hinder2017}) with $q,e_t (x=0.11)$ and $l(x=0.11)$ as its arguments. 
This $\Delta t$ fit is crucial to model the waveform in the blending region as we soon see 
below in Eqs.~\eqref{blending-eqns}.

Let us mention that there was a certain ``NR-PN'' fitting done in
Ref.~\cite{Hinder_10} which made it possible to assign PN
parameters ($x$ and $e$) to the 23 NR simulations in Table~I of Ref.~\cite{Hinder2017}; 
the top row of the table displays those PN parameters. Assignment of these
parameters was crucial to performing the above mentioned $\Delta t$ fit
and by extension, constructing the IMR waveform of Ref.~\cite{Hinder2017}. As expected,
this NR-PN fitting is sensitive to the PN model used and becomes less so if the fit is performed
earlier during the inspiral phase, as is evident in Figs.~4 and 5 of Ref.~\cite{Hinder_10}.
This is so because the PN parameter is smaller during earlier stages of inspiral.
Since Ref.~\cite{Hinder_10} explicitly mentions that the
authors chose the earliest possible fitting interval
to extract ``a unique set of PN parameters'', we don't feel the need to redo these NR-PN fits
and thereby the $\Delta t$ fit. Hence we don't perform them and stick with 
Eq.~(9) of Ref.~\cite{Hinder2017} for our IMR waveform.

Now we discuss how to deal with the blending region. 
The waveform is written as $h_{\text{sig}} = A e^{i \phi_w}$ with $\omega_w \equiv \dot{\phi_w}$.
The subscript `$w$' stand for `wave' and serves to distinguish $\phi_w$ and $\omega_w$ from the orbital phase $\phi$ and frequency $\omega$.
The IMR waveform in the blending region is then given by Eqs.~(10)-(17) of Ref.~\cite{Hinder2017}
which we reproduce almost exactly below (with the superscripts `PN' and `circ' standing for the
PN waveform model and the CMM)\footnote{Eqs.~(\ref{wrong_eqn_1}) and (\ref{wrong_eqn_2}) are different
from Eqs.~(14) and (15) of Ref.~\cite{Hinder2017} because there are typos in the latter set of equations.}
\begin{subequations}      \label{blending-eqns}
\begin{align}
t_{\mathrm{peak}} &= t_{\mathrm{ref}} + \Delta t   , \\
t_{\mathrm{circ}} &= t_{\mathrm{peak}} - 30  M   , \\
\alpha(t) &= \mathcal{T}(t;t_{\mathrm{blend}},t_{\mathrm{circ}})     ,  \\  
h_{\text{sig}}(t) &=  A(t) e^{i \phi_w(t)}       ,    \\
A(t) &= (1-\alpha(t)) A^{\text{PN}} + \alpha(t) A^{\text{circ}}(t  -t _{\mathrm{peak}})    ,    \label{wrong_eqn_1} \\
\omega_w(t) &=   (1-\alpha(t))  \omega_w^{\text{PN}} + \alpha(t)  \omega_w^{\text{circ}}(t  -t _{\mathrm{peak}})    ,     \label{wrong_eqn_2}  \\
\phi_w(t) &= \int^t \omega_w(t') dt'    .
\end{align}
\end{subequations}
\begin{equation}
\mathcal{T}(t,t_1,t_2) =
\begin{cases}
  0 & \text{ } t   \leq t_1    ,   \\
  \left [ \exp \left ( \frac{t_2-t_1}{t-t_1} + \frac{t_2-t_1}{t-t_2} \right ) + 1 \right ]^{-1} & \text{ } t_1 < t < t_2     ,  \\
  1 & \text{ } t \ge t_2     .
\end{cases} 
\end{equation}
A few remarks in regard to the above equations are in order.
$ A^{\text{PN}} ,  \omega_w^{\text{PN}} , A^{\text{circ}}$ and $\omega_w^{\text{circ}} $
refer to the values of the amplitude and frequency (time derivative of the phase) of the
PN and the CMM waveforms, respectively.
The function $\mathcal{T}(t,t_1,t_2)$ is a smooth function which goes from $0$ to $1$ between
$t_1$ to $t_2$ and essentially does the job of `blending' the PN waveform with the CMM 
waveform. The argument of $A^{\text{circ}}$ and $\omega_w^{\text{circ}}$ in the above equations is $(t  -t _{\mathrm{peak}})$ because
the CMM is time-shifted such that the peak occurs at $t=0$.

We are now in a position to plot the waveforms got from our extension of 
the eccentric IMR family of Ref.~\cite{Hinder2017} 
which we term as the 4-3PN (4PN conservative, 3PN reactive) IMR waveform in contrast
to the 3-2PN (3PN conservative, 2PN reactive) family of Ref.~\cite{Hinder2017}.
In Fig. \ref{fig:signal_7}, we provide a visual contrast  between these two IMR families 
for two mass ratios, namely $q=1$ and $3$.
Clearly, some dephasing is  evident near the late inspiral and it may be attributed to the use of 
3PN-accurate effects of GW emission in our approach.
Let us mention that we have added some missing terms in the 
expression of $\dot{x}$ in the \textsc{Mathematica} package accompanying Ref.~\cite{Hinder2017} prior
to plotting. Further, we have gauge-transformed
the eccentricity parameter before making the above comparison because 
the work of Ref.~\cite{Hinder2017} was done in the harmonic gauge. 
We now move on to explore preliminary data analysis implications of our time-domain 
IMR waveforms.

\subsection{Preliminary data analysis implications}


 It will be interesting to explore the de-phasing and possible data analysis implications of 
 our updated eccentric IMR waveform family in comparison with what is available in the literature.
A cursory look of the  Fig.~\ref{fig:signal_7} reveals that the effect of neglecting 3PN contributions to 
the GW emission can be substantial and 
we infer that 
this is mainly due to the late inspiral stage de-phasing of 4-3PN IMR 
waveform in contrast to the 3-2PN family.
Therefore, we here restrict our attention to explore the implications of our 4PN order 
contributions to the conservative dynamics while keeping the radiation reactions effects fixed 
at the 3PN order. Further, we restrict our attention to the inspiral domain only.
The plots in Fig.~\ref{fig:signal_6} provide visual comparisons between the 
members of the eccentric 
$4$-$3$PN and $3$-$3$PN  inspiral waveform families and the associated orbital phase differences 
for 3 different initial values of orbital eccentricity.
The plotted waveforms on the left panels of the first two rows are visually indistinguishable and their 
differences in the accumulated orbital phases are less than a fraction of a radian.
However, the plots in the bottom row suggests that the effects of 4PN contributions 
to the conservative dynamics can shift the location of periastron passages in the moderately high 
initial eccentricity scenario as is evident from the differences in the peaks of these two waveform families.
Additionally, the accumulated orbital phase difference can be roughly a radian for such eccentric binaries.
Therefore, it is reasonable to expect that 4PN order contributions to the conservative orbital 
dynamics should be relevant for high eccentric black hole binaries.

 We now move on to probe a preliminary GW data analysis implications of our 4PN order corrections to 
 the orbital dynamics by computing certain ``faithfulness'' estimates between the members of the above two inspiral 
 families.
The faithfulness between two GW signals $h_1(t)$ and $h_2(t)$ is defined as the following integral,
maximized over the time and phase of coalescence  $t_c$ and $\phi_c$ such that \cite{Hinder2017, 2019PhRvD.100f4006T, DIS}
\begin{align}
 F(h_1, h_2)  \equiv   \max_{t_c, \phi_c}    \frac{ (h_1 (t_c, \phi_c),h_2)    }{\sqrt{     (h_1,h_1)  (h_2, h_2)           }} \,,              \label{faithfulness}
\end{align} 
and  the above inner product between two waveforms  $(h_1, h_2)$ being defined as
\begin{align}
\label{Eq:innerproduct}
 (h_1,h_2)    &  \equiv  4\,   \Re   \,  \int_{f_{\rm min}}^{f_{\rm max}} \, 
\frac{\tilde h_1^*(f)\, \tilde h_2(f)}{S_{\rm h}(f)} df \,,
\end{align}
where $\tilde h_i(f)$ stands for the Fourier transform of $h_i(t)$ and $S_h(f)$ is the noise power spectral
density of the aLIGO detector.
For the present analysis, we employed 
the zero-detuned, high-power noise configuration
of aLIGO \cite{aLIGOpsd}.
The faithfulness plots  of  Fig. \ref{fig:match_7}  are for non-spinning compact binaries of different 
mass ratios. The faithfulness values are above the traditional 
$0.97$ value for equal mass binaries 
which suggest that   $3$-$3$PN  inspiral templates are `faithful' to the expected
$4$-$3$PN   inspiral GW signals \cite{DIS}.  Additionally,  the above  commonly accepted lower bound
ensures the recovery of $90 \%$ of the hidden signals in the noisy data \cite{1995PhRvD52605A, 1999PhRvD60b2002O}.
However, the $3$-$3$PN inspiral templates need not be `faithful' to their $4$-$3$PN  counterparts for 
moderately eccentric and unequal mass compact binaries as evident from the plots in the right panel of 
Fig.~\ref{fig:match_7}.  Therefore, it should be interesting to pursue computations to 
incorporate  the effects of 4PN hereditary contributions to the  conservative dynamics and model 
the resulting inspiral templates.

A true measure of validity of our 4-3PN IMR waveform 
can be had from comparisons with
the corresponding NR simulations
which we do not perform this paper. 
However it's reasonable to expect our IMR waveform to be 
valid across a slightly larger parameter range than that for the waveform of 
Ref.~\cite{Hinder2017} 
although we have not checked it.
As far as the  Ref.~\cite{Hinder2017} is concerned, the
faithfulness $F$ (defined in Eq.~\eqref{faithfulness})
between their IMR waveform and the NR counterparts
 $> 0.97 $ for systems with
$ M \leq 85 M_{\odot},~e_{t(7)} \leq 0.08,~q  \leq 3$
and $ M \geq 70 M_{\odot},~e_{t(7)} \lesssim 0.05,~q  \leq 3$ with $e_{t(7)}$
here standing for the eccentricity at $\sim 7$ cycles before the merger.
This faithfulness was computed in Ref.~\cite{Hinder2017} with the detector configuration 
of the first observing run (O1) of advanced LIGO with $f_{\text{min}}$ 
(lower frequency limit in Eq.~\eqref{faithfulness}) being 30 Hz.

It should be obvious that the validity range in eccentricity of our  inspiral prescription
(based on PN equations)
should be much larger than that for the entire IMR waveform.
 This is so because stitching the inspiral
part with the CMM part severely restricts the validity
 range of the IMR waveform in eccentricity
as the CMM assumes that the BBH has circularized towards the merger. 
We therefore also present a derived package which outputs the waveform by
only using the PN equations of motion \cite{MMA2, MMA3}
and is valid for much higher $e_t$'s.
Recall that most of our PN equations are exact in $e_t$, except
for the Pad{\'e} approximants, all of which do reasonably well up to $e_t \lesssim 0.85$.
Of course, this PN package can be trusted for the inspiral part only.

\section{Summary and Next Steps}

We incorporated the dynamics entailed by the 4PN-accurate Hamiltonian 
for non-spinning compact binaries, available in Ref.~\cite{JS_15}.
This was done using Keplerian type solution for the local part of the Hamiltonian
and canonical perturbation theory for the non-local part.
We ignored certain zero-average,
oscillatory terms from our solution arising due to 4PN tail effects.
  Additionally, we provided  consistency checks with existing results and detailed checks on the correctness of our lengthy results.
 Thereafter, we  employed our results to obtain an eccentric IMR family 
 by adapting  what is done in Ref.~\cite{Hinder2017}.
 This included providing an accurate and efficient implementation of GW phasing approach of 
 Ref.~\cite{Damour2004} to model the inspiral part of eccentric IMR family.
 A preliminary and quick study revealed that the conservative 4PN order contributions should be
 relevant for BH binaries with high mass ratios and moderately high initial eccentricities.

We modify the publicly available \textsc{Mathematica} package \cite{MMA1} of Ref.~\cite{Hinder2017} to obtain
our updated eccentric IMR family using our 4PN-order parametric solution~\cite{MMA2, MMA3}.
We also present a derived package which makes use of
only the PN equations of motion to produce the waveforms and 
hence can be trusted only for the 
inspiral part. Since this second package does not use the circular-merger-model, it should be
valid for a much higher range of $e_t$ ($e_t \lesssim 0.85$) than our former package.

  There are  a number of extensions that should be pursued in the near future to provide a ready-to-use 
  time domain eccentric IMR templates. Naturally, such a waveform family should be able to model eccentric mergers
  that should allow us to model BH binaries with moderately high initial eccentricities of $e_t \sim 0.5$.
  A very recent effort that developed a method to reconstruct eccentric merger waveforms from its circular
 counterparts, detailed in Ref.~\cite{2021arXiv210111033S}, should be helpful for such an effort.
It will be interesting to stitch what is done in Ref.~\cite{2021arXiv210111033S}  to  our inspiral approximant and
probe its validity with the full NR based eccentric IMR waveforms for various initial eccentricities and mass ratios.
 It will be also desirable to include the spin effects to our inspiral part 
as these effects are now fully computed to the next-to-leading order  
\cite{SO_1PN,SS_1PN}.
However, it will be desirable to develop Keplerian type parametric solution associated 
with these next-to-leading order effects 
by extending what is pursued in 
 Ref.~\cite{KBG18}.
 Additionally, efforts should be pursued to 
 extend our solution for non-spinning binaries to higher PN orders 
since the action and Hamiltonian computations of compact binaries have recently been pushed 
to 6PN \cite{2019PhRvL.123w1104B, 2020PhRvD.102b4062B, 2020PhRvD.102b4061B}. 
Very recent efforts suggest that it should be possible to include 4PN order contributions to the GW emission 
 in the coming years~\cite{4PN_QM_1,4PN_QM_2}.
However, this will require us to provide our improved Keplerian type parametric solution to 4PN 
order in the modified harmonic gauge with the help of 
Ref.~\cite{Marchand18}.
We plan to complete 4PN-accurate GW phasing for eccentric binaries with such inputs while incorporating 
the currently missing radiation reaction induced periodic terms in orbital elements that appear at 2.5, 3.5PN and 4PN orders 
\cite{Damour2004,KG06}. 


\section*{Acknowledgements}

We are grateful to Piotr Jaranowski, Gerhard Sch{\"a}fer and Donato Bini for their encouragements and 
illuminative suggestions.
We appreciate the detailed constructive feedback
from the anonymous referee.
We also thank Thibault Damour, Leo Stein, Jan Steinhoff, Antoni Ramos Buades, 
Sumeet Kulkarni, Anuradha Gupta and Justin Vines for helpful discussions and comments.
G.C is supported by the European Research Council (ERC) Consolidator Grant ``Precision Gravity: From the LHC to
LISA" provided by the ERC under the European Union's H2020 research
and innovation programme, grant agreement No. 817791.
A.G. acknowledges support of the Department of Atomic Energy, Government of India, under Project Identification No. RTI 4002. 
HML was supported by the National Research Foundation of Korea
 grant number NRF-2021R1A2C2012473 from Ministry of Science and ICT, Korea.

\appendix

\section{Obtaining factorized PN-accurate expressions for $dt/ds $ and $ d \phi/ds$}
\label{AppA} 

We provide steps that are required to obtain Eqs.~(\ref{Eq_4PN_t_t0_N}) and 
(\ref{dphids_4PN})
from PN-accurate expressions for   
$\dot r^2      =  \dot{s}^2 /s^4   $
and $ d \phi / ds=\dot \phi/\dot{s}$ while restricting our attention to
the 1PN order. The 1PN-accurate expression for $\dot r^2$ reads
\begin{widetext}
\bea\label{dotr2a}
\dot r ^2 \equiv  \frac{1}{s^4}
{\left({\frac {{ ds}}{{ dt}}}\right)}^{2}&=& 
\biggl ( 2E +{\frac {3{E}^{2} }{{c}^{2}}} (-1+3\eta)  \biggr )
+ 
\biggl ( 2+{\frac {2E }{{c}^{2}}} (-6+7\eta) \biggr )\, s
+
\biggl (
-{h}^{2} +{\frac{1}{{c}^{2}}} \bigg[5(-2+\eta)+2{h}^{2}E(1-3
\eta) \bigg]
\biggr )\, s^2 
\no&&
+ {\frac {{h}^{2}}{{c}^{2}}}(8-3\eta)^2\, s^3\,.
\eea
We may symbolically write 
$\dot r ^2 \equiv   \dot{s}^2 /s^4    $ expression as
$a_{{0}}+a_{{1}}s+
a_{{2}}{s}^{2}+a_{{3}}{s}^{3}$.
$a_{{0}}+a_{{1}}s+
a_{{2}}{s}^{2}+a_{{3}}{s}^{3}=0$ has 3 roots but  only two real roots $s_-$ and $s_+$
exist at the Newtonian order and therefore are of interest to us.
These two roots, expressed in terms of $h$, $E$ and $\eta$, have already been given in Eq.~(\ref{two_roots}).
It is possible to factorize in the following form
\begin{align}
 \dot r ^2 \equiv  \frac{1}{s^4}
{\left({\frac {{ ds}}{{ dt}}}\right)}^{2} &=
(s_-  \times s) \times (s-s_+) 
\bigg(\frac{h^2\,(-8+3\,\eta)}{c^2}s+h^2+\frac{1}{c^2}\big[-6+\eta+2\,h^2\,E\,(-1+3\,\eta)\big]\bigg)+ {\cal O} \left( \frac{1}{c^4}   \right)   \,.
\end{align}
It is now straightforward to write 
Eq.(\ref{dotr2a}) as 
\begin{align} 
\frac{dt}{ds}&=   \pm  \frac{1}{s^2}\frac{1}{\sqrt{a_{{0}}+a_{{1}}s+
a_{{2}}{s}^{2}+a_{{3}}{s}^{3}}}\,,\no
&=  \pm  \frac{1}{s^2}\frac{1}{\sqrt{(s_--s)(s-s_+)}}\left\{\frac{1}{h}+\frac{-\eta +h^2(2-6 \eta ) \,E+6}{2\,c^2\, h^3}+\frac{8-3 \eta }{2\,c^2\, h^3}s+{\cal O}(1/c^4)\right\}\,,\no
&=  \pm   \frac{1}{s^2}\frac{A_0+A_1\,s}{\sqrt{(s_--s)(s-s_+)}}\,.    \label{dtds}
\end{align}
It is now straightforward to extend the above computation to obtain to 4PN order 
\begin{align}
\frac{dt}{ds} = \frac{A_0+A_1\, s+A_2\, s^2
+A_3\,s^3+A_4 \,s^4+A_5\, s^5+A_6\,s^6+A_7 \,s^7}{\sqrt{(s_-- s)( s-s_+)}\, s^2}\,.
\label{Eq_4PN_t_t0}
\end{align}
\end{widetext}
This allows to write Eq.~(\ref{Eq_4PN_t_t0_N}) for $(t-t_0)$ at 4PN order.

In a similar fashion, we may start from 4PN order expression for $\dot \phi$, namely
$  \dot{\phi}=b_2\,s^2+b_3\,s^3+b_4\,s^4+b_5\,s^5+b_6\,s^6+b_7\,s^7+b_8\,s^8+b_9\,s^9$,
where again each of these coefficients are expressed in terms of $h$, $E$ and $\eta$, and extract a suitable expression for $ d\phi/ds$.
Restricting our attention to 1PN order, we find  
\bes
\bea
b_2&=&h -\frac{h\,E\,(1 + 3\,\eta)}{c^2}\,,\\
b_3&=&\frac{2\,h\,(-2+\eta)}{c^2}\,,
\eea
\ens
while all other coefficients have no contributions at 1PN order. 
Using Eq.(\ref{dtds}),
\begin{widetext}
\begin{align}
\frac{d\phi}{ds}&=\frac{d\phi}{dt}\times\frac{dt}{ds}\\\notag
&=\pm\frac{b_2+b_3\,s}{\sqrt{(s_--s)(s-s_+)}}\left\{\frac{1}{h}+\frac{-\eta +h^2(2-6 \eta ) \,E+6}{2\,c^2\, h^3}+\frac{8-3 \eta }{2\,c^2\, h^3}\,s\right\}+{\cal O}\left( \frac{1}{c^4} \right)\,,\\\notag
&=\pm\frac{B_0+B_1\,s}{\sqrt{(s_--s)(s-s_+)}}+{\cal O}\left( \frac{1}{c^4} \right)\,.
\end{align}
This procedure may be extended to 4PN order and this leads to
\begin{align}
\frac{d\phi}{ds}=\pm\frac{B_0+B_1\, s+B_2\, s^2+B_3\,s^3+B_4\, s^4+B_5\, s^5 +B_6\, s^6+B_7\,
s^7}{\sqrt{(s_-- s)( s-s_+)}}\,.
\end{align}
\end{widetext}
This helps us obtain the parametric solution for the angular part.

\section{Evaluating the temporal and angular integrals}
\label{AppB}

Here we sketch the steps required to obtain Eqs.~(\ref{Eq_t_t0_temp}) and (\ref{temp_phi}) from their integral forms, 
 namely Eqs.~(\ref{Eq_4PN_t_t0_N}) and (\ref{phi_3PN}). The plan is to provide details of the 1PN computations
as their PN extensions are fairly straightforward although 
computationally demanding. 
We begin by inspecting the 1PN-accurate version of Eq.~(\ref{Eq_4PN_t_t0_N}) which may be written as
\begin{align}
t - t_0 =& \int_s^{s_-}\frac{A_0+A_1\bar s}{\sqrt{(s_--\bar s)(\bar s-s_+)}\,\bar s^2}d \bar s \,,
\end{align}
A direct integration, if possible, should lead to an expression in terms of $E,h,s_-,s_+$ and $s$.
Therefore, certain additional relations are required to obtain Eq.~(\ref{Eq_t_t0_temp}) that employs the auxiliary true anomaly like variable.
It is obvious that we need to tackle integrals like
\begin{align}
 \int_s^{s_-}\frac{ds}{\sqrt{(s_--s)(s-s_+)}}\,f(s)
\end{align}
where $s=1/(a_r(1-e_r\,\cos u)) $ 
and 
 $ds=du\,s^2\,a_r\,e_r\,\sin u$. This leads to 
 \begin{align*}
 \int^u_0\frac{du \sqrt{1-e_r^2}}{a_r\,(1-e_r\,\sin u)}f(u)\,,
 \end{align*}
 where $f(u)= f(s(u))$.
 The structures of the temporal and angular integrals suggest that we require to tackle integrals of the form 
\begin{align}
I(u,n)=\int^u_0 \frac{d\bar{u}}{(1-e_r\,\cos \bar{u})^{n}}\,,
\end{align}
where $n$ is an integer. 
We may introduce an angular variable 
 $\tilde{\nu}(u)   \equiv  \sqrt{1-e_r^2}\,I(u,1)=2\,\arctan(\sqrt{{(1+e_r)}/{(1-e_r)}}\tan\frac{u}{2})$. 
It turns out that $d \tilde{\nu}/du=\sqrt{1-e_r^2}\,f(u)$ where $f(u)  \equiv  1/(1-e_r\,\cos u)$ such that 
\begin{align}
f(u)=\frac{1+e_r\,\cos(\tilde{\nu}(u))}{1-e_r^2}\,.
\end{align}
The above relation is influenced by the relation that connects eccentric and true anomalies of the Keplerian 
parametrization. We now introduce 
 a functional $G^{(n)}[\tilde{\nu}]$ such that $I(u,n)=G^{(n)}[\tilde{\nu}]$ and we  differentiate it with respect to $u$ to get
\begin{align}
\frac{d}{du}I(u,n)&=f(u)^n\,,\\\notag
\frac{d}{du}G^{(n)}[\tilde{\nu}]&=\frac{\delta G^{(n)}}{\delta \tilde{\nu}}\frac{d\tilde{\nu}}{du}=\sqrt{1-e_r^2}\,f(u)\frac{\delta G^{(n)}}{\delta \tilde{\nu}}\,.
\end{align}
We now equate the above 
 two expressions to obtain 
\begin{align}
\frac{\delta G^{(n)}}{\delta \tilde{\nu}}=\frac{f^{n-1}(u)}{\sqrt{1-e_r^2}}=\frac{1}{(1-e_r^2)^{n-\frac{1}{2}}}\,\big(1+e_r\,\cos(\tilde{\nu})\big)^{n-1}\,.
\end{align}
The above equation allows us to write integrals like $I(u,n)$ in terms of 
simple trigonometric integrations like  $\int\, \cos( n \tilde{\nu})  \, d\tilde{\nu}$.

 With these inputs, we may tackle the above 1PN accurate temporal integral 
\begin{align}
t - t_0 =& \int_s^{s_-}\frac{A_0+A_1\bar s}{\sqrt{(s_--\bar s)(\bar s-s_+)}\,\bar s^2}d \bar s \,,\\
=&\sqrt{1-e_r^2}\,\int^u_0\,d\bar{u}(\frac{A_0}{\bar{s}}+A_1)    \\
=& \sqrt{1-e_r^2}\,(A_0\,a_r\,(u-e_r\,\sin u)+A_1 \,u)    \\
\equiv &  \tilde{A}_0\,u+\tilde{A}_1\,\sin u.
\end{align}
 Similar arguments allow us to tackle  the 1PN-accurate orbital phase integral and it gives 
\begin{align}
\phi-\phi_0&= \int_s^{s_-}\frac{B_0+B_1\,\bar{s}}{\sqrt{(s_--\bar{s})(\bar{s}-s_+)}}\,,\\
&=\sqrt{1-e_r^2}\,\int^u_0\,d\bar{u}\,({B_0}\,\bar{s}+{B_1}\,\bar{s}^2)\,,\\
&=B_0\,\tilde\nu+\frac{B_1}{(1-e_r^2)}\int d\tilde\nu\,\big(1+e_r\,\cos(\tilde{\nu})\big)\,,\\
&=B_0\,\tilde\nu+\frac{B_1}{(1-e_r^2)}(\tilde{\nu}+e_r\sin(\tilde\nu))
\end{align}
It is not very difficult to figure out that 
the above expressions lead to 1PN versions of Eqs.~(\ref{Eq_t_t0_temp}) and (\ref{temp_phi}).

 Finally, we note that additional relations like
\bs
\bea
s& =& \frac{1}{a_r\,(1-e_r\cos(u))} =
\frac{1+e_r \cos \tilde v}{a(1-e_r^2)}\,,\\
u &=& \arccos \left( \frac {s_-+s_+}{s_--s_+}-2\frac{s_-  \times s_+}
{(s_--s_+)s} \right) \,,\\
\tilde v &=& 
\arccos \left(\frac {2s}{s_--s_+}-\frac{s_-+s_+}{s_--s_+}\right)
\,.
\eea
\es
are required to obtain 4PN order Kepler Equation and they are also heavily 
employed to obtain the parameteric solution to the angular integral at higher 
PN orders.

\section{Explicit 4PN order expressions for $e_t$ and its connection with $e_{r}$ and $e_{\phi}$}
\label{AppC} 
 We display below 4PN order expressions for the radial, time and 
 angular eccentricities in ADM-like gauge

 \begin{widetext}
\bes
\label{Eq_e4PN_AppC}
\bea
{\it e_t}^2 &=&1+{2\,E}\,{h}^{2}+
{\frac {{(-2\,E)}}{4\,{c}^{2}}}\, {\bigg\{ -8+8\,
\eta
- \left( -17+7\,\eta \right) {(-2\,E\, h^2)} \bigg\} }
\no&&
+
\frac{{{(-2\,E)}}^{2}}{8\,{c}^{4}} \bigg\{8+4\,\eta +20\,{\eta}^{2}
-  {(-2\,E\,h^2)}( 112-47\,\eta
+16\,{\eta}^{2} )
\no&&
-24\,\sqrt{(-2\,E\,h^2)}\, \left( -5+2\,\eta \right) 
+\frac{4}{(-2\,E\,h^2)} \left( 17 - 11\,\eta \right )
\no&&
-\frac{24}{ \sqrt{ (-2\,E\, h^2)}}
\, \left ( 5 -2\,\eta \right )
\bigg\}
\no &&
+{
\frac {{{(-2\,E)}}^{3}}{192\,c^{6}}}
\bigg\{ 24\, \left( -2+5\,\eta \right) 
 \left(-23+10\,\eta+ 4\,{\eta}^{2} \right) 
-15\,\biggl (-528
\no&&
+200\,\eta-77\,{\eta}^{2}
+ 24\,{
\eta}^{3} \biggr ) {(-2\,E\, h^2)}
-72\,( 265-193\,\eta
\no&&
+46\,{\eta}^{2} ) \sqrt {{(-2\,E\, h^2)}}
- \frac{2}{(-2\,E\,h^2)}
 \bigg(
6732 +117\,\eta\,{\pi }^{2}-12508\,\eta
\no&&
+2004\,{\eta}^{2}\bigg)
+ \frac{2}{\sqrt{ (-2\,E\,h^2)}}
\bigg(
16380-19964\,\eta+123\,\eta\,{\pi }^{2}
\no&&
+3240\,{\eta}^{2}
\bigg)
-\frac{2}{ (-2\,E\,h^2)^{3/2} }
\bigg( 10080+123\,\eta\,{
\pi }^{2}-13952\,\eta
\no&&
+1440\,{\eta}^{2}
\bigg)
+ \frac{96}{ (-2\,E\,h^2)^2}
\bigg(
134 -281\,\eta+5\,\eta\,{\pi }^{2}+16\,{\eta}^{2}
\bigg)
 \bigg\}\no&&
 +\frac{(-2\,E)^4}{460800 \,c^8}\,\biggl\{3600\,\big(-25500+25804 \,\eta -7267 \,\eta ^2+1236\, \eta ^3\no&&
 +280 \,\eta ^4\big)-28800 \,\big(1828-563 \,\eta +237 \,\eta ^2-84 \,\eta ^3
 \no&&+25 \,\eta ^4\big)\,(-2\,E\,h^2)-10800 \big(-18795+15344 \eta -6303 \eta ^2\no&&
 +1262 \eta ^3\Big)\,\sqrt{(-2\,E\,h^2)}+\frac{50}{\sqrt{(-2\,E\,h^2)}}\biggl(-9265320\no&&
 +\left(16589440-179031 \pi ^2\right) \,\eta +12 \left(-620854+4305 \pi ^2\right) \,\eta ^2\no&&
 +851472\, \eta ^3\biggr)+\frac{1}{(-2\,E\,h^2)}\,\biggl(215107200+6 \big(-58544128\no&&
 +595725 \,\pi ^2\big) \eta +\left(159273712-4409175 \,\pi ^2\right) \eta ^2-19425600 \eta ^3\biggr)\no&&
 -\frac{20}{(-2\,E\,h^2)^{3/2}}\,\biggl(-36918720+\left(84382336-1538535 \pi ^2\right) \eta \no&&
 +30 \left(-1188608+16359 \pi ^2\right) \eta ^2+2030400 \eta ^3\biggr)+\no&&
 +\frac{1}{(-2\,E\,h^2)^{2}}\,\biggl(-324777600+\left(765039936-16500150 \pi ^2\right) \eta \no&&
 +\left(-348347776+13645725 \pi ^2\right) \eta ^2+15062400 \eta ^3\biggr)\no&&
 +\frac{30}{(-2\,E\,h^2)^{5/2}}\,\biggl(-17297280+\left(37556864-771585 \,\pi ^2\right) \eta \no&&
 +1920 \left(-7013+123\, \pi ^2\right) \eta ^2+403200 \,\eta ^3\biggr)\no&&
-\frac{6}{(-2\,E\,h^2)^{3}}\,\biggl(-51187200-6 \left(-18365728+415175\, \pi ^2\right) \,\eta \no&&+\left(-34446144+1442225 \,\pi ^2\right) \,\eta ^2+576000 \,\eta ^3\biggr)\biggr\}  .
\eea
\ens
The three eccentricities
$e_r, e_t$ and $e_\phi$, which differ from each other
at PN orders, are related by
\begin{align}
\begin{autobreak}
\MoveEqLeft
\frac{e_r}{e_t}=
1
+\frac{(-2\,E)}{c^2} \big(4-\frac{3 \,\eta}{2}\big) 
+\frac{(-2\,E)}{8\,c^4\, h^2}\bigg(h^2 (-2\,E) \big(6 \,\eta^2-63 \,\eta+56\big)
+12 h \sqrt{(-2\,E)} (5-2 \,\eta)-22 \,\eta+34\bigg)
+\frac{(-2\,E) }{192 h^4} \bigg(-12 h^4 (-2\,E)^2 (2 \,\eta^3-98 \,\eta^2+299 \,\eta-192)
+36 h^3 (-2\,E)^{3/2} (6 \,\eta^2-69 \,\eta+65)
+h^2 (-2\,E) (492 \,\eta^2+(3 \pi ^2-2252) \,\eta+3828)
+h \sqrt{(-2\,E)} (1440 \,\eta^2+(123 \pi ^2-13952) \,\eta+10080)+48 (16 \,\eta^2+(5 \pi ^2-281) \,\eta+134)\bigg)
-\frac{(-2\,E)}{3686400\,c^8\, h^6} \bigg(14400 h^6 (-2\,E)^3 (603 \,\eta^3-5474 \,\eta^2+10176 \,\eta-5248)
+43200 h^5 (-2\,E)^{5/2} (6 \,\eta^3-779 \,\eta^2+2632 \,\eta-1775)
+h^4 (-2\,E)^2 (1152000 \,\eta^3+(4131675 \pi ^2-8360192) \,\eta^2
+18 (3594784+52725 \pi ^2) \,\eta-175104000)
-200 h^3 (-2\,E)^{3/2} (86400 \,\eta^3+324 (41 \pi ^2-2528) \,\eta^2+(410752-81615 \pi ^2) \,\eta+829440)
+8 h^2 (-2\,E) (705600 \,\eta^3-(25311328+2355525 \pi ^2) \,\eta^2
+(141245184-670500 \pi ^2) \,\eta-57484800)
+120 h \sqrt{(-2\,E)} (403200 \,\eta^3+1920 (123 \pi ^2-7013) \,\eta^2
+(37556864-771585 \pi ^2) \,\eta-17297280)+24 (576000 \,\eta^3
+(1442225 \pi ^2-34446144) \,\eta^2-6 (415175 \pi ^2-18365728) \,\eta-51187200)\bigg)\,,
\end{autobreak}
\end{align}
\begin{align}
\begin{autobreak}
\MoveEqLeft
\frac{e_\phi}{e_t}=
1
-\frac{(-2\,E)}{c^2} (\,\eta-4)
+\frac{(-2\,E)}{32\,c^4\, h^2} \bigg(h^2 (-2\,E) (11 \,\eta^2-168 \,\eta+224)
-48 h \sqrt{(-2\,E)} (2 \,\eta-5)-15 \,\eta^2-144 \,\eta+272\bigg)
-\frac{ (-2\,E)}{768 \,c^6\,h^4}\bigg(6 h^4 (-2\,E)^2 (3 \,\eta^3-370 \,\eta^2
+1488 \,\eta-1536)
+144 h^3 (-2\,E)^{3/2} (2 \,\eta^2+49 \,\eta-65)+h^2 (-2\,E) (-90 \,\eta^3+918 \,\eta^2
+(7420+3 \pi ^2) \,\eta-30624)-4 h \sqrt{(-2\,E)} (1440 \,\eta^2+(123 \pi ^2-13952) \,\eta+10080)
-3 (-70 \,\eta^3+2416 \,\eta^2+(1005 \pi ^2-47452) \,\eta+19200)\bigg)
+\frac{(-2\,E) }{29491200\,c^8\, h^6}\bigg(1800 h^6 (-2\,E)^3 (62 \,\eta^4-11607 \,\eta^3
+150368 \,\eta^2-381952 \,\eta+335872)-345600 h^5 (-2\,E)^{5/2} (46 \,\eta^3-287 \,\eta^2
+1712 \,\eta-1775)+h^4 (-2\,E)^2 (586800 \,\eta^4
-27495000 \,\eta^3+(445236992+2929725 \pi ^2) \,\eta^2
+1350 (1481 \pi ^2-291040) \,\eta+2801664000)
+1600 h^3 (-2\,E)^{3/2} (181440 \,\eta^3+12 (1599 \pi ^2-121400) \,\eta^2
+(417664-81615 \pi ^2) \,\eta+1416960)-12 h^2 (-2\,E) (207300 \,\eta^4-46247250 \,\eta^3+(930614528+892025 \pi ^2) \,\eta^2
+(454556608+23726550 \pi ^2) \,\eta-750336000)-960 h \sqrt{(-2\,E)} (403200 \,\eta^3+1920 (123 \pi ^2-7013) \,\eta^2
+(37556864-771585 \pi ^2) \,\eta-17297280)
-2300400 \,\eta^4-889504200 \,\eta^3+(38875770624-584054625 \pi ^2) \,\eta^2
+10 (223387845 \pi ^2-8201607776) \,\eta+23705395200\bigg)\,.
\end{autobreak}
\end{align}

\section{Energy and angular momentum in terms of $x$ and $e_t$}     \label{E-h}

\begin{align}
E&=\frac{-x\, c^2}{2}\Bigg(1+\bigg(\frac{5}{4}-2 {\zeta}^2-\frac{\eta}{12}\bigg) x+\bigg(\frac{5}{8}+5 {\zeta}+\frac{7 {\zeta}^2}{2}-\frac{25 {\zeta}^4}{2}+\big(-\frac{5}{8}-2 {\zeta}-4 {\zeta}^2+9
   {\zeta}^4\big) \eta-\frac{\eta^2}{24}\bigg) x^2\notag\\[0.5ex]
   &+\bigg(-\frac{185}{192}+\frac{5 {\zeta}}{2}+\frac{9 {\zeta}^2}{2}+30 {\zeta}^3+\frac{207 {\zeta}^4}{4}-30 {\zeta}^5-\frac{205
   {\zeta}^6}{3}\notag\\[0.5ex]
   &+\big(-\frac{75}{64}+\frac{23 {\zeta}}{6}+\frac{2 {\zeta}^2}{3}+12 {\zeta}^5+{\zeta}^6 \big(\frac{613}{3}-\frac{205 \pi ^2}{64}\big)+{\zeta}^3\big(-\frac{463}{9}+\frac{41 \pi ^2}{96}\big)\notag\\[0.5ex]
   &+{\zeta}^4 \big(-\frac{1301}{12}+\frac{41 \pi ^2}{64}\big)\big) \eta+\big(-\frac{25}{288}-\frac{10 {\zeta}}{3}-\frac{40 {\zeta}^2}{9}+7
   {\zeta}^3+36 {\zeta}^4-\frac{147 {\zeta}^6}{4}\big) \eta^2-\frac{35 \eta^3}{5184}\bigg) x^3\notag\\[0.5ex]
   &+\bigg(-\frac{931}{384}-\frac{85 {\zeta}}{8}+\frac{155 {\zeta}^2}{48}-\frac{435
   {\zeta}^3}{8}-\frac{559 {\zeta}^4}{32}+\frac{3693 {\zeta}^5}{8}+\frac{18551 {\zeta}^6}{48}-495 {\zeta}^7-\frac{29035 {\zeta}^8}{96}\notag\\[0.5ex]
   &+\big(\frac{245}{1152}+\frac{67
   {\zeta}}{6}+\frac{47 {\zeta}^2}{12}+{\zeta}^8 \big(\frac{847091}{432}-\frac{1964855 \pi ^2}{36864}\big)+{\zeta}^3 \big(\frac{14363}{54}-\frac{43441 \pi ^2}{9216}\big)+{\zeta}^4
   \big(\frac{19793}{72}-\frac{53281 \pi ^2}{12288}\big)\notag\\[0.5ex]
   &+{\zeta}^7 \big(\frac{2114}{3}-\frac{41 \pi ^2}{16}\big)+{\zeta}^5 \big(-\frac{124841}{120}+\frac{16709 \pi ^2}{1024}\big)+{\zeta}^6
   \big(-\frac{318437}{144}+\frac{334715 \pi ^2}{6144}\big)\big) \eta\notag\\[0.5ex]
   &+\big(\frac{245}{1152}-\frac{37 {\zeta}}{18}-\frac{8 {\zeta}^2}{3}-150 {\zeta}^7+\frac{{\zeta}^6 \big(2023688-35547 \pi
   ^2\big)}{1152}+\frac{1}{96} {\zeta}^5 \big(47668-615 \pi ^2\big)\notag\\[0.5ex]
   &+{\zeta}^4 \big(-\frac{22025}{72}+\frac{533 \pi ^2}{192}\big)+\frac{{\zeta}^3 \big(-370768+3567 \pi
   ^2\big)}{1728}+\frac{5 {\zeta}^8 \big(-396856+8733 \pi ^2\big)}{1152}\big) \eta^2\notag\\[0.5ex]
   &+\big(\frac{175}{5184}-\frac{28 {\zeta}}{9}-\frac{280 {\zeta}^2}{81}+\frac{82 {\zeta}^3}{3}+80
   {\zeta}^4-\frac{107 {\zeta}^5}{4}-\frac{431 {\zeta}^6}{2}+\frac{1133 {\zeta}^8}{8}\big) \eta^3+\frac{77 \eta^4}{31104}\bigg) x^4\Bigg)\,, \\[2ex]
h&=\frac{1}{c\,\sqrt{x}}\Bigg(\frac{1}{{\zeta}}+\bigg(\frac{3}{2 {\zeta}}+\big(-\frac{5}{6 {\zeta}}+{\zeta}\big) \eta\bigg) x+\bigg(5+\frac{11}{8 {\zeta}}-8 {\zeta}-\frac{15 {\zeta}^2}{2}+\frac{25
   {\zeta}^3}{2}\notag\\[0.5ex]
   &+\big(-2-\frac{73}{24 {\zeta}}+\frac{83 {\zeta}}{12}+3 {\zeta}^2-\frac{29 {\zeta}^3}{4}\big) \eta+\big(\frac{5}{24
   {\zeta}}+\frac{{\zeta}}{3}-\frac{{\zeta}^3}{2}\big) \eta^2\bigg) x^2\notag\\[0.5ex]
   &+\bigg(5+\frac{13}{48 {\zeta}}+\frac{39 {\zeta}}{4}+\frac{95 {\zeta}^2}{4}-97 {\zeta}^3-30 {\zeta}^4+\frac{290
   {\zeta}^5}{3}+\big(-\frac{14}{3}-\frac{283}{48 {\zeta}}-\frac{71 {\zeta}}{48}\notag\\[0.5ex]
   &+\frac{1}{384} {\zeta}^3 \big(58808-357 \pi ^2\big)+{\zeta}^4 \big(\frac{236}{3}-\frac{41 \pi
   ^2}{64}\big)+{\zeta}^2 \big(-\frac{2023}{36}+\frac{41 \pi ^2}{96}\big)+{\zeta}^5 \big(-\frac{2537}{12}+\frac{365 \pi ^2}{128}\big)\big) \eta\notag\\[0.5ex]
   &+\big(-\frac{1}{3}+\frac{71}{36
   {\zeta}}+\frac{223 {\zeta}}{36}+12 {\zeta}^2-\frac{241 {\zeta}^3}{6}-\frac{27 {\zeta}^4}{2}+\frac{281 {\zeta}^5}{8}\big)\, \eta^2\notag\\[0.5ex]
   &+\big(\frac{25}{1296 {\zeta}}+\frac{17
   {\zeta}}{72}-\frac{3 {\zeta}^3}{4}+\frac{{\zeta}^5}{2}\big) \eta^3\bigg) \,x^3+\bigg(-\frac{35}{8}-\frac{173}{128 {\zeta}}+\frac{93 {\zeta}}{8}-\frac{315 {\zeta}^2}{16}+\frac{21487
   {\zeta}^3}{64}\notag\\[0.5ex]
   &-\frac{709 {\zeta}^4}{16}-\frac{15235 {\zeta}^5}{16}+\frac{1581 {\zeta}^6}{16}+\frac{38265 {\zeta}^7}{64}+\big(-\frac{39}{4}-\frac{6991}{1152 {\zeta}}-\frac{2179
   {\zeta}}{192}+{\zeta}^5 \big(\frac{42531499}{14400}-\frac{684125 \pi ^2}{12288}\big)\notag\\[0.5ex]
   &+{\zeta}^6 \big(\frac{155773}{240}-\frac{47503 \pi ^2}{2048}\big)+{\zeta}^2
   \big(\frac{5344}{27}-\frac{41473 \pi ^2}{9216}\big)+{\zeta}^3 \big(-\frac{4869481}{7200}+\frac{165109 \pi ^2}{24576}\big)+{\zeta}^4 \big(-\frac{23789}{30}+\frac{47461 \pi
   ^2}{2048}\big)\notag\\[0.5ex]
   &+{\zeta}^7 \big(-\frac{24652879}{10800}+\frac{3632879 \pi ^2}{73728}\big)\big) \eta+\big(-\frac{5}{36}+\frac{2963}{384 {\zeta}}+\frac{11393 {\zeta}}{2304}+\frac{{\zeta}^3
   \big(407299312-6357975 \pi ^2\big)}{921600}\notag\\[0.5ex]
   &+{\zeta}^7 \big(\frac{11872813}{7200}-\frac{74123 \pi ^2}{2048}\big)+{\zeta}^4 \big(\frac{23909}{36}-\frac{3649 \pi ^2}{384}\big)+{\zeta}^6
   \big(-\frac{28367}{48}+\frac{41 \pi ^2}{4}\big)+{\zeta}^5 \big(-\frac{2319157}{1200}+\frac{454727 \pi ^2}{12288}\big)\notag\\[0.5ex]
   &+\frac{1}{432} {\zeta}^2 \big(-60904+615 \pi ^2\big)\big)
   \eta^2+\big(\frac{8}{9}-\frac{217}{576 {\zeta}}+\frac{10687 {\zeta}}{1728}+\frac{45 {\zeta}^2}{2}-\frac{56027 {\zeta}^3}{576}-\frac{309 {\zeta}^4}{4}+\frac{22117 {\zeta}^5}{96}\notag\\[0.5ex]&
   +\frac{441
   {\zeta}^6}{8}-\frac{2245 {\zeta}^7}{16}\big) \eta^3+\big(-\frac{175}{31104 {\zeta}}+\frac{35 {\zeta}}{324}-\frac{13 {\zeta}^3}{16}+\frac{4 {\zeta}^5}{3}-\frac{5 {\zeta}^7}{8}\big)
   \eta^4\bigg) x^4\Bigg)\,,
\end{align}
\end{widetext}
where $\zeta \equiv   {1}/{\sqrt{1-e_t^2}}$.

\section{Canonical perturbation theory}     \label{Delaunay technique}

Focusing our attention on an integrable and one degree of freedom\footnote{One
degree of freedom implies one position and 
one conjugate momentum variable. Integrability is equivalent
to the existence of action-angle variables \cite{jose1998classical}.} system for now, we have the 
total Hamiltonian (unperturbed plus the perturbation) written as
\begin{align}
H(\phi_0, J_0) =  H_0 (J_0)   +  \epsilon H_1 (\phi_0,  J_0) ,
\end{align}
where $(\phi_0, J_0)$ are the action-angles of the unperturbed system.
If the perturbed system is also integrable in the perturbative sense,
 then there exists a canonical transformation
to the new action-angles $(\phi_0 , J_0 ) \leftrightarrow  (\phi , J)$ such that $E(J)$ is the 
total Hamiltonian in terms of the new action $J$: $E(J) =H(\phi_0, J_0) $.
With a type-2 generator $S(\phi_0,J)$ of the form 
\begin{align}
S(\phi_0,J)   =  \phi_0 J +  \epsilon  S_1(\phi_0,J) +  \mathcal{O}(\epsilon^2),         \label{gen-func}
\end{align}
we have
\begin{align}
J_0 =  \frac{ \pd S}{\pd  \phi_0}  = J  + \epsilon  \frac{\pd S_1}{\pd \phi_0}  +  \mathcal{O}(\epsilon^2),          \label{new_AAVs_1}   \\
\phi =  \frac{ \pd S}{\pd  J}  =  \phi_0  + \epsilon  \frac{\pd S_1}{\pd J}  +  \mathcal{O}(\epsilon^2) .                \label{new_AAVs_2} 
\end{align}
One of the main results of the canonical perturbation theory is that the leading and the linear in $\epsilon$ order
contributions to $E(J) = E_0(J)  + \epsilon E_1(J)    + \mathcal{O}(\epsilon^2) $ read
\begin{align}
E_0(J)  & =  H_0 (J)             \label{pert-0}   \\
E_1(J)   &  =  H_1 (\phi_0, J)  +  \frac{\pd H_0}{\pd J} \frac{\pd S_1}{\pd \phi_0} .    \label{pert-1}
\end{align}
It is implied here that $H_0(J)$ and $H_1(\phi_0,J)$ have the same functional dependence on $J$
as $H_0(J_0)$ and $H_1(\phi_0, J_0)$ have on $J_0$.
We have borrowed the above concept and presentation largely from Sec.~6.3 of
Ref.~\cite{jose1998classical} ; the reader is referred to it for more details, including how to obtain $S_1$.
Evaluation of $S_1$ is required to get the perturbed action-angles from the 
unperturbed ones.

As explained in Ref.~\cite{DJS_15}, it is easy to see from Eq.~\eqref{pert-1} that a term of the form $ A(J) \cos n \phi_0$ in $H_1$
(with $n$ being a non-zero integer)
can be eliminated by a term $ -A(J)\sin n \phi_0/(n \Omega)  $ 
in $S_1$ where $\Omega(J) \equiv \pd H_0(J)/\pd J$.
And since all the sine terms in $H_1$ can also be eliminated
similarly, all there is left to deal with is the non-oscillatory part of $H_1$ possessing a non-zero average over $\phi_0$.
Hence, we can write
\begin{align}
E_1 (J) = \overline{H}_1       ,      \label{pert-1b}
\end{align}
where $\overline{H}_1$ denotes the average of $H_1(\phi_0, J)$ over $\phi_0$.

Actually, this technique of averaging the perturbation
is basically the von Zeipel-Brouwer technique applied to the Kepler problem
(also known as the Delaunay technique).
The von Zeipel-Brouwer technique is one of the many
\textit{degenerate} perturbation techniques \cite{ferraz2007canonical},
which differs from the non-degenerate one \cite{jose1998classical}
in one crucial aspect that the averaging is not performed over all the angles
(as in non-degenerate perturbation theory) but rather only a subset of them. This variation
of the non-degenerate method can cure the problem of vanishing denominators which occurs when one
tries to apply non-degenerate perturbation method to a degenerate system, such as the Newtonian Kepler system.

\section{The ``semi-perturbation'' scheme: neglecting oscillatory corrections to action-angles}     \label{neglect}

The main results of canonical perturbation theory are mainly
contained in Eqs.~(\ref{new_AAVs_1}), (\ref{new_AAVs_2}), (\ref{pert-0}), (\ref{pert-1}),
and (\ref{pert-1b}), along with an equation for $S_1$, which we don't present.
But often it is not necessary to retain all the information contained in these equations. 
We will elucidate this
with a 1 DOF example of a simple harmonic oscillator (SHO).

\begin{figure*} 
  \includegraphics[width=\linewidth, height=9cm]{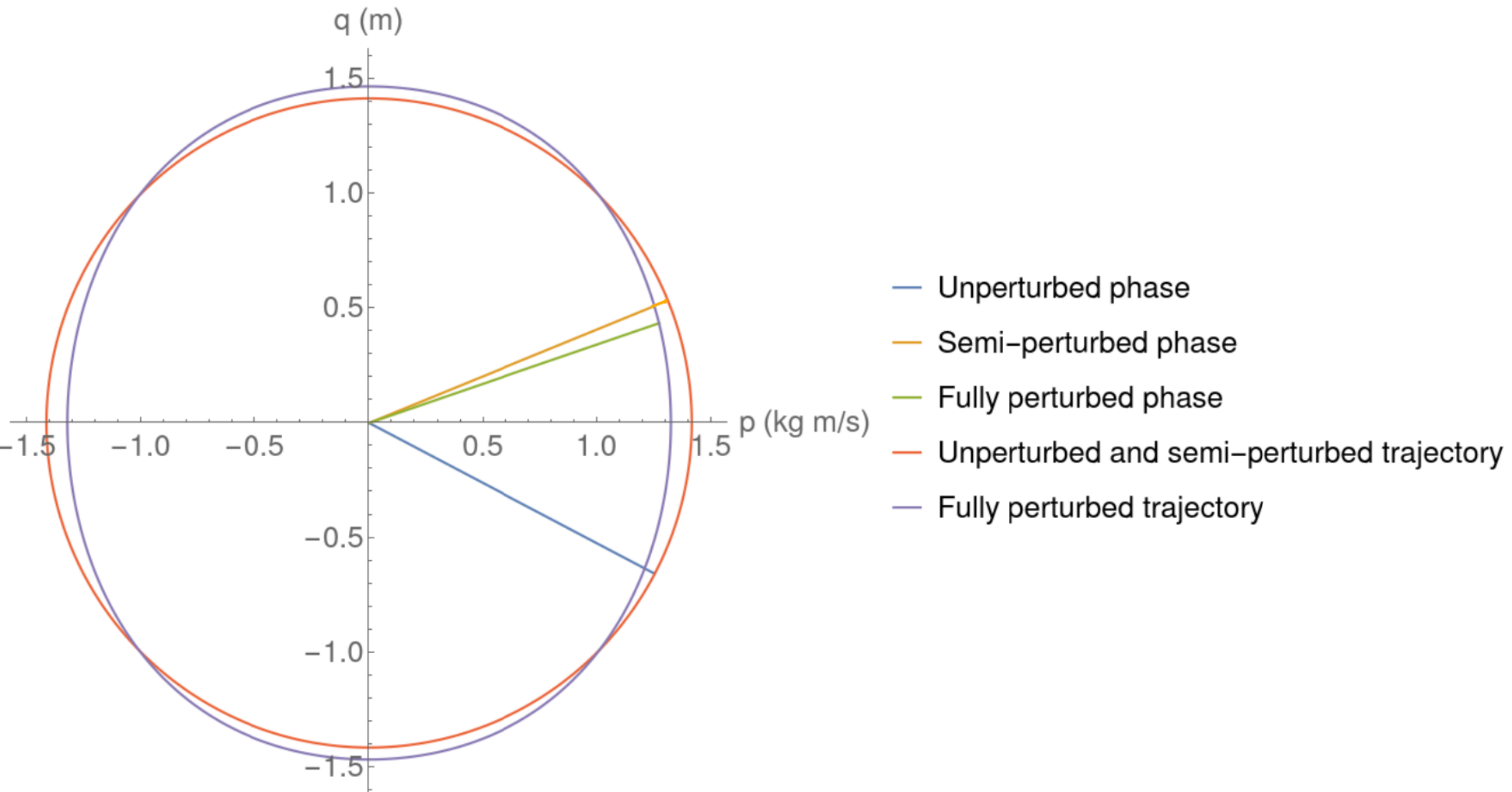}
  \caption{ The unperturbed, semi-perturbed and fully perturbed trajectories of a 
  harmonic oscillator with the perturbation being a quartic function 
  of the position coordinate $q$.
The unperturbed and semi-perturbed cases have the same
  trajectory which is an ``averaged'' version of the fully perturbed trajectory.  
In all three cases, we start from an initial phase of $0$ radians which corresponds to
being on the positive $p$-axis.
The three rays show the phase at $t = 5.8$ s (arbitrarily chosen).
  There is a large secular dephasing in the unperturbed case 
  with respect to the other two cases. The semi-perturbed and the fully
  perturbed cases have small zero-average, oscillatory dephasing
   between them. There exists no secular difference
   between the phase-space trajectories for these two cases.
}
  \label{fig:torus}
\end{figure*}

With $(q,p)$ being the pair of canonical variables
($q$ in this appendix does not stand for the mass ratio),
the full Hamiltonian of the perturbed SHO is
\begin{align}
H=\frac{p^{2}}{2 m}+\frac{1}{2} m \omega_{0}^{2} q^{2}+\frac{1}{4} \epsilon m q^{4} \equiv H_{0}+\epsilon H_{1},
\end{align}
where $\omega_0  = \sqrt{k/m}$ with $k$ and $m$ being
the spring constant and the mass of the oscillator respectively.
$\epsilon$ is a small perturbation parameter.
This is the 
subject of Worked Example 6.5 of Ref.~\cite{jose1998classical}.
We will simply use the results obtained there without showing the entire derivation.
$(q,p)$ as a function of unperturbed action-angles $(J_0, \phi_0)$ is 
\begin{align}                  \label{q-p}
\begin{aligned}
q &=\sqrt{\frac{2 J_0}{m   \omega_0  }}  \sin \phi_0    ,   \\
p &=\sqrt{2 J_0  m     \omega_0} \cos \phi_0     ,
\end{aligned}
\end{align}
whereas Eqs.~(\ref{new_AAVs_1}) and (\ref{new_AAVs_2}) connect the 
unperturbed action-angles with the perturbed ones as (with $\mu = 1/m\omega_0^2$)
\begin{align}            \label{J0-phi0}
\begin{aligned}
J_ 0  & =  J  +   \epsilon \frac{\mu J^{2}}{8  \omega_{0}}\left(4 \cos 2 \phi  - \cos 4 \phi  \right) + \mathcal{O}(\epsilon^2)    , \\
\phi_0  & =  \phi   - \epsilon \frac{\mu J}{4 \omega_{0}}\left(2 \sin ^{2} \phi +3\right) \sin \phi  \cos \phi  + \mathcal{O}(\epsilon^2)    .
\end{aligned}
\end{align}  
Note that Eqs. (\ref{J0-phi0}) look a little different than the ones in the
Worked Example 6.5 of Ref.~\cite{jose1998classical} because we have
swapped $\phi  \leftrightarrow \phi_0$
in linear in $\epsilon$ term. This is justified since it makes
a difference at the $\mathcal{O}(\epsilon^2)$ absolute order.
Finally, Eqs.~(\ref{pert-0}) and (\ref{pert-1}) give the perturbed Hamiltonian and
the perturbed frequency as
\begin{align}                  \label{E-omega}
\begin{aligned}
E(J)  &  =  \omega_{0} J+\epsilon \frac{3 \mu}{8} J^{2}   + \mathcal{O}(\epsilon^2)   , \\
\omega   = \frac{\partial E}{\partial J}  & = \omega_{0}+\epsilon \frac{3 \mu}{4} J  + \mathcal{O}(\epsilon^2)  .
\end{aligned}
\end{align}

We now try to describe this system at three varying levels of complication:
 (i) unperturbed (least accurate) (ii) fully perturbed (most accurate) (iii) semi-perturbed
 (slightly less accurate than the ``fully perturbed'' scheme). \\ \\
 \textbf{1. Unperturbed:} The equations of motion (EOMs) are
 Eqs.~(\ref{q-p}) with $\phi_ 0 = \omega_0 t$ and a given fixed value of $J_0$.
 The unperturbed action $J_0$ stays constant during evolution.\\   \\
 \textbf{2. Fully perturbed:}  The EOMs are Eqs.~(\ref{q-p}), (\ref{J0-phi0}) 
 and (\ref{E-omega}) with $\phi = \omega t$ with $J$ (rather than $J_0$) staying constant. \\  \\
 \textbf{3. Semi-perturbed:} The EOMs are
 \begin{align}               
\begin{aligned}
q &=  q(J_0, \phi_0) \vert_{ ( J_0, \phi_0)  \rightarrow (J , \phi )}  =  \sqrt{\frac{2 J}{m \omega_0}}   \sin \phi        ,   \\
p &=  p  (J_0, \phi_0) \vert_{ ( J_0, \phi_0)  \rightarrow (J , \phi )}    =  \sqrt{2 J m \omega_0}      \cos \phi      ,
\end{aligned}
\end{align}
again with $\phi = \omega t$.
In other words, in the semi-perturbed scheme, $(q,p)$ are taken to have
the same dependence on $(J, \phi)$ as they have on
$(J_0 , \phi_0)$. Along with this, we also have Eqs.~(\ref{E-omega}).
Thus, in the semi-perturbed scheme, the oscillatory corrections to action-angles 
and the generating
function $S$ (Eqs.~(\ref{new_AAVs_1}),
 (\ref{new_AAVs_2}) and (\ref{gen-func})) don't need to be computed
 and hence can be ignored.
This leads to significant simplifications when compared to the fully
perturbed case without much loss of information, as we will see below.
An example of these zero-average oscillatory corrections are the 
sinusoidal terms in Eqs.~\ref{J0-phi0}.

 To see what concrete effects the three above schemes
bring about, we plot the respective trajectories in the phase space for all the 
three cases in Fig.~\ref{fig:torus}, for the following numerical values: 
$k = 1$ N/m,$~m = 1 $ kg,~$J = J_0 = 1$ kg $\text{m}^2/$s,~$\epsilon = 0.2$.
 The closed contours denote the trajectories and the three rays denote the 
 phase swept ($ \arctan(m \omega_0 q/p)$) for all
 the above three cases from $t=0$ to $t = 5.8$ s (an arbitrarily chosen numerical value). 
Since $\phi_0 (t=0)  = \phi (t=0) = 0$, at $t=0$ we start from the points where
the closed trajectories of Fig.~\ref{fig:torus} intersect the
positive $p$-axis for all the above three cases. 
At the final time $t=5.8$ s,
 in the fully perturbed and semi-perturbed cases, we have swept a little
over a full revolution whereas in the unperturbed case,
 we have a significant phase lag compared with the other two.
The full effect of the perturbation $\epsilon H_{1}$
(corresponding to the ``fully perturbed'' case above) is that the perturbed trajectory
starts to deform around the unperturbed one and there is a dephasing which increases
with time. The dephasing is due to the frequency correction term (linear in $\epsilon$)
in Eq.~(\ref{E-omega}). 
In addition, note that the unperturbed trajectory (labeled by $J_0 = J_N$)
is the ``averaged version'' of the fully perturbed trajectory (labeled by $J = J_N$), with
$J_N$ standing for some numerical value of the action.

Now let's come to the semi-perturbed case.
The semi-perturbed approximation gives us the same trajectory as the
unperturbed one (see Fig.~\ref{fig:torus}), but with a phase
(yellow ray) that differs from the 
fully perturbed phase (green ray) in an oscillatory fashion, rather than secular.
This means that we get the averaged version of the fully perturbed trajectory
and a small amount of oscillatory dephasing with respect to the fully perturbed one,
which does not grow in time and whose time-average is zero.
The upshot is that instead of capturing the full perturbation effect,
 we can settle with the semi-perturbed case which has got no secular differences
(either in phase or amplitude) with respect to the fully perturbed case. This way we don't need to
compute $S$. Similar approaches have been adopted in some textbooks 
(Sec. 12.3 and 12.4 of Ref.~\cite{goldstein2002classical}).
Even from the GW data analysis point of view, it is the secular effects
that matter much more than the oscillatory ones because the method of matched filter 
employed in the analysis builds up signal-to-noise ratio by following the phase of
the two waveforms to be matched in a coherent manner \cite{maggiore2008gravitational,
creighton2012gravitational, Sathyaprakash:2009xs}.

It is important to note that to model the system using the semi-perturbed scheme, 
one needs to be given the numerical values of $(J,\phi)$ at some initial time; initial values of
$(q,p)$ or $(J_0, \phi_0)$ won't do.
On the other hand, 
initial numerical values of $(q,p)$ or $(J_0, \phi_0)$ are required 
to model the system as per the unperturbed
and the fully-perturbed schemes.
All this is not a cause for concern
because the 4-3PN IMR \textsc{Mathematica} 
package \cite{MMA2, MMA3} provided with this paper requires the user
to input the PN parameter $x$ and eccentricity which are ultimately related 
to the perturbed actions of the BBH system.

\bibliography{GTG_1020}

\begin{thebibliography}{109}%
\makeatletter
\providecommand \@ifxundefined [1]{%
 \@ifx{#1\undefined}
}%
\providecommand \@ifnum [1]{%
 \ifnum #1\expandafter \@firstoftwo
 \else \expandafter \@secondoftwo
 \fi
}%
\providecommand \@ifx [1]{%
 \ifx #1\expandafter \@firstoftwo
 \else \expandafter \@secondoftwo
 \fi
}%
\providecommand \natexlab [1]{#1}%
\providecommand \enquote  [1]{``#1''}%
\providecommand \bibnamefont  [1]{#1}%
\providecommand \bibfnamefont [1]{#1}%
\providecommand \citenamefont [1]{#1}%
\providecommand \href@noop [0]{\@secondoftwo}%
\providecommand \href [0]{\begingroup \@sanitize@url \@href}%
\providecommand \@href[1]{\@@startlink{#1}\@@href}%
\providecommand \@@href[1]{\endgroup#1\@@endlink}%
\providecommand \@sanitize@url [0]{\catcode `\\12\catcode `\$12\catcode
  `\&12\catcode `\#12\catcode `\^12\catcode `\_12\catcode `\%12\relax}%
\providecommand \@@startlink[1]{}%
\providecommand \@@endlink[0]{}%
\providecommand \url  [0]{\begingroup\@sanitize@url \@url }%
\providecommand \@url [1]{\endgroup\@href {#1}{\urlprefix }}%
\providecommand \urlprefix  [0]{URL }%
\providecommand \Eprint [0]{\href }%
\providecommand \doibase [0]{https://doi.org/}%
\providecommand \selectlanguage [0]{\@gobble}%
\providecommand \bibinfo  [0]{\@secondoftwo}%
\providecommand \bibfield  [0]{\@secondoftwo}%
\providecommand \translation [1]{[#1]}%
\providecommand \BibitemOpen [0]{}%
\providecommand \bibitemStop [0]{}%
\providecommand \bibitemNoStop [0]{.\EOS\space}%
\providecommand \EOS [0]{\spacefactor3000\relax}%
\providecommand \BibitemShut  [1]{\csname bibitem#1\endcsname}%
\let\auto@bib@innerbib\@empty
\bibitem [{\citenamefont {Abbott}\ \emph {et~al.}(2019)\citenamefont {Abbott}
  \emph {et~al.}}]{LIGOScientific:2018mvr}%
  \BibitemOpen
  \bibfield  {author} {\bibinfo {author} {\bibfnamefont {B.}~\bibnamefont
  {Abbott}} \emph {et~al.} (\bibinfo {collaboration} {LIGO Scientific,
  Virgo}),\ }\bibfield  {title} {\bibinfo {title} {{GWTC-1: A
  Gravitational-Wave Transient Catalog of Compact Binary Mergers Observed by
  LIGO and Virgo during the First and Second Observing Runs}},\ }\href
  {https://doi.org/10.1103/PhysRevX.9.031040} {\bibfield  {journal} {\bibinfo
  {journal} {Phys. Rev. X}\ }\textbf {\bibinfo {volume} {9}},\ \bibinfo {pages}
  {031040} (\bibinfo {year} {2019})},\ \Eprint
  {https://arxiv.org/abs/1811.12907} {arXiv:1811.12907 [astro-ph.HE]}
  \BibitemShut {NoStop}%
\bibitem [{\citenamefont {Abbott}\ \emph
  {et~al.}(2020{\natexlab{a}})\citenamefont {Abbott} \emph
  {et~al.}}]{Abbott:2020niy}%
  \BibitemOpen
  \bibfield  {author} {\bibinfo {author} {\bibfnamefont {R.}~\bibnamefont
  {Abbott}} \emph {et~al.} (\bibinfo {collaboration} {LIGO Scientific,
  Virgo}),\ }\bibfield  {title} {\bibinfo {title} {{GWTC-2: Compact Binary
  Coalescences Observed by LIGO and Virgo During the First Half of the Third
  Observing Run}},\ }\href@noop {} {\  (\bibinfo {year}
  {2020}{\natexlab{a}})},\ \Eprint {https://arxiv.org/abs/2010.14527}
  {arXiv:2010.14527 [gr-qc]} \BibitemShut {NoStop}%
\bibitem [{\citenamefont {Abbott}\ \emph
  {et~al.}(2017{\natexlab{a}})\citenamefont {Abbott} \emph
  {et~al.}}]{TheLIGOScientific:2017qsa}%
  \BibitemOpen
  \bibfield  {author} {\bibinfo {author} {\bibfnamefont {B.}~\bibnamefont
  {Abbott}} \emph {et~al.} (\bibinfo {collaboration} {LIGO Scientific,
  Virgo}),\ }\bibfield  {title} {\bibinfo {title} {{GW170817: Observation of
  Gravitational Waves from a Binary Neutron Star Inspiral}},\ }\href
  {https://doi.org/10.1103/PhysRevLett.119.161101} {\bibfield  {journal}
  {\bibinfo  {journal} {Phys. Rev. Lett.}\ }\textbf {\bibinfo {volume} {119}},\
  \bibinfo {pages} {161101} (\bibinfo {year} {2017}{\natexlab{a}})},\ \Eprint
  {https://arxiv.org/abs/1710.05832} {arXiv:1710.05832 [gr-qc]} \BibitemShut
  {NoStop}%
\bibitem [{\citenamefont {{Poggiani}}\ \emph {et~al.}(2019)\citenamefont
  {{Poggiani}}, \citenamefont {{LIGO Scientific Collaboration}},\ and\
  \citenamefont {{Virgo Collaboration}}}]{2019mbhe.confE..19P}%
  \BibitemOpen
  \bibfield  {author} {\bibinfo {author} {\bibfnamefont {R.}~\bibnamefont
  {{Poggiani}}}, \bibinfo {author} {\bibnamefont {{LIGO Scientific
  Collaboration}}},\ and\ \bibinfo {author} {\bibnamefont {{Virgo
  Collaboration}}},\ }\bibfield  {title} {\bibinfo {title} {{Multi-messenger
  Observations of a Binary Neutron Star Merger}},\ }in\ \href@noop {} {\emph
  {\bibinfo {booktitle} {Multifrequency Behaviour of High Energy Cosmic Sources
  - XIII. 3-8 June 2019. Palermo}}}\ (\bibinfo {year} {2019})\ p.~\bibinfo
  {pages} {19}\BibitemShut {NoStop}%
\bibitem [{\citenamefont {Venumadhav}\ \emph {et~al.}(2020)\citenamefont
  {Venumadhav}, \citenamefont {Zackay}, \citenamefont {Roulet}, \citenamefont
  {Dai},\ and\ \citenamefont {Zaldarriaga}}]{Venu}%
  \BibitemOpen
  \bibfield  {author} {\bibinfo {author} {\bibfnamefont {T.}~\bibnamefont
  {Venumadhav}}, \bibinfo {author} {\bibfnamefont {B.}~\bibnamefont {Zackay}},
  \bibinfo {author} {\bibfnamefont {J.}~\bibnamefont {Roulet}}, \bibinfo
  {author} {\bibfnamefont {L.}~\bibnamefont {Dai}},\ and\ \bibinfo {author}
  {\bibfnamefont {M.}~\bibnamefont {Zaldarriaga}},\ }\bibfield  {title}
  {\bibinfo {title} {{New binary black hole mergers in the second observing run
  of Advanced LIGO and Advanced Virgo}},\ }\href
  {https://doi.org/10.1103/PhysRevD.101.083030} {\bibfield  {journal} {\bibinfo
   {journal} {Phys. Rev. D}\ }\textbf {\bibinfo {volume} {101}},\ \bibinfo
  {pages} {083030} (\bibinfo {year} {2020})},\ \Eprint
  {https://arxiv.org/abs/1904.07214} {arXiv:1904.07214 [astro-ph.HE]}
  \BibitemShut {NoStop}%
\bibitem [{\citenamefont {Abbott}\ \emph
  {et~al.}(2020{\natexlab{b}})\citenamefont {Abbott} \emph
  {et~al.}}]{Abbott:2020gyp}%
  \BibitemOpen
  \bibfield  {author} {\bibinfo {author} {\bibfnamefont {R.}~\bibnamefont
  {Abbott}} \emph {et~al.} (\bibinfo {collaboration} {LIGO Scientific,
  Virgo}),\ }\bibfield  {title} {\bibinfo {title} {{Population Properties of
  Compact Objects from the Second LIGO-Virgo Gravitational-Wave Transient
  Catalog}},\ }\href@noop {} {\  (\bibinfo {year} {2020}{\natexlab{b}})},\
  \Eprint {https://arxiv.org/abs/2010.14533} {arXiv:2010.14533 [astro-ph.HE]}
  \BibitemShut {NoStop}%
\bibitem [{\citenamefont {Abbott}\ \emph
  {et~al.}(2017{\natexlab{b}})\citenamefont {Abbott} \emph
  {et~al.}}]{2017Natur.551...85A}%
  \BibitemOpen
  \bibfield  {author} {\bibinfo {author} {\bibfnamefont {B.}~\bibnamefont
  {Abbott}} \emph {et~al.},\ }\bibfield  {title} {\bibinfo {title} {{A
  gravitational-wave standard siren measurement of the Hubble constant}},\
  }\href {https://doi.org/10.1038/nature24471} {\bibfield  {journal} {\bibinfo
  {journal} {\nat}\ }\textbf {\bibinfo {volume} {551}},\ \bibinfo {pages} {85}
  (\bibinfo {year} {2017}{\natexlab{b}})},\ \Eprint
  {https://arxiv.org/abs/1710.05835} {arXiv:1710.05835 [astro-ph.CO]}
  \BibitemShut {NoStop}%
\bibitem [{\citenamefont {Abbott}\ \emph
  {et~al.}(2020{\natexlab{c}})\citenamefont {Abbott} \emph
  {et~al.}}]{Abbott:2020jks}%
  \BibitemOpen
  \bibfield  {author} {\bibinfo {author} {\bibfnamefont {R.}~\bibnamefont
  {Abbott}} \emph {et~al.} (\bibinfo {collaboration} {LIGO Scientific,
  Virgo}),\ }\bibfield  {title} {\bibinfo {title} {{Tests of General Relativity
  with Binary Black Holes from the second LIGO-Virgo Gravitational-Wave
  Transient Catalog}},\ }\href@noop {} {\  (\bibinfo {year}
  {2020}{\natexlab{c}})},\ \Eprint {https://arxiv.org/abs/2010.14529}
  {arXiv:2010.14529 [gr-qc]} \BibitemShut {NoStop}%
\bibitem [{\citenamefont {Aasi}\ \emph {et~al.}(2015)\citenamefont {Aasi} \emph
  {et~al.}}]{TheLIGOScientific:2014jea}%
  \BibitemOpen
  \bibfield  {author} {\bibinfo {author} {\bibfnamefont {J.}~\bibnamefont
  {Aasi}} \emph {et~al.} (\bibinfo {collaboration} {LIGO Scientific}),\
  }\bibfield  {title} {\bibinfo {title} {{Advanced LIGO}},\ }\href
  {https://doi.org/10.1088/0264-9381/32/7/074001} {\bibfield  {journal}
  {\bibinfo  {journal} {Class. Quant. Grav.}\ }\textbf {\bibinfo {volume}
  {32}},\ \bibinfo {pages} {074001} (\bibinfo {year} {2015})},\ \Eprint
  {https://arxiv.org/abs/1411.4547} {arXiv:1411.4547 [gr-qc]} \BibitemShut
  {NoStop}%
\bibitem [{\citenamefont {Acernese}\ \emph {et~al.}(2015)\citenamefont
  {Acernese} \emph {et~al.}}]{TheVirgo:2014hva}%
  \BibitemOpen
  \bibfield  {author} {\bibinfo {author} {\bibfnamefont {F.}~\bibnamefont
  {Acernese}} \emph {et~al.} (\bibinfo {collaboration} {VIRGO}),\ }\bibfield
  {title} {\bibinfo {title} {{Advanced Virgo: a second-generation
  interferometric gravitational wave detector}},\ }\href
  {https://doi.org/10.1088/0264-9381/32/2/024001} {\bibfield  {journal}
  {\bibinfo  {journal} {Class. Quant. Grav.}\ }\textbf {\bibinfo {volume}
  {32}},\ \bibinfo {pages} {024001} (\bibinfo {year} {2015})},\ \Eprint
  {https://arxiv.org/abs/1408.3978} {arXiv:1408.3978 [gr-qc]} \BibitemShut
  {NoStop}%
\bibitem [{\citenamefont {Akutsu}\ \emph {et~al.}(2019)\citenamefont {Akutsu}
  \emph {et~al.}}]{Akutsu:2018axf}%
  \BibitemOpen
  \bibfield  {author} {\bibinfo {author} {\bibfnamefont {T.}~\bibnamefont
  {Akutsu}} \emph {et~al.} (\bibinfo {collaboration} {KAGRA}),\ }\bibfield
  {title} {\bibinfo {title} {{KAGRA: 2.5 Generation Interferometric
  Gravitational Wave Detector}},\ }\href
  {https://doi.org/10.1038/s41550-018-0658-y} {\bibfield  {journal} {\bibinfo
  {journal} {Nature Astron.}\ }\textbf {\bibinfo {volume} {3}},\ \bibinfo
  {pages} {35} (\bibinfo {year} {2019})},\ \Eprint
  {https://arxiv.org/abs/1811.08079} {arXiv:1811.08079 [gr-qc]} \BibitemShut
  {NoStop}%
\bibitem [{\citenamefont {Mapelli}(2020)}]{Mapelli:2018uds}%
  \BibitemOpen
  \bibfield  {author} {\bibinfo {author} {\bibfnamefont {M.}~\bibnamefont
  {Mapelli}},\ }\bibfield  {title} {\bibinfo {title} {{Astrophysics of stellar
  black holes}},\ }\href {https://doi.org/10.3254/ENFI200005} {\bibfield
  {journal} {\bibinfo  {journal} {Proc. Int. Sch. Phys. Fermi}\ }\textbf
  {\bibinfo {volume} {200}},\ \bibinfo {pages} {87} (\bibinfo {year} {2020})},\
  \Eprint {https://arxiv.org/abs/1809.09130} {arXiv:1809.09130 [astro-ph.HE]}
  \BibitemShut {NoStop}%
\bibitem [{\citenamefont {{Mandel}}\ and\ \citenamefont
  {{Farmer}}(2018)}]{2018arXiv180605820M}%
  \BibitemOpen
  \bibfield  {author} {\bibinfo {author} {\bibfnamefont {I.}~\bibnamefont
  {{Mandel}}}\ and\ \bibinfo {author} {\bibfnamefont {A.}~\bibnamefont
  {{Farmer}}},\ }\bibfield  {title} {\bibinfo {title} {{Merging stellar-mass
  binary black holes}},\ }\href@noop {} {\bibfield  {journal} {\bibinfo
  {journal} {arXiv e-prints}\ ,\ \bibinfo {eid} {arXiv:1806.05820}} (\bibinfo
  {year} {2018})},\ \Eprint {https://arxiv.org/abs/1806.05820}
  {arXiv:1806.05820 [astro-ph.HE]} \BibitemShut {NoStop}%
\bibitem [{\citenamefont {{Postnov}}\ and\ \citenamefont
  {{Yungelson}}(2006)}]{2006LRR96P}%
  \BibitemOpen
  \bibfield  {author} {\bibinfo {author} {\bibfnamefont {K.~A.}\ \bibnamefont
  {{Postnov}}}\ and\ \bibinfo {author} {\bibfnamefont {L.~R.}\ \bibnamefont
  {{Yungelson}}},\ }\bibfield  {title} {\bibinfo {title} {{The Evolution of
  Compact Binary Star Systems}},\ }\href {https://doi.org/10.12942/lrr-2006-6}
  {\bibfield  {journal} {\bibinfo  {journal} {Living Reviews in Relativity}\
  }\textbf {\bibinfo {volume} {9}},\ \bibinfo {eid} {6} (\bibinfo {year}
  {2006})},\ \Eprint {https://arxiv.org/abs/astro-ph/0701059}
  {arXiv:astro-ph/0701059 [astro-ph]} \BibitemShut {NoStop}%
\bibitem [{\citenamefont {{Kruckow}}\ \emph {et~al.}(2018)\citenamefont
  {{Kruckow}}, \citenamefont {{Tauris}}, \citenamefont {{Langer}},
  \citenamefont {{Kramer}},\ and\ \citenamefont
  {{Izzard}}}]{2018MNRAS.481.1908K}%
  \BibitemOpen
  \bibfield  {author} {\bibinfo {author} {\bibfnamefont {M.~U.}\ \bibnamefont
  {{Kruckow}}}, \bibinfo {author} {\bibfnamefont {T.~M.}\ \bibnamefont
  {{Tauris}}}, \bibinfo {author} {\bibfnamefont {N.}~\bibnamefont {{Langer}}},
  \bibinfo {author} {\bibfnamefont {M.}~\bibnamefont {{Kramer}}},\ and\
  \bibinfo {author} {\bibfnamefont {R.~G.}\ \bibnamefont {{Izzard}}},\
  }\bibfield  {title} {\bibinfo {title} {{Progenitors of gravitational wave
  mergers: binary evolution with the stellar grid-based code COMBINE}},\ }\href
  {https://doi.org/10.1093/mnras/sty2190} {\bibfield  {journal} {\bibinfo
  {journal} {\mnras}\ }\textbf {\bibinfo {volume} {481}},\ \bibinfo {pages}
  {1908} (\bibinfo {year} {2018})},\ \Eprint {https://arxiv.org/abs/1801.05433}
  {arXiv:1801.05433 [astro-ph.SR]} \BibitemShut {NoStop}%
\bibitem [{\citenamefont {Kowalska}\ \emph {et~al.}(2011)\citenamefont
  {Kowalska}, \citenamefont {Bulik}, \citenamefont {Belczynski}, \citenamefont
  {Dominik},\ and\ \citenamefont {Gondek-Rosinska}}]{Kowalska:2010qg}%
  \BibitemOpen
  \bibfield  {author} {\bibinfo {author} {\bibfnamefont {I.}~\bibnamefont
  {Kowalska}}, \bibinfo {author} {\bibfnamefont {T.}~\bibnamefont {Bulik}},
  \bibinfo {author} {\bibfnamefont {K.}~\bibnamefont {Belczynski}}, \bibinfo
  {author} {\bibfnamefont {M.}~\bibnamefont {Dominik}},\ and\ \bibinfo {author}
  {\bibfnamefont {D.}~\bibnamefont {Gondek-Rosinska}},\ }\bibfield  {title}
  {\bibinfo {title} {{The eccentricity distribution of compact binaries}},\
  }\href {https://doi.org/10.1051/0004-6361/201015777} {\bibfield  {journal}
  {\bibinfo  {journal} {Astron. Astrophys.}\ }\textbf {\bibinfo {volume}
  {527}},\ \bibinfo {pages} {A70} (\bibinfo {year} {2011})},\ \Eprint
  {https://arxiv.org/abs/1010.0511} {arXiv:1010.0511 [astro-ph.CO]}
  \BibitemShut {NoStop}%
\bibitem [{\citenamefont {{Fragione}}\ and\ \citenamefont
  {{Bromberg}}(2019)}]{2019MNRAS.488.4370F}%
  \BibitemOpen
  \bibfield  {author} {\bibinfo {author} {\bibfnamefont {G.}~\bibnamefont
  {{Fragione}}}\ and\ \bibinfo {author} {\bibfnamefont {O.}~\bibnamefont
  {{Bromberg}}},\ }\bibfield  {title} {\bibinfo {title} {{Eccentric binary
  black hole mergers in globular clusters hosting intermediate-mass black
  holes}},\ }\href {https://doi.org/10.1093/mnras/stz2024} {\bibfield
  {journal} {\bibinfo  {journal} {\mnras}\ }\textbf {\bibinfo {volume} {488}},\
  \bibinfo {pages} {4370} (\bibinfo {year} {2019})},\ \Eprint
  {https://arxiv.org/abs/1903.09659} {arXiv:1903.09659 [astro-ph.GA]}
  \BibitemShut {NoStop}%
\bibitem [{\citenamefont {Samsing}(2018)}]{Samsing:2017xmd}%
  \BibitemOpen
  \bibfield  {author} {\bibinfo {author} {\bibfnamefont {J.}~\bibnamefont
  {Samsing}},\ }\bibfield  {title} {\bibinfo {title} {{Eccentric Black Hole
  Mergers Forming in Globular Clusters}},\ }\href
  {https://doi.org/10.1103/PhysRevD.97.103014} {\bibfield  {journal} {\bibinfo
  {journal} {Phys. Rev. D}\ }\textbf {\bibinfo {volume} {97}},\ \bibinfo
  {pages} {103014} (\bibinfo {year} {2018})},\ \Eprint
  {https://arxiv.org/abs/1711.07452} {arXiv:1711.07452 [astro-ph.HE]}
  \BibitemShut {NoStop}%
\bibitem [{\citenamefont {{Kumamoto}}\ \emph {et~al.}(2020)\citenamefont
  {{Kumamoto}}, \citenamefont {{Fujii}},\ and\ \citenamefont
  {{Tanikawa}}}]{Kumamoto:2020wqr}%
  \BibitemOpen
  \bibfield  {author} {\bibinfo {author} {\bibfnamefont {J.}~\bibnamefont
  {{Kumamoto}}}, \bibinfo {author} {\bibfnamefont {M.~S.}\ \bibnamefont
  {{Fujii}}},\ and\ \bibinfo {author} {\bibfnamefont {A.}~\bibnamefont
  {{Tanikawa}}},\ }\bibfield  {title} {\bibinfo {title} {{Merger rate density
  of binary black holes formed in open clusters}},\ }\href
  {https://doi.org/10.1093/mnras/staa1440} {\bibfield  {journal} {\bibinfo
  {journal} {\mnras}\ }\textbf {\bibinfo {volume} {495}},\ \bibinfo {pages}
  {4268} (\bibinfo {year} {2020})},\ \Eprint {https://arxiv.org/abs/2001.10690}
  {arXiv:2001.10690 [astro-ph.HE]} \BibitemShut {NoStop}%
\bibitem [{\citenamefont {O'Leary}\ \emph {et~al.}(2009)\citenamefont
  {O'Leary}, \citenamefont {Kocsis},\ and\ \citenamefont
  {Loeb}}]{OLeary:2008myb}%
  \BibitemOpen
  \bibfield  {author} {\bibinfo {author} {\bibfnamefont {R.~M.}\ \bibnamefont
  {O'Leary}}, \bibinfo {author} {\bibfnamefont {B.}~\bibnamefont {Kocsis}},\
  and\ \bibinfo {author} {\bibfnamefont {A.}~\bibnamefont {Loeb}},\ }\bibfield
  {title} {\bibinfo {title} {{Gravitational waves from scattering of
  stellar-mass black holes in galactic nuclei}},\ }\href
  {https://doi.org/10.1111/j.1365-2966.2009.14653.x} {\bibfield  {journal}
  {\bibinfo  {journal} {Mon. Not. Roy. Astron. Soc.}\ }\textbf {\bibinfo
  {volume} {395}},\ \bibinfo {pages} {2127} (\bibinfo {year} {2009})},\ \Eprint
  {https://arxiv.org/abs/0807.2638} {arXiv:0807.2638 [astro-ph]} \BibitemShut
  {NoStop}%
\bibitem [{\citenamefont {{Farr}}\ \emph {et~al.}(2017)\citenamefont {{Farr}},
  \citenamefont {{Stevenson}}, \citenamefont {{Miller}}, \citenamefont
  {{Mandel}}, \citenamefont {{Farr}},\ and\ \citenamefont
  {{Vecchio}}}]{Farr:2017uvj}%
  \BibitemOpen
  \bibfield  {author} {\bibinfo {author} {\bibfnamefont {W.~M.}\ \bibnamefont
  {{Farr}}}, \bibinfo {author} {\bibfnamefont {S.}~\bibnamefont {{Stevenson}}},
  \bibinfo {author} {\bibfnamefont {M.~C.}\ \bibnamefont {{Miller}}}, \bibinfo
  {author} {\bibfnamefont {I.}~\bibnamefont {{Mandel}}}, \bibinfo {author}
  {\bibfnamefont {B.}~\bibnamefont {{Farr}}},\ and\ \bibinfo {author}
  {\bibfnamefont {A.}~\bibnamefont {{Vecchio}}},\ }\bibfield  {title} {\bibinfo
  {title} {{Distinguishing spin-aligned and isotropic black hole populations
  with gravitational waves}},\ }\href {https://doi.org/10.1038/nature23453}
  {\bibfield  {journal} {\bibinfo  {journal} {\nat}\ }\textbf {\bibinfo
  {volume} {548}},\ \bibinfo {pages} {426} (\bibinfo {year} {2017})},\ \Eprint
  {https://arxiv.org/abs/1706.01385} {arXiv:1706.01385 [astro-ph.HE]}
  \BibitemShut {NoStop}%
\bibitem [{\citenamefont {Sedda}\ \emph {et~al.}(2020)\citenamefont {Sedda},
  \citenamefont {Mapelli}, \citenamefont {Spera}, \citenamefont {Benacquista},\
  and\ \citenamefont {Giacobbo}}]{Sedda:2020vwo}%
  \BibitemOpen
  \bibfield  {author} {\bibinfo {author} {\bibfnamefont {M.~A.}\ \bibnamefont
  {Sedda}}, \bibinfo {author} {\bibfnamefont {M.}~\bibnamefont {Mapelli}},
  \bibinfo {author} {\bibfnamefont {M.}~\bibnamefont {Spera}}, \bibinfo
  {author} {\bibfnamefont {M.}~\bibnamefont {Benacquista}},\ and\ \bibinfo
  {author} {\bibfnamefont {N.}~\bibnamefont {Giacobbo}},\ }\bibfield  {title}
  {\bibinfo {title} {{Fingerprints of binary black hole formation channels
  encoded in the mass and spin of merger remnants}},\ }\href
  {https://doi.org/10.3847/1538-4357/ab88b2} {\bibfield  {journal} {\bibinfo
  {journal} {Astrophys. J.}\ }\textbf {\bibinfo {volume} {894}},\ \bibinfo
  {pages} {133} (\bibinfo {year} {2020})},\ \Eprint
  {https://arxiv.org/abs/2003.07409} {arXiv:2003.07409 [astro-ph.GA]}
  \BibitemShut {NoStop}%
\bibitem [{\citenamefont {{Park}}\ \emph {et~al.}(2017)\citenamefont {{Park}},
  \citenamefont {{Kim}}, \citenamefont {{Lee}}, \citenamefont {{Bae}},\ and\
  \citenamefont {{Belczynski}}}]{Park_2017}%
  \BibitemOpen
  \bibfield  {author} {\bibinfo {author} {\bibfnamefont {D.}~\bibnamefont
  {{Park}}}, \bibinfo {author} {\bibfnamefont {C.}~\bibnamefont {{Kim}}},
  \bibinfo {author} {\bibfnamefont {H.~M.}\ \bibnamefont {{Lee}}}, \bibinfo
  {author} {\bibfnamefont {Y.-B.}\ \bibnamefont {{Bae}}},\ and\ \bibinfo
  {author} {\bibfnamefont {K.}~\bibnamefont {{Belczynski}}},\ }\bibfield
  {title} {\bibinfo {title} {{Black hole binaries dynamically formed in
  globular clusters}},\ }\href {https://doi.org/10.1093/mnras/stx1015}
  {\bibfield  {journal} {\bibinfo  {journal} {\mnras}\ }\textbf {\bibinfo
  {volume} {469}},\ \bibinfo {pages} {4665} (\bibinfo {year} {2017})},\ \Eprint
  {https://arxiv.org/abs/1703.01568} {arXiv:1703.01568 [astro-ph.HE]}
  \BibitemShut {NoStop}%
\bibitem [{\citenamefont {{Hong}}\ and\ \citenamefont
  {{Lee}}(2015)}]{Hong_2015}%
  \BibitemOpen
  \bibfield  {author} {\bibinfo {author} {\bibfnamefont {J.}~\bibnamefont
  {{Hong}}}\ and\ \bibinfo {author} {\bibfnamefont {H.~M.}\ \bibnamefont
  {{Lee}}},\ }\bibfield  {title} {\bibinfo {title} {{Black hole binaries in
  galactic nuclei and gravitational wave sources}},\ }\href
  {https://doi.org/10.1093/mnras/stv035} {\bibfield  {journal} {\bibinfo
  {journal} {\mnras}\ }\textbf {\bibinfo {volume} {448}},\ \bibinfo {pages}
  {754} (\bibinfo {year} {2015})},\ \Eprint {https://arxiv.org/abs/1501.02717}
  {arXiv:1501.02717 [astro-ph.GA]} \BibitemShut {NoStop}%
\bibitem [{\citenamefont {{Rodriguez}}\ \emph {et~al.}(2018)\citenamefont
  {{Rodriguez}}, \citenamefont {{Amaro-Seoane}}, \citenamefont {{Chatterjee}},
  \citenamefont {{Kremer}}, \citenamefont {{Rasio}}, \citenamefont {{Samsing}},
  \citenamefont {{Ye}},\ and\ \citenamefont {{Zevin}}}]{2018PhRvD..98l3005R}%
  \BibitemOpen
  \bibfield  {author} {\bibinfo {author} {\bibfnamefont {C.~L.}\ \bibnamefont
  {{Rodriguez}}}, \bibinfo {author} {\bibfnamefont {P.}~\bibnamefont
  {{Amaro-Seoane}}}, \bibinfo {author} {\bibfnamefont {S.}~\bibnamefont
  {{Chatterjee}}}, \bibinfo {author} {\bibfnamefont {K.}~\bibnamefont
  {{Kremer}}}, \bibinfo {author} {\bibfnamefont {F.~A.}\ \bibnamefont
  {{Rasio}}}, \bibinfo {author} {\bibfnamefont {J.}~\bibnamefont {{Samsing}}},
  \bibinfo {author} {\bibfnamefont {C.~S.}\ \bibnamefont {{Ye}}},\ and\
  \bibinfo {author} {\bibfnamefont {M.}~\bibnamefont {{Zevin}}},\ }\bibfield
  {title} {\bibinfo {title} {{Post-Newtonian dynamics in dense star clusters:
  Formation, masses, and merger rates of highly-eccentric black hole
  binaries}},\ }\href {https://doi.org/10.1103/PhysRevD.98.123005} {\bibfield
  {journal} {\bibinfo  {journal} {\prd}\ }\textbf {\bibinfo {volume} {98}},\
  \bibinfo {eid} {123005} (\bibinfo {year} {2018})},\ \Eprint
  {https://arxiv.org/abs/1811.04926} {arXiv:1811.04926 [astro-ph.HE]}
  \BibitemShut {NoStop}%
\bibitem [{\citenamefont {{Romero-Shaw}}\ \emph {et~al.}(2019)\citenamefont
  {{Romero-Shaw}}, \citenamefont {{Lasky}},\ and\ \citenamefont
  {{Thrane}}}]{2019MNRAS.490.5210R}%
  \BibitemOpen
  \bibfield  {author} {\bibinfo {author} {\bibfnamefont {I.~M.}\ \bibnamefont
  {{Romero-Shaw}}}, \bibinfo {author} {\bibfnamefont {P.~D.}\ \bibnamefont
  {{Lasky}}},\ and\ \bibinfo {author} {\bibfnamefont {E.}~\bibnamefont
  {{Thrane}}},\ }\bibfield  {title} {\bibinfo {title} {{Searching for
  eccentricity: signatures of dynamical formation in the first
  gravitational-wave transient catalogue of LIGO and Virgo}},\ }\href
  {https://doi.org/10.1093/mnras/stz2996} {\bibfield  {journal} {\bibinfo
  {journal} {\mnras}\ }\textbf {\bibinfo {volume} {490}},\ \bibinfo {pages}
  {5210} (\bibinfo {year} {2019})},\ \Eprint {https://arxiv.org/abs/1909.05466}
  {arXiv:1909.05466 [astro-ph.HE]} \BibitemShut {NoStop}%
\bibitem [{\citenamefont {{Gond{\'a}n}}\ and\ \citenamefont
  {{Kocsis}}(2020)}]{2020arXiv201102507G}%
  \BibitemOpen
  \bibfield  {author} {\bibinfo {author} {\bibfnamefont {L.}~\bibnamefont
  {{Gond{\'a}n}}}\ and\ \bibinfo {author} {\bibfnamefont {B.}~\bibnamefont
  {{Kocsis}}},\ }\bibfield  {title} {\bibinfo {title} {{High Eccentricities and
  High Masses Characterize Gravitational-wave Captures in Galactic Nuclei as
  Seen by Earth-based Detectors}},\ }\href@noop {} {\bibfield  {journal}
  {\bibinfo  {journal} {arXiv e-prints}\ ,\ \bibinfo {eid} {arXiv:2011.02507}}
  (\bibinfo {year} {2020})},\ \Eprint {https://arxiv.org/abs/2011.02507}
  {arXiv:2011.02507 [astro-ph.HE]} \BibitemShut {NoStop}%
\bibitem [{\citenamefont {Hinder}\ \emph {et~al.}(2018)\citenamefont {Hinder},
  \citenamefont {Kidder},\ and\ \citenamefont {Pfeiffer}}]{Hinder2017}%
  \BibitemOpen
  \bibfield  {author} {\bibinfo {author} {\bibfnamefont {I.}~\bibnamefont
  {Hinder}}, \bibinfo {author} {\bibfnamefont {L.~E.}\ \bibnamefont {Kidder}},\
  and\ \bibinfo {author} {\bibfnamefont {H.~P.}\ \bibnamefont {Pfeiffer}},\
  }\bibfield  {title} {\bibinfo {title} {{Eccentric binary black hole
  inspiral-merger-ringdown gravitational waveform model from numerical
  relativity and post-Newtonian theory}},\ }\href
  {https://doi.org/10.1103/PhysRevD.98.044015} {\bibfield  {journal} {\bibinfo
  {journal} {Phys. Rev.}\ }\textbf {\bibinfo {volume} {D98}},\ \bibinfo {pages}
  {044015} (\bibinfo {year} {2018})},\ \Eprint
  {https://arxiv.org/abs/1709.02007} {arXiv:1709.02007 [gr-qc]} \BibitemShut
  {NoStop}%
\bibitem [{\citenamefont {{Huerta}}\ \emph {et~al.}(2018)\citenamefont
  {{Huerta}}, \citenamefont {{Moore}}, \citenamefont {{Kumar}}, \citenamefont
  {{George}}, \citenamefont {{Chua}}, \citenamefont {{Haas}}, \citenamefont
  {{Wessel}}, \citenamefont {{Johnson}}, \citenamefont {{Glennon}},
  \citenamefont {{Rebei}}, \citenamefont {{Holgado}}, \citenamefont {{Gair}},\
  and\ \citenamefont {{Pfeiffer}}}]{Huerta_18}%
  \BibitemOpen
  \bibfield  {author} {\bibinfo {author} {\bibfnamefont {E.~A.}\ \bibnamefont
  {{Huerta}}}, \bibinfo {author} {\bibfnamefont {C.~J.}\ \bibnamefont
  {{Moore}}}, \bibinfo {author} {\bibfnamefont {P.}~\bibnamefont {{Kumar}}},
  \bibinfo {author} {\bibfnamefont {D.}~\bibnamefont {{George}}}, \bibinfo
  {author} {\bibfnamefont {A.~J.~K.}\ \bibnamefont {{Chua}}}, \bibinfo {author}
  {\bibfnamefont {R.}~\bibnamefont {{Haas}}}, \bibinfo {author} {\bibfnamefont
  {E.}~\bibnamefont {{Wessel}}}, \bibinfo {author} {\bibfnamefont
  {D.}~\bibnamefont {{Johnson}}}, \bibinfo {author} {\bibfnamefont
  {D.}~\bibnamefont {{Glennon}}}, \bibinfo {author} {\bibfnamefont
  {A.}~\bibnamefont {{Rebei}}}, \bibinfo {author} {\bibfnamefont {A.~M.}\
  \bibnamefont {{Holgado}}}, \bibinfo {author} {\bibfnamefont {J.~R.}\
  \bibnamefont {{Gair}}},\ and\ \bibinfo {author} {\bibfnamefont {H.~P.}\
  \bibnamefont {{Pfeiffer}}},\ }\bibfield  {title} {\bibinfo {title}
  {{Eccentric, nonspinning, inspiral, Gaussian-process merger approximant for
  the detection and characterization of eccentric binary black hole mergers}},\
  }\href {https://doi.org/10.1103/PhysRevD.97.024031} {\bibfield  {journal}
  {\bibinfo  {journal} {\prd}\ }\textbf {\bibinfo {volume} {97}},\ \bibinfo
  {eid} {024031} (\bibinfo {year} {2018})},\ \Eprint
  {https://arxiv.org/abs/1711.06276} {arXiv:1711.06276 [gr-qc]} \BibitemShut
  {NoStop}%
\bibitem [{\citenamefont {{Hinderer}}\ and\ \citenamefont
  {{Babak}}(2017)}]{2017PhRvD..96j4048H}%
  \BibitemOpen
  \bibfield  {author} {\bibinfo {author} {\bibfnamefont {T.}~\bibnamefont
  {{Hinderer}}}\ and\ \bibinfo {author} {\bibfnamefont {S.}~\bibnamefont
  {{Babak}}},\ }\bibfield  {title} {\bibinfo {title} {{Foundations of an
  effective-one-body model for coalescing binaries on eccentric orbits}},\
  }\href {https://doi.org/10.1103/PhysRevD.96.104048} {\bibfield  {journal}
  {\bibinfo  {journal} {\prd}\ }\textbf {\bibinfo {volume} {96}},\ \bibinfo
  {eid} {104048} (\bibinfo {year} {2017})},\ \Eprint
  {https://arxiv.org/abs/1707.08426} {arXiv:1707.08426 [gr-qc]} \BibitemShut
  {NoStop}%
\bibitem [{\citenamefont {{Cao}}\ and\ \citenamefont
  {{Han}}(2017)}]{2017PhRvD..96d4028C}%
  \BibitemOpen
  \bibfield  {author} {\bibinfo {author} {\bibfnamefont {Z.}~\bibnamefont
  {{Cao}}}\ and\ \bibinfo {author} {\bibfnamefont {W.-B.}\ \bibnamefont
  {{Han}}},\ }\bibfield  {title} {\bibinfo {title} {{Waveform model for an
  eccentric binary black hole based on the
  effective-one-body-numerical-relativity formalism}},\ }\href
  {https://doi.org/10.1103/PhysRevD.96.044028} {\bibfield  {journal} {\bibinfo
  {journal} {\prd}\ }\textbf {\bibinfo {volume} {96}},\ \bibinfo {eid} {044028}
  (\bibinfo {year} {2017})},\ \Eprint {https://arxiv.org/abs/1708.00166}
  {arXiv:1708.00166 [gr-qc]} \BibitemShut {NoStop}%
\bibitem [{\citenamefont {{Chiaramello}}\ and\ \citenamefont
  {{Nagar}}(2020)}]{CN20}%
  \BibitemOpen
  \bibfield  {author} {\bibinfo {author} {\bibfnamefont {D.}~\bibnamefont
  {{Chiaramello}}}\ and\ \bibinfo {author} {\bibfnamefont {A.}~\bibnamefont
  {{Nagar}}},\ }\bibfield  {title} {\bibinfo {title} {{Faithful analytical
  effective-one-body waveform model for spin-aligned, moderately eccentric,
  coalescing black hole binaries}},\ }\href
  {https://doi.org/10.1103/PhysRevD.101.101501} {\bibfield  {journal} {\bibinfo
   {journal} {\prd}\ }\textbf {\bibinfo {volume} {101}},\ \bibinfo {eid}
  {101501} (\bibinfo {year} {2020})},\ \Eprint
  {https://arxiv.org/abs/2001.11736} {arXiv:2001.11736 [gr-qc]} \BibitemShut
  {NoStop}%
\bibitem [{\citenamefont {{Tiwari}}\ \emph {et~al.}(2019)\citenamefont
  {{Tiwari}}, \citenamefont {{Gopakumar}}, \citenamefont {{Haney}},\ and\
  \citenamefont {{Hemantakumar}}}]{2019PhRvD..99l4008T}%
  \BibitemOpen
  \bibfield  {author} {\bibinfo {author} {\bibfnamefont {S.}~\bibnamefont
  {{Tiwari}}}, \bibinfo {author} {\bibfnamefont {A.}~\bibnamefont
  {{Gopakumar}}}, \bibinfo {author} {\bibfnamefont {M.}~\bibnamefont
  {{Haney}}},\ and\ \bibinfo {author} {\bibfnamefont {P.}~\bibnamefont
  {{Hemantakumar}}},\ }\bibfield  {title} {\bibinfo {title} {{Ready-to-use
  Fourier domain templates for compact binaries inspiraling along moderately
  eccentric orbits}},\ }\href {https://doi.org/10.1103/PhysRevD.99.124008}
  {\bibfield  {journal} {\bibinfo  {journal} {\prd}\ }\textbf {\bibinfo
  {volume} {99}},\ \bibinfo {eid} {124008} (\bibinfo {year} {2019})},\ \Eprint
  {https://arxiv.org/abs/1905.07956} {arXiv:1905.07956 [gr-qc]} \BibitemShut
  {NoStop}%
\bibitem [{\citenamefont {{Ramos-Buades}}\ \emph {et~al.}(2020)\citenamefont
  {{Ramos-Buades}}, \citenamefont {{Tiwari}}, \citenamefont {{Haney}},\ and\
  \citenamefont {{Husa}}}]{RTHH}%
  \BibitemOpen
  \bibfield  {author} {\bibinfo {author} {\bibfnamefont {A.}~\bibnamefont
  {{Ramos-Buades}}}, \bibinfo {author} {\bibfnamefont {S.}~\bibnamefont
  {{Tiwari}}}, \bibinfo {author} {\bibfnamefont {M.}~\bibnamefont {{Haney}}},\
  and\ \bibinfo {author} {\bibfnamefont {S.}~\bibnamefont {{Husa}}},\
  }\bibfield  {title} {\bibinfo {title} {{Impact of eccentricity on the
  gravitational-wave searches for binary black holes: High mass case}},\ }\href
  {https://doi.org/10.1103/PhysRevD.102.043005} {\bibfield  {journal} {\bibinfo
   {journal} {\prd}\ }\textbf {\bibinfo {volume} {102}},\ \bibinfo {eid}
  {043005} (\bibinfo {year} {2020})},\ \Eprint
  {https://arxiv.org/abs/2005.14016} {arXiv:2005.14016 [gr-qc]} \BibitemShut
  {NoStop}%
\bibitem [{\citenamefont {{Nagar}}\ \emph {et~al.}(2021)\citenamefont
  {{Nagar}}, \citenamefont {{Bonino}},\ and\ \citenamefont
  {{Rettegno}}}]{2021arXiv210108624N}%
  \BibitemOpen
  \bibfield  {author} {\bibinfo {author} {\bibfnamefont {A.}~\bibnamefont
  {{Nagar}}}, \bibinfo {author} {\bibfnamefont {A.}~\bibnamefont {{Bonino}}},\
  and\ \bibinfo {author} {\bibfnamefont {P.}~\bibnamefont {{Rettegno}}},\
  }\bibfield  {title} {\bibinfo {title} {{All in one: effective one body
  multipolar waveform model for spin-aligned, quasi-circular, eccentric,
  hyperbolic black hole binaries}},\ }\href@noop {} {\bibfield  {journal}
  {\bibinfo  {journal} {arXiv e-prints}\ ,\ \bibinfo {eid} {arXiv:2101.08624}}
  (\bibinfo {year} {2021})},\ \Eprint {https://arxiv.org/abs/2101.08624}
  {arXiv:2101.08624 [gr-qc]} \BibitemShut {NoStop}%
\bibitem [{\citenamefont {{Romero-Shaw}}\ \emph
  {et~al.}(2020{\natexlab{a}})\citenamefont {{Romero-Shaw}}, \citenamefont
  {{Farrow}}, \citenamefont {{Stevenson}}, \citenamefont {{Thrane}},\ and\
  \citenamefont {{Zhu}}}]{2020MNRAS.496L..64R}%
  \BibitemOpen
  \bibfield  {author} {\bibinfo {author} {\bibfnamefont {I.~M.}\ \bibnamefont
  {{Romero-Shaw}}}, \bibinfo {author} {\bibfnamefont {N.}~\bibnamefont
  {{Farrow}}}, \bibinfo {author} {\bibfnamefont {S.}~\bibnamefont
  {{Stevenson}}}, \bibinfo {author} {\bibfnamefont {E.}~\bibnamefont
  {{Thrane}}},\ and\ \bibinfo {author} {\bibfnamefont {X.-J.}\ \bibnamefont
  {{Zhu}}},\ }\bibfield  {title} {\bibinfo {title} {{On the origin of
  GW190425}},\ }\href {https://doi.org/10.1093/mnrasl/slaa084} {\bibfield
  {journal} {\bibinfo  {journal} {\mnras}\ }\textbf {\bibinfo {volume} {496}},\
  \bibinfo {pages} {L64} (\bibinfo {year} {2020}{\natexlab{a}})},\ \Eprint
  {https://arxiv.org/abs/2001.06492} {arXiv:2001.06492 [astro-ph.HE]}
  \BibitemShut {NoStop}%
\bibitem [{\citenamefont {{Romero-Shaw}}\ \emph
  {et~al.}(2020{\natexlab{b}})\citenamefont {{Romero-Shaw}}, \citenamefont
  {{Lasky}}, \citenamefont {{Thrane}},\ and\ \citenamefont {{Calder{\'o}n
  Bustillo}}}]{2020ApJ...903L...5R}%
  \BibitemOpen
  \bibfield  {author} {\bibinfo {author} {\bibfnamefont {I.}~\bibnamefont
  {{Romero-Shaw}}}, \bibinfo {author} {\bibfnamefont {P.~D.}\ \bibnamefont
  {{Lasky}}}, \bibinfo {author} {\bibfnamefont {E.}~\bibnamefont {{Thrane}}},\
  and\ \bibinfo {author} {\bibfnamefont {J.}~\bibnamefont {{Calder{\'o}n
  Bustillo}}},\ }\bibfield  {title} {\bibinfo {title} {{GW190521: Orbital
  Eccentricity and Signatures of Dynamical Formation in a Binary Black Hole
  Merger Signal}},\ }\href {https://doi.org/10.3847/2041-8213/abbe26}
  {\bibfield  {journal} {\bibinfo  {journal} {\apjl}\ }\textbf {\bibinfo
  {volume} {903}},\ \bibinfo {eid} {L5} (\bibinfo {year}
  {2020}{\natexlab{b}})},\ \Eprint {https://arxiv.org/abs/2009.04771}
  {arXiv:2009.04771 [astro-ph.HE]} \BibitemShut {NoStop}%
\bibitem [{\citenamefont {{Huerta}}\ \emph {et~al.}(2014)\citenamefont
  {{Huerta}}, \citenamefont {{Kumar}}, \citenamefont {{McWilliams}},
  \citenamefont {{O'Shaughnessy}},\ and\ \citenamefont
  {{Yunes}}}]{2014PhRvD..90h4016H}%
  \BibitemOpen
  \bibfield  {author} {\bibinfo {author} {\bibfnamefont {E.~A.}\ \bibnamefont
  {{Huerta}}}, \bibinfo {author} {\bibfnamefont {P.}~\bibnamefont {{Kumar}}},
  \bibinfo {author} {\bibfnamefont {S.~T.}\ \bibnamefont {{McWilliams}}},
  \bibinfo {author} {\bibfnamefont {R.}~\bibnamefont {{O'Shaughnessy}}},\ and\
  \bibinfo {author} {\bibfnamefont {N.}~\bibnamefont {{Yunes}}},\ }\bibfield
  {title} {\bibinfo {title} {{Accurate and efficient waveforms for compact
  binaries on eccentric orbits}},\ }\href
  {https://doi.org/10.1103/PhysRevD.90.084016} {\bibfield  {journal} {\bibinfo
  {journal} {\prd}\ }\textbf {\bibinfo {volume} {90}},\ \bibinfo {eid} {084016}
  (\bibinfo {year} {2014})},\ \Eprint {https://arxiv.org/abs/1408.3406}
  {arXiv:1408.3406 [gr-qc]} \BibitemShut {NoStop}%
\bibitem [{\citenamefont {{Moore}}\ \emph {et~al.}(2016)\citenamefont
  {{Moore}}, \citenamefont {{Favata}}, \citenamefont {{Arun}},\ and\
  \citenamefont {{Mishra}}}]{2016PhRvD..93l4061M}%
  \BibitemOpen
  \bibfield  {author} {\bibinfo {author} {\bibfnamefont {B.}~\bibnamefont
  {{Moore}}}, \bibinfo {author} {\bibfnamefont {M.}~\bibnamefont {{Favata}}},
  \bibinfo {author} {\bibfnamefont {K.~G.}\ \bibnamefont {{Arun}}},\ and\
  \bibinfo {author} {\bibfnamefont {C.~K.}\ \bibnamefont {{Mishra}}},\
  }\bibfield  {title} {\bibinfo {title} {{Gravitational-wave phasing for
  low-eccentricity inspiralling compact binaries to 3PN order}},\ }\href
  {https://doi.org/10.1103/PhysRevD.93.124061} {\bibfield  {journal} {\bibinfo
  {journal} {\prd}\ }\textbf {\bibinfo {volume} {93}},\ \bibinfo {eid} {124061}
  (\bibinfo {year} {2016})},\ \Eprint {https://arxiv.org/abs/1605.00304}
  {arXiv:1605.00304 [gr-qc]} \BibitemShut {NoStop}%
\bibitem [{\citenamefont {{Setyawati}}\ and\ \citenamefont
  {{Ohme}}(2021)}]{2021arXiv210111033S}%
  \BibitemOpen
  \bibfield  {author} {\bibinfo {author} {\bibfnamefont {Y.}~\bibnamefont
  {{Setyawati}}}\ and\ \bibinfo {author} {\bibfnamefont {F.}~\bibnamefont
  {{Ohme}}},\ }\bibfield  {title} {\bibinfo {title} {{Adding eccentricity to
  quasi-circular binary-black-hole waveform models}},\ }\href@noop {}
  {\bibfield  {journal} {\bibinfo  {journal} {arXiv e-prints}\ ,\ \bibinfo
  {eid} {arXiv:2101.11033}} (\bibinfo {year} {2021})},\ \Eprint
  {https://arxiv.org/abs/2101.11033} {arXiv:2101.11033 [gr-qc]} \BibitemShut
  {NoStop}%
\bibitem [{\citenamefont {Huerta}\ \emph {et~al.}(2017)\citenamefont {Huerta}
  \emph {et~al.}}]{Huerta:2016rwp}%
  \BibitemOpen
  \bibfield  {author} {\bibinfo {author} {\bibfnamefont {E.~A.}\ \bibnamefont
  {Huerta}} \emph {et~al.},\ }\bibfield  {title} {\bibinfo {title} {{Complete
  waveform model for compact binaries on eccentric orbits}},\ }\href
  {https://doi.org/10.1103/PhysRevD.95.024038} {\bibfield  {journal} {\bibinfo
  {journal} {Phys. Rev. D}\ }\textbf {\bibinfo {volume} {95}},\ \bibinfo
  {pages} {024038} (\bibinfo {year} {2017})},\ \Eprint
  {https://arxiv.org/abs/1609.05933} {arXiv:1609.05933 [gr-qc]} \BibitemShut
  {NoStop}%
\bibitem [{\citenamefont {{Damour}}\ \emph {et~al.}(2004)\citenamefont
  {{Damour}}, \citenamefont {{Gopakumar}},\ and\ \citenamefont
  {{Iyer}}}]{Damour2004}%
  \BibitemOpen
  \bibfield  {author} {\bibinfo {author} {\bibfnamefont {T.}~\bibnamefont
  {{Damour}}}, \bibinfo {author} {\bibfnamefont {A.}~\bibnamefont
  {{Gopakumar}}},\ and\ \bibinfo {author} {\bibfnamefont {B.~R.}\ \bibnamefont
  {{Iyer}}},\ }\bibfield  {title} {\bibinfo {title} {{Phasing of gravitational
  waves from inspiralling eccentric binaries}},\ }\href
  {https://doi.org/10.1103/PhysRevD.70.064028} {\bibfield  {journal} {\bibinfo
  {journal} {\prd}\ }\textbf {\bibinfo {volume} {70}},\ \bibinfo {eid} {064028}
  (\bibinfo {year} {2004})},\ \Eprint {https://arxiv.org/abs/gr-qc/0404128}
  {gr-qc/0404128} \BibitemShut {NoStop}%
\bibitem [{\citenamefont {{K{\"o}nigsd{\"o}rffer}}\ and\ \citenamefont
  {{Gopakumar}}(2006)}]{KG06}%
  \BibitemOpen
  \bibfield  {author} {\bibinfo {author} {\bibfnamefont {C.}~\bibnamefont
  {{K{\"o}nigsd{\"o}rffer}}}\ and\ \bibinfo {author} {\bibfnamefont
  {A.}~\bibnamefont {{Gopakumar}}},\ }\bibfield  {title} {\bibinfo {title}
  {{Phasing of gravitational waves from inspiralling eccentric binaries at the
  third-and-a-half post-Newtonian order}},\ }\href
  {https://doi.org/10.1103/PhysRevD.73.124012} {\bibfield  {journal} {\bibinfo
  {journal} {\prd}\ }\textbf {\bibinfo {volume} {73}},\ \bibinfo {eid} {124012}
  (\bibinfo {year} {2006})},\ \Eprint {https://arxiv.org/abs/gr-qc/0603056}
  {gr-qc/0603056} \BibitemShut {NoStop}%
\bibitem [{\citenamefont {{Sch{\"a}fer}}\ and\ \citenamefont
  {{Wex}}(1993)}]{SW93}%
  \BibitemOpen
  \bibfield  {author} {\bibinfo {author} {\bibfnamefont {G.}~\bibnamefont
  {{Sch{\"a}fer}}}\ and\ \bibinfo {author} {\bibfnamefont {N.}~\bibnamefont
  {{Wex}}},\ }\bibfield  {title} {\bibinfo {title} {{Second post-Newtonian
  motion of compact binaries}},\ }\href
  {https://doi.org/10.1016/0375-9601(93)90758-R} {\bibfield  {journal}
  {\bibinfo  {journal} {Physics Letters A}\ }\textbf {\bibinfo {volume}
  {174}},\ \bibinfo {pages} {196} (\bibinfo {year} {1993})}\BibitemShut
  {NoStop}%
\bibitem [{\citenamefont {{Memmesheimer}}\ \emph {et~al.}(2004)\citenamefont
  {{Memmesheimer}}, \citenamefont {{Gopakumar}},\ and\ \citenamefont
  {{Sch{\"a}fer}}}]{MGS}%
  \BibitemOpen
  \bibfield  {author} {\bibinfo {author} {\bibfnamefont {R.-M.}\ \bibnamefont
  {{Memmesheimer}}}, \bibinfo {author} {\bibfnamefont {A.}~\bibnamefont
  {{Gopakumar}}},\ and\ \bibinfo {author} {\bibfnamefont {G.}~\bibnamefont
  {{Sch{\"a}fer}}},\ }\bibfield  {title} {\bibinfo {title} {{Third
  post-Newtonian accurate generalized quasi-Keplerian parametrization for
  compact binaries in eccentric orbits}},\ }\href
  {https://doi.org/10.1103/PhysRevD.70.104011} {\bibfield  {journal} {\bibinfo
  {journal} {\prd}\ }\textbf {\bibinfo {volume} {70}},\ \bibinfo {eid} {104011}
  (\bibinfo {year} {2004})},\ \Eprint {https://arxiv.org/abs/gr-qc/0407049}
  {gr-qc/0407049} \BibitemShut {NoStop}%
\bibitem [{\citenamefont {{Sch{\"a}fer}}\ and\ \citenamefont
  {{Jaranowski}}(2018)}]{JS_LR}%
  \BibitemOpen
  \bibfield  {author} {\bibinfo {author} {\bibfnamefont {G.}~\bibnamefont
  {{Sch{\"a}fer}}}\ and\ \bibinfo {author} {\bibfnamefont {P.}~\bibnamefont
  {{Jaranowski}}},\ }\bibfield  {title} {\bibinfo {title} {{Hamiltonian
  formulation of general relativity and post-Newtonian dynamics of compact
  binaries}},\ }\href {https://doi.org/10.1007/s41114-018-0016-5} {\bibfield
  {journal} {\bibinfo  {journal} {Living Reviews in Relativity}\ }\textbf
  {\bibinfo {volume} {21}},\ \bibinfo {eid} {7} (\bibinfo {year} {2018})},\
  \Eprint {https://arxiv.org/abs/1805.07240} {arXiv:1805.07240 [gr-qc]}
  \BibitemShut {NoStop}%
\bibitem [{\citenamefont {{Blanchet}}(2014)}]{LB_LR}%
  \BibitemOpen
  \bibfield  {author} {\bibinfo {author} {\bibfnamefont {L.}~\bibnamefont
  {{Blanchet}}},\ }\bibfield  {title} {\bibinfo {title} {{Gravitational
  Radiation from Post-Newtonian Sources and Inspiralling Compact Binaries}},\
  }\href {https://doi.org/10.12942/lrr-2014-2} {\bibfield  {journal} {\bibinfo
  {journal} {Living Reviews in Relativity}\ }\textbf {\bibinfo {volume} {17}},\
  \bibinfo {eid} {2} (\bibinfo {year} {2014})},\ \Eprint
  {https://arxiv.org/abs/1310.1528} {arXiv:1310.1528 [gr-qc]} \BibitemShut
  {NoStop}%
\bibitem [{\citenamefont {{Blanchet}}\ and\ \citenamefont
  {{Schaefer}}(1989{\natexlab{a}})}]{1989MNRAS.239..845B}%
  \BibitemOpen
  \bibfield  {author} {\bibinfo {author} {\bibfnamefont {L.}~\bibnamefont
  {{Blanchet}}}\ and\ \bibinfo {author} {\bibfnamefont {G.}~\bibnamefont
  {{Schaefer}}},\ }\bibfield  {title} {\bibinfo {title} {{Higher order
  gravitational radiation losses in binary systems.}},\ }\href
  {https://doi.org/10.1093/mnras/239.3.845} {\bibfield  {journal} {\bibinfo
  {journal} {\mnras}\ }\textbf {\bibinfo {volume} {239}},\ \bibinfo {pages}
  {845} (\bibinfo {year} {1989}{\natexlab{a}})}\BibitemShut {NoStop}%
\bibitem [{\citenamefont {{Junker}}\ and\ \citenamefont
  {{Schaefer}}(1992)}]{JS92}%
  \BibitemOpen
  \bibfield  {author} {\bibinfo {author} {\bibfnamefont {W.}~\bibnamefont
  {{Junker}}}\ and\ \bibinfo {author} {\bibfnamefont {G.}~\bibnamefont
  {{Schaefer}}},\ }\bibfield  {title} {\bibinfo {title} {{Binary systems -
  Higher order gravitational radiation damping and wave emission}},\ }\href
  {https://doi.org/10.1093/mnras/254.1.146} {\bibfield  {journal} {\bibinfo
  {journal} {\mnras}\ }\textbf {\bibinfo {volume} {254}},\ \bibinfo {pages}
  {146} (\bibinfo {year} {1992})}\BibitemShut {NoStop}%
\bibitem [{\citenamefont {{Gopakumar}}\ and\ \citenamefont
  {{Iyer}}(1997{\natexlab{a}})}]{1997PhRvD..56.7708G}%
  \BibitemOpen
  \bibfield  {author} {\bibinfo {author} {\bibfnamefont {A.}~\bibnamefont
  {{Gopakumar}}}\ and\ \bibinfo {author} {\bibfnamefont {B.~R.}\ \bibnamefont
  {{Iyer}}},\ }\bibfield  {title} {\bibinfo {title} {{Gravitational waves from
  inspiraling compact binaries: Angular momentum flux, evolution of the orbital
  elements, and the waveform to the second post-Newtonian order}},\ }\href
  {https://doi.org/10.1103/PhysRevD.56.7708} {\bibfield  {journal} {\bibinfo
  {journal} {\prd}\ }\textbf {\bibinfo {volume} {56}},\ \bibinfo {pages} {7708}
  (\bibinfo {year} {1997}{\natexlab{a}})},\ \Eprint
  {https://arxiv.org/abs/gr-qc/9710075} {arXiv:gr-qc/9710075 [gr-qc]}
  \BibitemShut {NoStop}%
\bibitem [{\citenamefont {{Arun}}\ \emph {et~al.}(2009)\citenamefont {{Arun}},
  \citenamefont {{Blanchet}}, \citenamefont {{Iyer}},\ and\ \citenamefont
  {{Sinha}}}]{Arun2009}%
  \BibitemOpen
  \bibfield  {author} {\bibinfo {author} {\bibfnamefont {K.~G.}\ \bibnamefont
  {{Arun}}}, \bibinfo {author} {\bibfnamefont {L.}~\bibnamefont {{Blanchet}}},
  \bibinfo {author} {\bibfnamefont {B.~R.}\ \bibnamefont {{Iyer}}},\ and\
  \bibinfo {author} {\bibfnamefont {S.}~\bibnamefont {{Sinha}}},\ }\bibfield
  {title} {\bibinfo {title} {{Third post-Newtonian angular momentum flux and
  the secular evolution of orbital elements for inspiralling compact binaries
  in quasi-elliptical orbits}},\ }\href
  {https://doi.org/10.1103/PhysRevD.80.124018} {\bibfield  {journal} {\bibinfo
  {journal} {\prd}\ }\textbf {\bibinfo {volume} {80}},\ \bibinfo {eid} {124018}
  (\bibinfo {year} {2009})},\ \Eprint {https://arxiv.org/abs/0908.3854}
  {arXiv:0908.3854 [gr-qc]} \BibitemShut {NoStop}%
\bibitem [{\citenamefont {{Jaranowski}}\ and\ \citenamefont
  {{Sch{\"a}fer}}(2015)}]{JS_15}%
  \BibitemOpen
  \bibfield  {author} {\bibinfo {author} {\bibfnamefont {P.}~\bibnamefont
  {{Jaranowski}}}\ and\ \bibinfo {author} {\bibfnamefont {G.}~\bibnamefont
  {{Sch{\"a}fer}}},\ }\bibfield  {title} {\bibinfo {title} {{Derivation of
  local-in-time fourth post-Newtonian ADM Hamiltonian for spinless compact
  binaries}},\ }\href {https://doi.org/10.1103/PhysRevD.92.124043} {\bibfield
  {journal} {\bibinfo  {journal} {\prd}\ }\textbf {\bibinfo {volume} {92}},\
  \bibinfo {eid} {124043} (\bibinfo {year} {2015})},\ \Eprint
  {https://arxiv.org/abs/1508.01016} {arXiv:1508.01016 [gr-qc]} \BibitemShut
  {NoStop}%
\bibitem [{\citenamefont {{Damour}}\ \emph {et~al.}(2014)\citenamefont
  {{Damour}}, \citenamefont {{Jaranowski}},\ and\ \citenamefont
  {{Sch{\"a}fer}}}]{DJS_14}%
  \BibitemOpen
  \bibfield  {author} {\bibinfo {author} {\bibfnamefont {T.}~\bibnamefont
  {{Damour}}}, \bibinfo {author} {\bibfnamefont {P.}~\bibnamefont
  {{Jaranowski}}},\ and\ \bibinfo {author} {\bibfnamefont {G.}~\bibnamefont
  {{Sch{\"a}fer}}},\ }\bibfield  {title} {\bibinfo {title} {{Nonlocal-in-time
  action for the fourth post-Newtonian conservative dynamics of two-body
  systems}},\ }\href {https://doi.org/10.1103/PhysRevD.89.064058} {\bibfield
  {journal} {\bibinfo  {journal} {\prd}\ }\textbf {\bibinfo {volume} {89}},\
  \bibinfo {eid} {064058} (\bibinfo {year} {2014})},\ \Eprint
  {https://arxiv.org/abs/1401.4548} {arXiv:1401.4548 [gr-qc]} \BibitemShut
  {NoStop}%
\bibitem [{\citenamefont {{Bini}}\ \emph
  {et~al.}(2020{\natexlab{a}})\citenamefont {{Bini}}, \citenamefont
  {{Damour}},\ and\ \citenamefont {{Geralico}}}]{Bini2020}%
  \BibitemOpen
  \bibfield  {author} {\bibinfo {author} {\bibfnamefont {D.}~\bibnamefont
  {{Bini}}}, \bibinfo {author} {\bibfnamefont {T.}~\bibnamefont {{Damour}}},\
  and\ \bibinfo {author} {\bibfnamefont {A.}~\bibnamefont {{Geralico}}},\
  }\bibfield  {title} {\bibinfo {title} {{Sixth post-Newtonian nonlocal-in-time
  dynamics of binary systems}},\ }\href
  {https://doi.org/10.1103/PhysRevD.102.084047} {\bibfield  {journal} {\bibinfo
   {journal} {\prd}\ }\textbf {\bibinfo {volume} {102}},\ \bibinfo {eid}
  {084047} (\bibinfo {year} {2020}{\natexlab{a}})},\ \Eprint
  {https://arxiv.org/abs/2007.11239} {arXiv:2007.11239 [gr-qc]} \BibitemShut
  {NoStop}%
\bibitem [{\citenamefont {{Bl{\"u}mlein}}\ \emph {et~al.}(2020)\citenamefont
  {{Bl{\"u}mlein}}, \citenamefont {{Maier}}, \citenamefont {{Marquard}},\ and\
  \citenamefont {{Sch{\"a}fer}}}]{BMMS}%
  \BibitemOpen
  \bibfield  {author} {\bibinfo {author} {\bibfnamefont {J.}~\bibnamefont
  {{Bl{\"u}mlein}}}, \bibinfo {author} {\bibfnamefont {A.}~\bibnamefont
  {{Maier}}}, \bibinfo {author} {\bibfnamefont {P.}~\bibnamefont
  {{Marquard}}},\ and\ \bibinfo {author} {\bibfnamefont {G.}~\bibnamefont
  {{Sch{\"a}fer}}},\ }\bibfield  {title} {\bibinfo {title} {{Fourth
  post-Newtonian Hamiltonian dynamics of two-body systems from an effective
  field theory approach}},\ }\href
  {https://doi.org/10.1016/j.nuclphysb.2020.115041} {\bibfield  {journal}
  {\bibinfo  {journal} {Nuclear Physics B}\ }\textbf {\bibinfo {volume}
  {955}},\ \bibinfo {eid} {115041} (\bibinfo {year} {2020})},\ \Eprint
  {https://arxiv.org/abs/2003.01692} {arXiv:2003.01692 [gr-qc]} \BibitemShut
  {NoStop}%
\bibitem [{MMA(2017)}]{MMA1}%
  \BibitemOpen
  \href@noop {} {}\bibinfo {howpublished}
  {\url{https://github.com/ianhinder/EccentricIMR}} (\bibinfo {year}
  {2017})\BibitemShut {NoStop}%
\bibitem [{MMA(2021)}]{MMA2}%
  \BibitemOpen
  \href@noop {} {}\bibinfo {howpublished}
  {\url{https://github.com/sashwattanay/EccentricIMR}} (\bibinfo {year}
  {2021})\BibitemShut {NoStop}%
\bibitem [{MMA()}]{MMA3}%
  \BibitemOpen
  \href@noop {} {}\bibinfo {howpublished} {See Supplemental Material at [URL
  will be inserted by publisher] for the \textsc{Mathematica} packages,
  notebook and pdf files that produce the gravitational waveforms (IMR and
  inspiral) and contain various lengthy expressions.}\BibitemShut {Stop}%
\bibitem [{\citenamefont {{Damour}}\ and\ \citenamefont
  {{Deruelle}}(1985)}]{DD85}%
  \BibitemOpen
  \bibfield  {author} {\bibinfo {author} {\bibfnamefont {T.}~\bibnamefont
  {{Damour}}}\ and\ \bibinfo {author} {\bibfnamefont {N.}~\bibnamefont
  {{Deruelle}}},\ }\bibfield  {title} {\bibinfo {title} {{General relativistic
  celestial mechanics of binary systems. I. The post-Newtonian motion.}},\
  }\href@noop {} {\bibfield  {journal} {\bibinfo  {journal} {Ann.~Inst.~Henri
  Poincar{\'e} Phys.~Th{\'e}or., Vol.~43, No.~1, p.~107 - 132}\ }\textbf
  {\bibinfo {volume} {43}},\ \bibinfo {pages} {107} (\bibinfo {year}
  {1985})}\BibitemShut {NoStop}%
\bibitem [{\citenamefont {{Damour}}\ and\ \citenamefont
  {{Schafer}}(1988)}]{DS88}%
  \BibitemOpen
  \bibfield  {author} {\bibinfo {author} {\bibfnamefont {T.}~\bibnamefont
  {{Damour}}}\ and\ \bibinfo {author} {\bibfnamefont {G.}~\bibnamefont
  {{Schafer}}},\ }\bibfield  {title} {\bibinfo {title} {{Higher-order
  relativistic periastron advances and binary pulsars.}},\ }\href
  {https://doi.org/10.1007/BF02828697} {\bibfield  {journal} {\bibinfo
  {journal} {Nuovo Cimento B Serie}\ }\textbf {\bibinfo {volume} {101}},\
  \bibinfo {pages} {127} (\bibinfo {year} {1988})}\BibitemShut {NoStop}%
\bibitem [{\citenamefont {{Klioner}}(2016)}]{CM_SK}%
  \BibitemOpen
  \bibfield  {author} {\bibinfo {author} {\bibfnamefont {S.~A.}\ \bibnamefont
  {{Klioner}}},\ }\bibfield  {title} {\bibinfo {title} {{Basic Celestial
  Mechanics}},\ }\href@noop {} {\bibfield  {journal} {\bibinfo  {journal}
  {arXiv e-prints}\ ,\ \bibinfo {eid} {arXiv:1609.00915}} (\bibinfo {year}
  {2016})},\ \Eprint {https://arxiv.org/abs/1609.00915} {arXiv:1609.00915
  [astro-ph.IM]} \BibitemShut {NoStop}%
\bibitem [{\citenamefont {Colwell}(1993)}]{Colwell92}%
  \BibitemOpen
  \bibfield  {author} {\bibinfo {author} {\bibfnamefont {P.}~\bibnamefont
  {Colwell}},\ }\href {https://books.google.com/books?id=ynjvAAAAMAAJ} {\emph
  {\bibinfo {title} {Solving Kepler's Equation Over Three Centuries}}}\
  (\bibinfo  {publisher} {Willmann-Bell},\ \bibinfo {year} {1993})\BibitemShut
  {NoStop}%
\bibitem [{\citenamefont {{Boetzel}}\ \emph {et~al.}(2017)\citenamefont
  {{Boetzel}}, \citenamefont {{Susobhanan}}, \citenamefont {{Gopakumar}},
  \citenamefont {{Klein}},\ and\ \citenamefont {{Jetzer}}}]{Boetzel_17}%
  \BibitemOpen
  \bibfield  {author} {\bibinfo {author} {\bibfnamefont {Y.}~\bibnamefont
  {{Boetzel}}}, \bibinfo {author} {\bibfnamefont {A.}~\bibnamefont
  {{Susobhanan}}}, \bibinfo {author} {\bibfnamefont {A.}~\bibnamefont
  {{Gopakumar}}}, \bibinfo {author} {\bibfnamefont {A.}~\bibnamefont
  {{Klein}}},\ and\ \bibinfo {author} {\bibfnamefont {P.}~\bibnamefont
  {{Jetzer}}},\ }\bibfield  {title} {\bibinfo {title} {{Solving post-Newtonian
  accurate Kepler equation}},\ }\href
  {https://doi.org/10.1103/PhysRevD.96.044011} {\bibfield  {journal} {\bibinfo
  {journal} {\prd}\ }\textbf {\bibinfo {volume} {96}},\ \bibinfo {eid} {044011}
  (\bibinfo {year} {2017})},\ \Eprint {https://arxiv.org/abs/1707.02088}
  {arXiv:1707.02088 [gr-qc]} \BibitemShut {NoStop}%
\bibitem [{\citenamefont {{Cho}}\ \emph {et~al.}(2018)\citenamefont {{Cho}},
  \citenamefont {{Gopakumar}}, \citenamefont {{Haney}},\ and\ \citenamefont
  {{Lee}}}]{GG_18}%
  \BibitemOpen
  \bibfield  {author} {\bibinfo {author} {\bibfnamefont {G.}~\bibnamefont
  {{Cho}}}, \bibinfo {author} {\bibfnamefont {A.}~\bibnamefont {{Gopakumar}}},
  \bibinfo {author} {\bibfnamefont {M.}~\bibnamefont {{Haney}}},\ and\ \bibinfo
  {author} {\bibfnamefont {H.~M.}\ \bibnamefont {{Lee}}},\ }\bibfield  {title}
  {\bibinfo {title} {{Gravitational waves from compact binaries in
  post-Newtonian accurate hyperbolic orbits}},\ }\href
  {https://doi.org/10.1103/PhysRevD.98.024039} {\bibfield  {journal} {\bibinfo
  {journal} {\prd}\ }\textbf {\bibinfo {volume} {98}},\ \bibinfo {eid} {024039}
  (\bibinfo {year} {2018})},\ \Eprint {https://arxiv.org/abs/1807.02380}
  {arXiv:1807.02380 [gr-qc]} \BibitemShut {NoStop}%
\bibitem [{\citenamefont {{Damour}}\ and\ \citenamefont
  {{Deruelle}}(1986)}]{DD_86}%
  \BibitemOpen
  \bibfield  {author} {\bibinfo {author} {\bibfnamefont {T.}~\bibnamefont
  {{Damour}}}\ and\ \bibinfo {author} {\bibfnamefont {N.}~\bibnamefont
  {{Deruelle}}},\ }\bibfield  {title} {\bibinfo {title} {{General relativistic
  celestial mechanics of binary systems. II. The post-Newtonian timing
  formula.}},\ }\href@noop {} {\bibfield  {journal} {\bibinfo  {journal} {Ann.
  Inst. Henri Poincar{\'e} Phys. Th{\'e}or}\ }\textbf {\bibinfo {volume}
  {44}},\ \bibinfo {pages} {263} (\bibinfo {year} {1986})}\BibitemShut
  {NoStop}%
\bibitem [{\citenamefont {{Damour}}\ and\ \citenamefont
  {{Taylor}}(1992)}]{DT92}%
  \BibitemOpen
  \bibfield  {author} {\bibinfo {author} {\bibfnamefont {T.}~\bibnamefont
  {{Damour}}}\ and\ \bibinfo {author} {\bibfnamefont {J.~H.}\ \bibnamefont
  {{Taylor}}},\ }\bibfield  {title} {\bibinfo {title} {{Strong-field tests of
  relativistic gravity and binary pulsars}},\ }\href
  {https://doi.org/10.1103/PhysRevD.45.1840} {\bibfield  {journal} {\bibinfo
  {journal} {\prd}\ }\textbf {\bibinfo {volume} {45}},\ \bibinfo {pages} {1840}
  (\bibinfo {year} {1992})}\BibitemShut {NoStop}%
\bibitem [{\citenamefont {{Susobhanan}}\ \emph {et~al.}(2020)\citenamefont
  {{Susobhanan}}, \citenamefont {{Gopakumar}}, \citenamefont {{Hobbs}},\ and\
  \citenamefont {{Taylor}}}]{SGHT_20}%
  \BibitemOpen
  \bibfield  {author} {\bibinfo {author} {\bibfnamefont {A.}~\bibnamefont
  {{Susobhanan}}}, \bibinfo {author} {\bibfnamefont {A.}~\bibnamefont
  {{Gopakumar}}}, \bibinfo {author} {\bibfnamefont {G.}~\bibnamefont
  {{Hobbs}}},\ and\ \bibinfo {author} {\bibfnamefont {S.~R.}\ \bibnamefont
  {{Taylor}}},\ }\bibfield  {title} {\bibinfo {title} {{Pulsar timing array
  signals induced by black hole binaries in relativistic eccentric orbits}},\
  }\href {https://doi.org/10.1103/PhysRevD.101.043022} {\bibfield  {journal}
  {\bibinfo  {journal} {\prd}\ }\textbf {\bibinfo {volume} {101}},\ \bibinfo
  {eid} {043022} (\bibinfo {year} {2020})},\ \Eprint
  {https://arxiv.org/abs/2002.03285} {arXiv:2002.03285 [gr-qc]} \BibitemShut
  {NoStop}%
\bibitem [{\citenamefont {{Damour}}\ \emph {et~al.}(2015)\citenamefont
  {{Damour}}, \citenamefont {{Jaranowski}},\ and\ \citenamefont
  {{Sch{\"a}fer}}}]{DJS_15}%
  \BibitemOpen
  \bibfield  {author} {\bibinfo {author} {\bibfnamefont {T.}~\bibnamefont
  {{Damour}}}, \bibinfo {author} {\bibfnamefont {P.}~\bibnamefont
  {{Jaranowski}}},\ and\ \bibinfo {author} {\bibfnamefont {G.}~\bibnamefont
  {{Sch{\"a}fer}}},\ }\bibfield  {title} {\bibinfo {title} {{Fourth
  post-Newtonian effective one-body dynamics}},\ }\href
  {https://doi.org/10.1103/PhysRevD.91.084024} {\bibfield  {journal} {\bibinfo
  {journal} {\prd}\ }\textbf {\bibinfo {volume} {91}},\ \bibinfo {eid} {084024}
  (\bibinfo {year} {2015})},\ \Eprint {https://arxiv.org/abs/1502.07245}
  {arXiv:1502.07245 [gr-qc]} \BibitemShut {NoStop}%
\bibitem [{\citenamefont {{Galley}}\ \emph {et~al.}(2016)\citenamefont
  {{Galley}}, \citenamefont {{Leibovich}}, \citenamefont {{Porto}},\ and\
  \citenamefont {{Ross}}}]{Galley_15}%
  \BibitemOpen
  \bibfield  {author} {\bibinfo {author} {\bibfnamefont {C.~R.}\ \bibnamefont
  {{Galley}}}, \bibinfo {author} {\bibfnamefont {A.~K.}\ \bibnamefont
  {{Leibovich}}}, \bibinfo {author} {\bibfnamefont {R.~A.}\ \bibnamefont
  {{Porto}}},\ and\ \bibinfo {author} {\bibfnamefont {A.}~\bibnamefont
  {{Ross}}},\ }\bibfield  {title} {\bibinfo {title} {{Tail effect in
  gravitational radiation reaction: Time nonlocality and renormalization group
  evolution}},\ }\href {https://doi.org/10.1103/PhysRevD.93.124010} {\bibfield
  {journal} {\bibinfo  {journal} {\prd}\ }\textbf {\bibinfo {volume} {93}},\
  \bibinfo {eid} {124010} (\bibinfo {year} {2016})},\ \Eprint
  {https://arxiv.org/abs/1511.07379} {arXiv:1511.07379 [gr-qc]} \BibitemShut
  {NoStop}%
\bibitem [{\citenamefont {{Damour}}\ \emph {et~al.}(2001)\citenamefont
  {{Damour}}, \citenamefont {{Jaranowski}},\ and\ \citenamefont
  {{Sch{\"a}fer}}}]{DJS_01}%
  \BibitemOpen
  \bibfield  {author} {\bibinfo {author} {\bibfnamefont {T.}~\bibnamefont
  {{Damour}}}, \bibinfo {author} {\bibfnamefont {P.}~\bibnamefont
  {{Jaranowski}}},\ and\ \bibinfo {author} {\bibfnamefont {G.}~\bibnamefont
  {{Sch{\"a}fer}}},\ }\bibfield  {title} {\bibinfo {title} {{Dimensional
  regularization of the gravitational interaction of point masses}},\ }\href
  {https://doi.org/10.1016/S0370-2693(01)00642-6} {\bibfield  {journal}
  {\bibinfo  {journal} {Physics Letters B}\ }\textbf {\bibinfo {volume}
  {513}},\ \bibinfo {pages} {147} (\bibinfo {year} {2001})},\ \Eprint
  {https://arxiv.org/abs/gr-qc/0105038} {arXiv:gr-qc/0105038 [gr-qc]}
  \BibitemShut {NoStop}%
\bibitem [{\citenamefont {{Jaranowski}}\ and\ \citenamefont
  {{Sch{\"a}fer}}(2012)}]{JS_12}%
  \BibitemOpen
  \bibfield  {author} {\bibinfo {author} {\bibfnamefont {P.}~\bibnamefont
  {{Jaranowski}}}\ and\ \bibinfo {author} {\bibfnamefont {G.}~\bibnamefont
  {{Sch{\"a}fer}}},\ }\bibfield  {title} {\bibinfo {title} {{Towards the fourth
  post-Newtonian Hamiltonian for two-point-mass systems}},\ }\href
  {https://doi.org/10.1103/PhysRevD.86.061503} {\bibfield  {journal} {\bibinfo
  {journal} {\prd}\ }\textbf {\bibinfo {volume} {86}},\ \bibinfo {eid} {061503}
  (\bibinfo {year} {2012})},\ \Eprint {https://arxiv.org/abs/1207.5448}
  {arXiv:1207.5448 [gr-qc]} \BibitemShut {NoStop}%
\bibitem [{\citenamefont {{Jaranowski}}\ and\ \citenamefont
  {{Sch{\"a}fer}}(2013)}]{JS_13}%
  \BibitemOpen
  \bibfield  {author} {\bibinfo {author} {\bibfnamefont {P.}~\bibnamefont
  {{Jaranowski}}}\ and\ \bibinfo {author} {\bibfnamefont {G.}~\bibnamefont
  {{Sch{\"a}fer}}},\ }\bibfield  {title} {\bibinfo {title} {{Dimensional
  regularization of local singularities in the fourth post-Newtonian
  two-point-mass Hamiltonian}},\ }\href
  {https://doi.org/10.1103/PhysRevD.87.081503} {\bibfield  {journal} {\bibinfo
  {journal} {\prd}\ }\textbf {\bibinfo {volume} {87}},\ \bibinfo {eid} {081503}
  (\bibinfo {year} {2013})},\ \Eprint {https://arxiv.org/abs/1303.3225}
  {arXiv:1303.3225 [gr-qc]} \BibitemShut {NoStop}%
\bibitem [{\citenamefont {Foffa}\ \emph {et~al.}(2019)\citenamefont {Foffa},
  \citenamefont {Porto}, \citenamefont {Rothstein},\ and\ \citenamefont
  {Sturani}}]{Foffa2019}%
  \BibitemOpen
  \bibfield  {author} {\bibinfo {author} {\bibfnamefont {S.}~\bibnamefont
  {Foffa}}, \bibinfo {author} {\bibfnamefont {R.~A.}\ \bibnamefont {Porto}},
  \bibinfo {author} {\bibfnamefont {I.}~\bibnamefont {Rothstein}},\ and\
  \bibinfo {author} {\bibfnamefont {R.}~\bibnamefont {Sturani}},\ }\bibfield
  {title} {\bibinfo {title} {{Conservative dynamics of binary systems to fourth
  post-Newtonian order in the EFT approach. II. Renormalized Lagrangian}},\
  }\href {https://doi.org/10.1103/physrevd.100.024048} {\bibfield  {journal}
  {\bibinfo  {journal} {Phys. Rev. D}\ }\textbf {\bibinfo {volume} {100}},\
  \bibinfo {pages} {024048} (\bibinfo {year} {2019})},\ \Eprint
  {https://arxiv.org/abs/1903.05118} {arXiv:1903.05118} \BibitemShut {NoStop}%
\bibitem [{\citenamefont {Foffa}\ and\ \citenamefont
  {Sturani}(2019)}]{Foffa2019a}%
  \BibitemOpen
  \bibfield  {author} {\bibinfo {author} {\bibfnamefont {S.}~\bibnamefont
  {Foffa}}\ and\ \bibinfo {author} {\bibfnamefont {R.}~\bibnamefont
  {Sturani}},\ }\bibfield  {title} {\bibinfo {title} {{Conservative dynamics of
  binary systems to fourth post-Newtonian order in the EFT approach. I.
  Regularized Lagrangian}},\ }\href
  {https://doi.org/10.1103/physrevd.100.024047} {\bibfield  {journal} {\bibinfo
   {journal} {Phys. Rev. D}\ }\textbf {\bibinfo {volume} {100}},\ \bibinfo
  {pages} {024047} (\bibinfo {year} {2019})},\ \Eprint
  {https://arxiv.org/abs/1903.05113} {arXiv:1903.05113} \BibitemShut {NoStop}%
\bibitem [{\citenamefont {{Marchand}}\ \emph {et~al.}(2018)\citenamefont
  {{Marchand}}, \citenamefont {{Bernard}}, \citenamefont {{Blanchet}},\ and\
  \citenamefont {{Faye}}}]{Marchand18}%
  \BibitemOpen
  \bibfield  {author} {\bibinfo {author} {\bibfnamefont {T.}~\bibnamefont
  {{Marchand}}}, \bibinfo {author} {\bibfnamefont {L.}~\bibnamefont
  {{Bernard}}}, \bibinfo {author} {\bibfnamefont {L.}~\bibnamefont
  {{Blanchet}}},\ and\ \bibinfo {author} {\bibfnamefont {G.}~\bibnamefont
  {{Faye}}},\ }\bibfield  {title} {\bibinfo {title} {{Ambiguity-free completion
  of the equations of motion of compact binary systems at the fourth
  post-Newtonian order}},\ }\href {https://doi.org/10.1103/PhysRevD.97.044023}
  {\bibfield  {journal} {\bibinfo  {journal} {\prd}\ }\textbf {\bibinfo
  {volume} {97}},\ \bibinfo {eid} {044023} (\bibinfo {year} {2018})},\ \Eprint
  {https://arxiv.org/abs/1707.09289} {arXiv:1707.09289 [gr-qc]} \BibitemShut
  {NoStop}%
\bibitem [{\citenamefont {Jos{\'e}}\ and\ \citenamefont
  {Saletan}(1998)}]{jose1998classical}%
  \BibitemOpen
  \bibfield  {author} {\bibinfo {author} {\bibfnamefont {J.}~\bibnamefont
  {Jos{\'e}}}\ and\ \bibinfo {author} {\bibfnamefont {E.}~\bibnamefont
  {Saletan}},\ }\href {https://books.google.com/books?id=Eql9dRQDgvQC} {\emph
  {\bibinfo {title} {Classical Dynamics: A Contemporary Approach}}}\ (\bibinfo
  {publisher} {Cambridge University Press},\ \bibinfo {year}
  {1998})\BibitemShut {NoStop}%
\bibitem [{\citenamefont {Goldstein}\ \emph {et~al.}(2002)\citenamefont
  {Goldstein}, \citenamefont {Poole},\ and\ \citenamefont
  {Safko}}]{goldstein2002classical}%
  \BibitemOpen
  \bibfield  {author} {\bibinfo {author} {\bibfnamefont {H.}~\bibnamefont
  {Goldstein}}, \bibinfo {author} {\bibfnamefont {C.}~\bibnamefont {Poole}},\
  and\ \bibinfo {author} {\bibfnamefont {J.}~\bibnamefont {Safko}},\ }\href
  {https://books.google.com/books?id=tJCuQgAACAAJ} {\emph {\bibinfo {title}
  {Classical Mechanics}}}\ (\bibinfo  {publisher} {Addison Wesley},\ \bibinfo
  {year} {2002})\BibitemShut {NoStop}%
\bibitem [{\citenamefont {Ferraz-Mello}(2007)}]{ferraz2007canonical}%
  \BibitemOpen
  \bibfield  {author} {\bibinfo {author} {\bibfnamefont {S.}~\bibnamefont
  {Ferraz-Mello}},\ }\href {https://books.google.com/books?id=TYVCAAAAQBAJ}
  {\emph {\bibinfo {title} {Canonical Perturbation Theories: Degenerate Systems
  and Resonance}}},\ Astrophysics and Space Science Library\ (\bibinfo
  {publisher} {Springer New York},\ \bibinfo {year} {2007})\BibitemShut
  {NoStop}%
\bibitem [{\citenamefont {{Arun}}\ \emph {et~al.}(2008)\citenamefont {{Arun}},
  \citenamefont {{Blanchet}}, \citenamefont {{Iyer}},\ and\ \citenamefont
  {{Qusailah}}}]{2008PhRvD..77f4034A}%
  \BibitemOpen
  \bibfield  {author} {\bibinfo {author} {\bibfnamefont {K.~G.}\ \bibnamefont
  {{Arun}}}, \bibinfo {author} {\bibfnamefont {L.}~\bibnamefont {{Blanchet}}},
  \bibinfo {author} {\bibfnamefont {B.~R.}\ \bibnamefont {{Iyer}}},\ and\
  \bibinfo {author} {\bibfnamefont {M.~S.~S.}\ \bibnamefont {{Qusailah}}},\
  }\bibfield  {title} {\bibinfo {title} {{Tail effects in the third
  post-Newtonian gravitational wave energy flux of compact binaries in
  quasi-elliptical orbits}},\ }\href
  {https://doi.org/10.1103/PhysRevD.77.064034} {\bibfield  {journal} {\bibinfo
  {journal} {\prd}\ }\textbf {\bibinfo {volume} {77}},\ \bibinfo {eid} {064034}
  (\bibinfo {year} {2008})},\ \Eprint {https://arxiv.org/abs/0711.0250}
  {arXiv:0711.0250 [gr-qc]} \BibitemShut {NoStop}%
\bibitem [{\citenamefont {Damour}\ and\ \citenamefont
  {Schaefer}(1991)}]{Damour:1990jh}%
  \BibitemOpen
  \bibfield  {author} {\bibinfo {author} {\bibfnamefont {T.}~\bibnamefont
  {Damour}}\ and\ \bibinfo {author} {\bibfnamefont {G.}~\bibnamefont
  {Schaefer}},\ }\bibfield  {title} {\bibinfo {title} {{Redefinition of
  position variables and the reduction of higher order Lagrangians}},\ }\href
  {https://doi.org/10.1063/1.529135} {\bibfield  {journal} {\bibinfo  {journal}
  {J. Math. Phys.}\ }\textbf {\bibinfo {volume} {32}},\ \bibinfo {pages} {127}
  (\bibinfo {year} {1991})}\BibitemShut {NoStop}%
\bibitem [{\citenamefont {Damour}\ \emph {et~al.}(2016)\citenamefont {Damour},
  \citenamefont {Jaranowski},\ and\ \citenamefont
  {Sch\"afer}}]{Damour:2016abl}%
  \BibitemOpen
  \bibfield  {author} {\bibinfo {author} {\bibfnamefont {T.}~\bibnamefont
  {Damour}}, \bibinfo {author} {\bibfnamefont {P.}~\bibnamefont {Jaranowski}},\
  and\ \bibinfo {author} {\bibfnamefont {G.}~\bibnamefont {Sch\"afer}},\
  }\bibfield  {title} {\bibinfo {title} {{Conservative dynamics of two-body
  systems at the fourth post-Newtonian approximation of general relativity}},\
  }\href {https://doi.org/10.1103/PhysRevD.93.084014} {\bibfield  {journal}
  {\bibinfo  {journal} {Phys. Rev. D}\ }\textbf {\bibinfo {volume} {93}},\
  \bibinfo {pages} {084014} (\bibinfo {year} {2016})},\ \Eprint
  {https://arxiv.org/abs/1601.01283} {arXiv:1601.01283 [gr-qc]} \BibitemShut
  {NoStop}%
\bibitem [{\citenamefont {Damour}\ \emph
  {et~al.}(2000{\natexlab{a}})\citenamefont {Damour}, \citenamefont
  {Jaranowski},\ and\ \citenamefont {Schaefer}}]{Damour:1999cr}%
  \BibitemOpen
  \bibfield  {author} {\bibinfo {author} {\bibfnamefont {T.}~\bibnamefont
  {Damour}}, \bibinfo {author} {\bibfnamefont {P.}~\bibnamefont {Jaranowski}},\
  and\ \bibinfo {author} {\bibfnamefont {G.}~\bibnamefont {Schaefer}},\
  }\bibfield  {title} {\bibinfo {title} {{Dynamical invariants for general
  relativistic two-body systems at the third postNewtonian approximation}},\
  }\href {https://doi.org/10.1103/PhysRevD.62.044024} {\bibfield  {journal}
  {\bibinfo  {journal} {Phys. Rev. D}\ }\textbf {\bibinfo {volume} {62}},\
  \bibinfo {pages} {044024} (\bibinfo {year} {2000}{\natexlab{a}})},\ \Eprint
  {https://arxiv.org/abs/gr-qc/9912092} {arXiv:gr-qc/9912092} \BibitemShut
  {NoStop}%
\bibitem [{\citenamefont {{Dlapa}}\ \emph {et~al.}(2021)\citenamefont
  {{Dlapa}}, \citenamefont {{K{\"a}lin}}, \citenamefont {{Liu}},\ and\
  \citenamefont {{Porto}}}]{2021arXiv210608276D}%
  \BibitemOpen
  \bibfield  {author} {\bibinfo {author} {\bibfnamefont {C.}~\bibnamefont
  {{Dlapa}}}, \bibinfo {author} {\bibfnamefont {G.}~\bibnamefont
  {{K{\"a}lin}}}, \bibinfo {author} {\bibfnamefont {Z.}~\bibnamefont {{Liu}}},\
  and\ \bibinfo {author} {\bibfnamefont {R.~A.}\ \bibnamefont {{Porto}}},\
  }\bibfield  {title} {\bibinfo {title} {{Dynamics of Binary Systems to Fourth
  Post-Minkowskian Order from the Effective Field Theory Approach}},\
  }\href@noop {} {\bibfield  {journal} {\bibinfo  {journal} {arXiv e-prints}\
  ,\ \bibinfo {eid} {arXiv:2106.08276}} (\bibinfo {year} {2021})},\ \Eprint
  {https://arxiv.org/abs/2106.08276} {arXiv:2106.08276 [hep-th]} \BibitemShut
  {NoStop}%
\bibitem [{\citenamefont {{Peters}}\ and\ \citenamefont
  {{Mathews}}(1963)}]{PM63}%
  \BibitemOpen
  \bibfield  {author} {\bibinfo {author} {\bibfnamefont {P.~C.}\ \bibnamefont
  {{Peters}}}\ and\ \bibinfo {author} {\bibfnamefont {J.}~\bibnamefont
  {{Mathews}}},\ }\bibfield  {title} {\bibinfo {title} {{Gravitational
  Radiation from Point Masses in a Keplerian Orbit}},\ }\href
  {https://doi.org/10.1103/PhysRev.131.435} {\bibfield  {journal} {\bibinfo
  {journal} {Physical Review}\ }\textbf {\bibinfo {volume} {131}},\ \bibinfo
  {pages} {435} (\bibinfo {year} {1963})}\BibitemShut {NoStop}%
\bibitem [{\citenamefont {Moulton}(1970)}]{moulton1970introduction}%
  \BibitemOpen
  \bibfield  {author} {\bibinfo {author} {\bibfnamefont {F.}~\bibnamefont
  {Moulton}},\ }\href {https://books.google.com/books?id=URPSrBntwdAC} {\emph
  {\bibinfo {title} {An Introduction to Celestial Mechanics}}},\ Dover books in
  astronomy\ (\bibinfo  {publisher} {Dover Publications},\ \bibinfo {year}
  {1970})\BibitemShut {NoStop}%
\bibitem [{\citenamefont {Finch}\ and\ \citenamefont
  {Rota}(2003)}]{finch2003mathematical}%
  \BibitemOpen
  \bibfield  {author} {\bibinfo {author} {\bibfnamefont {S.}~\bibnamefont
  {Finch}}\ and\ \bibinfo {author} {\bibfnamefont {G.}~\bibnamefont {Rota}},\
  }\href {https://books.google.com/books?id=Pl5I2ZSI6uAC} {\emph {\bibinfo
  {title} {Mathematical Constants}}},\ Encyclopedia of Mathematics and its
  Applications\ (\bibinfo  {publisher} {Cambridge University Press},\ \bibinfo
  {year} {2003})\BibitemShut {NoStop}%
\bibitem [{\citenamefont {{Hinder}}\ \emph {et~al.}(2010)\citenamefont
  {{Hinder}}, \citenamefont {{Herrmann}}, \citenamefont {{Laguna}},\ and\
  \citenamefont {{Shoemaker}}}]{Hinder_10}%
  \BibitemOpen
  \bibfield  {author} {\bibinfo {author} {\bibfnamefont {I.}~\bibnamefont
  {{Hinder}}}, \bibinfo {author} {\bibfnamefont {F.}~\bibnamefont
  {{Herrmann}}}, \bibinfo {author} {\bibfnamefont {P.}~\bibnamefont
  {{Laguna}}},\ and\ \bibinfo {author} {\bibfnamefont {D.}~\bibnamefont
  {{Shoemaker}}},\ }\bibfield  {title} {\bibinfo {title} {{Comparisons of
  eccentric binary black hole simulations with post-Newtonian models}},\ }\href
  {https://doi.org/10.1103/PhysRevD.82.024033} {\bibfield  {journal} {\bibinfo
  {journal} {\prd}\ }\textbf {\bibinfo {volume} {82}},\ \bibinfo {eid} {024033}
  (\bibinfo {year} {2010})},\ \Eprint {https://arxiv.org/abs/0806.1037}
  {arXiv:0806.1037 [gr-qc]} \BibitemShut {NoStop}%
\bibitem [{\citenamefont {{Klein}}\ \emph {et~al.}(2018)\citenamefont
  {{Klein}}, \citenamefont {{Boetzel}}, \citenamefont {{Gopakumar}},
  \citenamefont {{Jetzer}},\ and\ \citenamefont {{de Vittori}}}]{KBG18}%
  \BibitemOpen
  \bibfield  {author} {\bibinfo {author} {\bibfnamefont {A.}~\bibnamefont
  {{Klein}}}, \bibinfo {author} {\bibfnamefont {Y.}~\bibnamefont {{Boetzel}}},
  \bibinfo {author} {\bibfnamefont {A.}~\bibnamefont {{Gopakumar}}}, \bibinfo
  {author} {\bibfnamefont {P.}~\bibnamefont {{Jetzer}}},\ and\ \bibinfo
  {author} {\bibfnamefont {L.}~\bibnamefont {{de Vittori}}},\ }\bibfield
  {title} {\bibinfo {title} {{Fourier domain gravitational waveforms for
  precessing eccentric binaries}},\ }\href
  {https://doi.org/10.1103/PhysRevD.98.104043} {\bibfield  {journal} {\bibinfo
  {journal} {\prd}\ }\textbf {\bibinfo {volume} {98}},\ \bibinfo {eid} {104043}
  (\bibinfo {year} {2018})},\ \Eprint {https://arxiv.org/abs/1801.08542}
  {arXiv:1801.08542 [gr-qc]} \BibitemShut {NoStop}%
\bibitem [{\citenamefont {Poisson}\ and\ \citenamefont
  {Will}(2014)}]{poisson2014gravity}%
  \BibitemOpen
  \bibfield  {author} {\bibinfo {author} {\bibfnamefont {E.}~\bibnamefont
  {Poisson}}\ and\ \bibinfo {author} {\bibfnamefont {C.}~\bibnamefont {Will}},\
  }\href {https://books.google.com/books?id=PZ5cAwAAQBAJ} {\emph {\bibinfo
  {title} {Gravity: Newtonian, Post-Newtonian, Relativistic}}}\ (\bibinfo
  {publisher} {Cambridge University Press},\ \bibinfo {year}
  {2014})\BibitemShut {NoStop}%
\bibitem [{\citenamefont {{Blanchet}}\ and\ \citenamefont
  {{Schaefer}}(1989{\natexlab{b}})}]{BS89}%
  \BibitemOpen
  \bibfield  {author} {\bibinfo {author} {\bibfnamefont {L.}~\bibnamefont
  {{Blanchet}}}\ and\ \bibinfo {author} {\bibfnamefont {G.}~\bibnamefont
  {{Schaefer}}},\ }\bibfield  {title} {\bibinfo {title} {{Higher order
  gravitational radiation losses in binary systems.}},\ }\href
  {https://doi.org/10.1093/mnras/239.3.845} {\bibfield  {journal} {\bibinfo
  {journal} {\mnras}\ }\textbf {\bibinfo {volume} {239}},\ \bibinfo {pages}
  {845} (\bibinfo {year} {1989}{\natexlab{b}})}\BibitemShut {NoStop}%
\bibitem [{\citenamefont {{Gopakumar}}\ and\ \citenamefont
  {{Iyer}}(1997{\natexlab{b}})}]{gopu1997}%
  \BibitemOpen
  \bibfield  {author} {\bibinfo {author} {\bibfnamefont {A.}~\bibnamefont
  {{Gopakumar}}}\ and\ \bibinfo {author} {\bibfnamefont {B.~R.}\ \bibnamefont
  {{Iyer}}},\ }\bibfield  {title} {\bibinfo {title} {{Gravitational waves from
  inspiraling compact binaries: Angular momentum flux, evolution of the orbital
  elements, and the waveform to the second post-Newtonian order}},\ }\href
  {https://doi.org/10.1103/PhysRevD.56.7708} {\bibfield  {journal} {\bibinfo
  {journal} {\prd}\ }\textbf {\bibinfo {volume} {56}},\ \bibinfo {pages} {7708}
  (\bibinfo {year} {1997}{\natexlab{b}})},\ \Eprint
  {https://arxiv.org/abs/gr-qc/9710075} {gr-qc/9710075} \BibitemShut {NoStop}%
\bibitem [{\citenamefont {{Mikkola}}(1987)}]{Mikkola}%
  \BibitemOpen
  \bibfield  {author} {\bibinfo {author} {\bibfnamefont {S.}~\bibnamefont
  {{Mikkola}}},\ }\bibfield  {title} {\bibinfo {title} {{A cubic approximation
  for Kepler's equation}},\ }\href {https://doi.org/10.1007/BF01235850}
  {\bibfield  {journal} {\bibinfo  {journal} {Celestial Mechanics}\ }\textbf
  {\bibinfo {volume} {40}},\ \bibinfo {pages} {329} (\bibinfo {year}
  {1987})}\BibitemShut {NoStop}%
\bibitem [{\citenamefont {{Tanay}}\ \emph {et~al.}(2016)\citenamefont
  {{Tanay}}, \citenamefont {{Haney}},\ and\ \citenamefont
  {{Gopakumar}}}]{Tanay2016}%
  \BibitemOpen
  \bibfield  {author} {\bibinfo {author} {\bibfnamefont {S.}~\bibnamefont
  {{Tanay}}}, \bibinfo {author} {\bibfnamefont {M.}~\bibnamefont {{Haney}}},\
  and\ \bibinfo {author} {\bibfnamefont {A.}~\bibnamefont {{Gopakumar}}},\
  }\bibfield  {title} {\bibinfo {title} {{Frequency and time-domain inspiral
  templates for comparable mass compact binaries in eccentric orbits}},\ }\href
  {https://doi.org/10.1103/PhysRevD.93.064031} {\bibfield  {journal} {\bibinfo
  {journal} {\prd}\ }\textbf {\bibinfo {volume} {93}},\ \bibinfo {eid} {064031}
  (\bibinfo {year} {2016})},\ \Eprint {https://arxiv.org/abs/1602.03081}
  {arXiv:1602.03081 [gr-qc]} \BibitemShut {NoStop}%
\bibitem [{\citenamefont {Mroue}\ \emph {et~al.}(2013)\citenamefont {Mroue}
  \emph {et~al.}}]{Mroue:2013xna}%
  \BibitemOpen
  \bibfield  {author} {\bibinfo {author} {\bibfnamefont {A.~H.}\ \bibnamefont
  {Mroue}} \emph {et~al.},\ }\bibfield  {title} {\bibinfo {title} {{Catalog of
  174 Binary Black Hole Simulations for Gravitational Wave Astronomy}},\ }\href
  {https://doi.org/10.1103/PhysRevLett.111.241104} {\bibfield  {journal}
  {\bibinfo  {journal} {Phys. Rev. Lett.}\ }\textbf {\bibinfo {volume} {111}},\
  \bibinfo {pages} {241104} (\bibinfo {year} {2013})},\ \Eprint
  {https://arxiv.org/abs/1304.6077} {arXiv:1304.6077 [gr-qc]} \BibitemShut
  {NoStop}%
\bibitem [{\citenamefont {{Tanay}}\ \emph {et~al.}(2019)\citenamefont
  {{Tanay}}, \citenamefont {{Klein}}, \citenamefont {{Berti}},\ and\
  \citenamefont {{Nishizawa}}}]{2019PhRvD.100f4006T}%
  \BibitemOpen
  \bibfield  {author} {\bibinfo {author} {\bibfnamefont {S.}~\bibnamefont
  {{Tanay}}}, \bibinfo {author} {\bibfnamefont {A.}~\bibnamefont {{Klein}}},
  \bibinfo {author} {\bibfnamefont {E.}~\bibnamefont {{Berti}}},\ and\ \bibinfo
  {author} {\bibfnamefont {A.}~\bibnamefont {{Nishizawa}}},\ }\bibfield
  {title} {\bibinfo {title} {{Convergence of Fourier-domain templates for
  inspiraling eccentric compact binaries}},\ }\href
  {https://doi.org/10.1103/PhysRevD.100.064006} {\bibfield  {journal} {\bibinfo
   {journal} {\prd}\ }\textbf {\bibinfo {volume} {100}},\ \bibinfo {eid}
  {064006} (\bibinfo {year} {2019})},\ \Eprint
  {https://arxiv.org/abs/1905.08811} {arXiv:1905.08811 [gr-qc]} \BibitemShut
  {NoStop}%
\bibitem [{\citenamefont {Damour}\ \emph
  {et~al.}(2000{\natexlab{b}})\citenamefont {Damour}, \citenamefont {Iyer},\
  and\ \citenamefont {Sathyaprakash}}]{DIS}%
  \BibitemOpen
  \bibfield  {author} {\bibinfo {author} {\bibfnamefont {T.}~\bibnamefont
  {Damour}}, \bibinfo {author} {\bibfnamefont {B.~R.}\ \bibnamefont {Iyer}},\
  and\ \bibinfo {author} {\bibfnamefont {B.~S.}\ \bibnamefont
  {Sathyaprakash}},\ }\bibfield  {title} {\bibinfo {title} {{Frequency domain P
  approximant filters for time truncated inspiral gravitational wave signals
  from compact binaries}},\ }\href {https://doi.org/10.1103/PhysRevD.62.084036}
  {\bibfield  {journal} {\bibinfo  {journal} {Phys. Rev. D}\ }\textbf {\bibinfo
  {volume} {62}},\ \bibinfo {pages} {084036} (\bibinfo {year}
  {2000}{\natexlab{b}})},\ \Eprint {https://arxiv.org/abs/gr-qc/0001023}
  {arXiv:gr-qc/0001023} \BibitemShut {NoStop}%
\bibitem [{aLI()}]{aLIGOpsd}%
  \BibitemOpen
  \href@noop {} {}\bibinfo {howpublished}
  {\url{https://dcc.ligo.org/public/0002/T0900288/003/ZERO_DET_high_P.txt/}}\BibitemShut
  {NoStop}%
\bibitem [{\citenamefont {{Apostolatos}}(1995)}]{1995PhRvD52605A}%
  \BibitemOpen
  \bibfield  {author} {\bibinfo {author} {\bibfnamefont {T.~A.}\ \bibnamefont
  {{Apostolatos}}},\ }\bibfield  {title} {\bibinfo {title} {{Search templates
  for gravitational waves from precessing, inspiraling binaries}},\ }\href
  {https://doi.org/10.1103/PhysRevD.52.605} {\bibfield  {journal} {\bibinfo
  {journal} {\prd}\ }\textbf {\bibinfo {volume} {52}},\ \bibinfo {pages} {605}
  (\bibinfo {year} {1995})}\BibitemShut {NoStop}%
\bibitem [{\citenamefont {{Owen}}\ and\ \citenamefont
  {{Sathyaprakash}}(1999)}]{1999PhRvD60b2002O}%
  \BibitemOpen
  \bibfield  {author} {\bibinfo {author} {\bibfnamefont {B.~J.}\ \bibnamefont
  {{Owen}}}\ and\ \bibinfo {author} {\bibfnamefont {B.~S.}\ \bibnamefont
  {{Sathyaprakash}}},\ }\bibfield  {title} {\bibinfo {title} {{Matched
  filtering of gravitational waves from inspiraling compact binaries:
  Computational cost and template placement}},\ }\href
  {https://doi.org/10.1103/PhysRevD.60.022002} {\bibfield  {journal} {\bibinfo
  {journal} {\prd}\ }\textbf {\bibinfo {volume} {60}},\ \bibinfo {eid} {022002}
  (\bibinfo {year} {1999})},\ \Eprint {https://arxiv.org/abs/gr-qc/9808076}
  {arXiv:gr-qc/9808076 [gr-qc]} \BibitemShut {NoStop}%
\bibitem [{\citenamefont {{Blanchet}}\ \emph {et~al.}(2006)\citenamefont
  {{Blanchet}}, \citenamefont {{Buonanno}},\ and\ \citenamefont
  {{Faye}}}]{SO_1PN}%
  \BibitemOpen
  \bibfield  {author} {\bibinfo {author} {\bibfnamefont {L.}~\bibnamefont
  {{Blanchet}}}, \bibinfo {author} {\bibfnamefont {A.}~\bibnamefont
  {{Buonanno}}},\ and\ \bibinfo {author} {\bibfnamefont {G.}~\bibnamefont
  {{Faye}}},\ }\bibfield  {title} {\bibinfo {title} {{Higher-order spin effects
  in the dynamics of compact binaries. II. Radiation field}},\ }\href
  {https://doi.org/10.1103/PhysRevD.74.104034} {\bibfield  {journal} {\bibinfo
  {journal} {\prd}\ }\textbf {\bibinfo {volume} {74}},\ \bibinfo {eid} {104034}
  (\bibinfo {year} {2006})},\ \Eprint {https://arxiv.org/abs/gr-qc/0605140}
  {arXiv:gr-qc/0605140 [gr-qc]} \BibitemShut {NoStop}%
\bibitem [{\citenamefont {{Cho}}\ \emph {et~al.}(2021)\citenamefont {{Cho}},
  \citenamefont {{Pardo}},\ and\ \citenamefont {{Porto}}}]{SS_1PN}%
  \BibitemOpen
  \bibfield  {author} {\bibinfo {author} {\bibfnamefont {G.}~\bibnamefont
  {{Cho}}}, \bibinfo {author} {\bibfnamefont {B.}~\bibnamefont {{Pardo}}},\
  and\ \bibinfo {author} {\bibfnamefont {R.~A.}\ \bibnamefont {{Porto}}},\
  }\bibfield  {title} {\bibinfo {title} {{Gravitational radiation from
  inspiralling compact objects: Spin-spin effects completed at the
  next-to-leading post-Newtonian order}},\ }\href
  {https://doi.org/10.1103/PhysRevD.104.024037} {\bibfield  {journal} {\bibinfo
   {journal} {\prd}\ }\textbf {\bibinfo {volume} {104}},\ \bibinfo {eid}
  {024037} (\bibinfo {year} {2021})},\ \Eprint
  {https://arxiv.org/abs/2103.14612} {arXiv:2103.14612 [gr-qc]} \BibitemShut
  {NoStop}%
\bibitem [{\citenamefont {{Bini}}\ \emph {et~al.}(2019)\citenamefont {{Bini}},
  \citenamefont {{Damour}},\ and\ \citenamefont
  {{Geralico}}}]{2019PhRvL.123w1104B}%
  \BibitemOpen
  \bibfield  {author} {\bibinfo {author} {\bibfnamefont {D.}~\bibnamefont
  {{Bini}}}, \bibinfo {author} {\bibfnamefont {T.}~\bibnamefont {{Damour}}},\
  and\ \bibinfo {author} {\bibfnamefont {A.}~\bibnamefont {{Geralico}}},\
  }\bibfield  {title} {\bibinfo {title} {{Novel Approach to Binary Dynamics:
  Application to the Fifth Post-Newtonian Level}},\ }\href
  {https://doi.org/10.1103/PhysRevLett.123.231104} {\bibfield  {journal}
  {\bibinfo  {journal} {\prl}\ }\textbf {\bibinfo {volume} {123}},\ \bibinfo
  {eid} {231104} (\bibinfo {year} {2019})},\ \Eprint
  {https://arxiv.org/abs/1909.02375} {arXiv:1909.02375 [gr-qc]} \BibitemShut
  {NoStop}%
\bibitem [{\citenamefont {{Bini}}\ \emph
  {et~al.}(2020{\natexlab{b}})\citenamefont {{Bini}}, \citenamefont
  {{Damour}},\ and\ \citenamefont {{Geralico}}}]{2020PhRvD.102b4062B}%
  \BibitemOpen
  \bibfield  {author} {\bibinfo {author} {\bibfnamefont {D.}~\bibnamefont
  {{Bini}}}, \bibinfo {author} {\bibfnamefont {T.}~\bibnamefont {{Damour}}},\
  and\ \bibinfo {author} {\bibfnamefont {A.}~\bibnamefont {{Geralico}}},\
  }\bibfield  {title} {\bibinfo {title} {{Binary dynamics at the fifth and
  fifth-and-a-half post-Newtonian orders}},\ }\href
  {https://doi.org/10.1103/PhysRevD.102.024062} {\bibfield  {journal} {\bibinfo
   {journal} {\prd}\ }\textbf {\bibinfo {volume} {102}},\ \bibinfo {eid}
  {024062} (\bibinfo {year} {2020}{\natexlab{b}})},\ \Eprint
  {https://arxiv.org/abs/2003.11891} {arXiv:2003.11891 [gr-qc]} \BibitemShut
  {NoStop}%
\bibitem [{\citenamefont {{Bini}}\ \emph
  {et~al.}(2020{\natexlab{c}})\citenamefont {{Bini}}, \citenamefont
  {{Damour}},\ and\ \citenamefont {{Geralico}}}]{2020PhRvD.102b4061B}%
  \BibitemOpen
  \bibfield  {author} {\bibinfo {author} {\bibfnamefont {D.}~\bibnamefont
  {{Bini}}}, \bibinfo {author} {\bibfnamefont {T.}~\bibnamefont {{Damour}}},\
  and\ \bibinfo {author} {\bibfnamefont {A.}~\bibnamefont {{Geralico}}},\
  }\bibfield  {title} {\bibinfo {title} {{Sixth post-Newtonian local-in-time
  dynamics of binary systems}},\ }\href
  {https://doi.org/10.1103/PhysRevD.102.024061} {\bibfield  {journal} {\bibinfo
   {journal} {\prd}\ }\textbf {\bibinfo {volume} {102}},\ \bibinfo {eid}
  {024061} (\bibinfo {year} {2020}{\natexlab{c}})},\ \Eprint
  {https://arxiv.org/abs/2004.05407} {arXiv:2004.05407 [gr-qc]} \BibitemShut
  {NoStop}%
\bibitem [{\citenamefont {{Larrouturou}}\ \emph
  {et~al.}(2021{\natexlab{a}})\citenamefont {{Larrouturou}}, \citenamefont
  {{Henry}}, \citenamefont {{Blanchet}},\ and\ \citenamefont
  {{Faye}}}]{4PN_QM_1}%
  \BibitemOpen
  \bibfield  {author} {\bibinfo {author} {\bibfnamefont {F.}~\bibnamefont
  {{Larrouturou}}}, \bibinfo {author} {\bibfnamefont {Q.}~\bibnamefont
  {{Henry}}}, \bibinfo {author} {\bibfnamefont {L.}~\bibnamefont
  {{Blanchet}}},\ and\ \bibinfo {author} {\bibfnamefont {G.}~\bibnamefont
  {{Faye}}},\ }\bibfield  {title} {\bibinfo {title} {{The Quadrupole Moment of
  Compact Binaries to the Fourth post-Newtonian Order: I. Non-Locality in Time
  and Infra-Red Divergencies}},\ }\href@noop {} {\bibfield  {journal} {\bibinfo
   {journal} {arXiv e-prints}\ ,\ \bibinfo {eid} {arXiv:2110.02240}} (\bibinfo
  {year} {2021}{\natexlab{a}})},\ \Eprint {https://arxiv.org/abs/2110.02240}
  {arXiv:2110.02240 [gr-qc]} \BibitemShut {NoStop}%
\bibitem [{\citenamefont {{Larrouturou}}\ \emph
  {et~al.}(2021{\natexlab{b}})\citenamefont {{Larrouturou}}, \citenamefont
  {{Blanchet}}, \citenamefont {{Henry}},\ and\ \citenamefont
  {{Faye}}}]{4PN_QM_2}%
  \BibitemOpen
  \bibfield  {author} {\bibinfo {author} {\bibfnamefont {F.}~\bibnamefont
  {{Larrouturou}}}, \bibinfo {author} {\bibfnamefont {L.}~\bibnamefont
  {{Blanchet}}}, \bibinfo {author} {\bibfnamefont {Q.}~\bibnamefont
  {{Henry}}},\ and\ \bibinfo {author} {\bibfnamefont {G.}~\bibnamefont
  {{Faye}}},\ }\bibfield  {title} {\bibinfo {title} {{The Quadrupole Moment of
  Compact Binaries to the Fourth post-Newtonian Order: II. Dimensional
  Regularization and Renormalization}},\ }\href@noop {} {\bibfield  {journal}
  {\bibinfo  {journal} {arXiv e-prints}\ ,\ \bibinfo {eid} {arXiv:2110.02243}}
  (\bibinfo {year} {2021}{\natexlab{b}})},\ \Eprint
  {https://arxiv.org/abs/2110.02243} {arXiv:2110.02243 [gr-qc]} \BibitemShut
  {NoStop}%
\bibitem [{\citenamefont {Maggiore}(2008)}]{maggiore2008gravitational}%
  \BibitemOpen
  \bibfield  {author} {\bibinfo {author} {\bibfnamefont {M.}~\bibnamefont
  {Maggiore}},\ }\href {https://books.google.com/books?id=mk-1DAAAQBAJ} {\emph
  {\bibinfo {title} {Gravitational Waves: Volume 1: Theory and Experiments}}},\
  Gravitational Waves\ (\bibinfo  {publisher} {OUP Oxford},\ \bibinfo {year}
  {2008})\BibitemShut {NoStop}%
\bibitem [{\citenamefont {Creighton}\ and\ \citenamefont
  {Anderson}(2012)}]{creighton2012gravitational}%
  \BibitemOpen
  \bibfield  {author} {\bibinfo {author} {\bibfnamefont {J.}~\bibnamefont
  {Creighton}}\ and\ \bibinfo {author} {\bibfnamefont {W.}~\bibnamefont
  {Anderson}},\ }\href {https://books.google.com/books?id=W\_TVS\_6JYJcC}
  {\emph {\bibinfo {title} {Gravitational-Wave Physics and Astronomy: An
  Introduction to Theory, Experiment and Data Analysis}}},\ Wiley Series in
  Cosmology\ (\bibinfo  {publisher} {Wiley},\ \bibinfo {year}
  {2012})\BibitemShut {NoStop}%
\bibitem [{\citenamefont {Sathyaprakash}\ and\ \citenamefont
  {Schutz}(2009)}]{Sathyaprakash:2009xs}%
  \BibitemOpen
  \bibfield  {author} {\bibinfo {author} {\bibfnamefont {B.~S.}\ \bibnamefont
  {Sathyaprakash}}\ and\ \bibinfo {author} {\bibfnamefont {B.~F.}\ \bibnamefont
  {Schutz}},\ }\bibfield  {title} {\bibinfo {title} {{Physics, Astrophysics and
  Cosmology with Gravitational Waves}},\ }\href
  {https://doi.org/10.12942/lrr-2009-2} {\bibfield  {journal} {\bibinfo
  {journal} {Living Rev. Rel.}\ }\textbf {\bibinfo {volume} {12}},\ \bibinfo
  {pages} {2} (\bibinfo {year} {2009})},\ \Eprint
  {https://arxiv.org/abs/0903.0338} {arXiv:0903.0338 [gr-qc]} \BibitemShut
  {NoStop}%
\end{thebibliography}%
\end{document}